\documentclass[a4paper,UKenglish]{lipicsnew}

\usepackage{epic,eepic,color,latexsym,amsmath,amssymb}

\def\<<{\ouvreguillemet\everypar={\ouvreguillemet\ }}
\def\ouvreguillemet{\leavevmode\hbox{(\kern-0.20em(\kern+0.20em}\nobreak}
\def\>>{\fermeguillemet\everypar={}}
\def\fermeguillemet{\nobreak\leavevmode\hbox{\kern+0.20em)\kern-0.20em)}}

\newcommand{\haut}[1]{\mbox{\raisebox{1ex}{\scriptsize \it #1}}}
\newcommand{\bBas}[1]{\mbox{\raisebox{-0.5ex}{\tiny \rm #1}}}
\newcommand{\arrow}{\mbox{\small$\longrightarrow$}}
\newcommand{\arrowleft}{\mbox{\small$\longleftarrow$}}

\newcommand{\norme}[1]{\mbox{$\mathop{\parallel}{#1}\mathop{\parallel}$}}
\newcommand{\fleche}[1]{\mbox{$\mathop{\arrow}\limits^{#1}$}}
\newcommand{\nofleche}[2]{\mbox{$\ {\tiny/}\hspace*{-1.1em}\mathop{\arrow}
                                           \limits^{#1}_{\mbox{\tiny$#2$}}$}}
\newcommand{\inverse}[1]{\mbox{$\mathop{\arrowleft}\limits^{#1}$}}
\newcommand{\croix}{\mbox{\scriptsize $\times$}}
\newcommand{\inter}[1]{\hbox{{\rm [}\hskip -2pt {\rm [}{$#1$}{\rm ]}\hskip -2pt {\rm ]}}}
\newcommand{\interInd}[1]{\tiny\hbox{{\rm [}\hskip -3.15pt{\rm [}{$#1$}{\rm ]}
\hskip -3.5pt {\rm ]}}}
\newcommand{\interFootnote}[1]{\footnotesize\hbox{{\rm [}\hskip -1.5pt
{\rm [}{$#1$}{\rm ]}\hskip -1.5pt {\rm ]}}}
\newcommand{\relatif}{\mbox{$\mathbb{Z}$}}
\newcommand{\entier}{\mbox{$\mathbb{N}$}}
\newcommand{\reel}{\mbox{$\mathbb{R}$}}
\newcommand{\smallentier}{\hbox{\scriptsize\rm I\hskip -1.5pt N}}
\newcommand{\InfSup}[1]{\mbox{$\mathop{<}\!{#1}\!\mathop{>}$}}
\newcommand{\InfSupInd}[1]{\tiny\mbox{$\mathop{<}\!{#1}\!\mathop{>}$}}
\newcommand{\Infsup}[1]{\mbox{$\mathop{\ll}\!{#1}\!\mathop{\gg}$}}
\newcommand{\InfsupInd}[1]{\tiny\mbox{$\mathop{\ll}\!{#1}\!\mathop{\gg}$}}

\title{On Cayley graphs of algebraic structures}
\titlerunning{Cayley graphs}
\author[1]{Didier Caucal}
\affil[1]{CNRS, LIGM, France\\
  \texttt{didier.caucal@univ-mlv.fr}}
\authorrunning{D. Caucal}
\begin{document}

\maketitle

\begin{abstract}
{\noindent}We present simple graph-theoretic characterizations of Cayley graphs 
for left-cancellative monoids, groups, left-quasigroups and quasigroups. 
We show that these characterizations are effective for the end-regular graphs 
of finite degree.
\end{abstract}

\section{Introduction}

\hspace*{1.5em}To describe the structure of a group, Cayley introduced in 1878 
\cite{Ca} the concept of graph for any group \,$(G,\cdot)$ \,according to any 
generating subset \,$S$. 
This is simply the set of labeled oriented edges \,$g\ \fleche{s}\ g{\cdot}s$ 
\,for every \,$g$ \,of \,$G$ \,and \,$s$ \,of \,$S$. 
Such a graph, called Cayley graph, is directed and labeled in \,$S$ \,(or an 
encoding of \,$S$ \,by symbols called letters or colors). 
The study of groups by their Cayley graphs is a main topic of algebraic graph 
theory \cite{Bi,GR,BW}. 
A characterization of unlabeled and undirected Cayley graphs was given by 
Sabidussi in 1958 \cite{Sa}\,: an unlabeled and undirected graph is a 
Cayley graph if and only if we can find a group with a free and transitive 
action on the graph. 
However, this algebraic characterization is not well suited for deciding 
whether a possibly infinite graph is a Cayley graph.
It is pertinent to look for characterizations by graph-theoretic conditions. 
This approach was clearly stated by Hamkins in 2010: Which graphs are 
Cayley graphs? \cite{Ha}. 
In this paper, we present simple graph-theoretic characterizations of 
Cayley graphs for firstly left-cancellative and cancellative monoids, and then 
for groups. 
These characterizations are then extended to any subset \,$S$ \,of 
left-cancellative magmas, left-quasigroups, quasigroups, and groups. 
Finally, we show that these characterizations are effective for the end-regular 
graphs of finite degree \cite{MS} which are the graphs finitely decomposable 
by distance from a(ny) vertex or equivalently are the suffix transition graphs 
of labeled word rewriting systems.

Let us present the main structural characterizations starting with the Cayley 
graphs of left-cancellative monoids. 
Among many properties of these graphs, we retain only three basic ones.
First and by definition, any Cayley graph is deterministic: there are no two 
arcs of the same source and label.
Furthermore, the left-cancellative condition implies that any Cayley graph is 
simple: there are no two arcs of the same source and goal. 
Finally, any Cayley graph is rooted: there is a path from the identity 
element to any vertex. 
To these three necessary basic conditions is added a structural property, 
called arc-symmetric: all the vertices are accessible-isomorphic 
\,{\it i.e.} \,the induced subgraphs by vertex accessibility are isomorphic. 
These four properties characterize the Cayley graphs of left-cancellative 
monoids. 
To describe exactly the Cayley graphs of cancellative monoids, we just have to 
add the co-determinism: there are no two arcs of the same target and label. 
This characterization is strengthened for the Cayley graphs of groups using the 
same properties but expressed in both arc directions: these are the graphs 
that are connected, deterministic, co-deterministic, and symmetric: all the 
vertices are isomorphic.

We also consider the Cayley graph of a magma \,$G$ \,according to any subset 
\,$S$ \,and that we called generalized. The characterizations obtained require 
the assumption of the axiom of choice. First, a graph is a generalized Cayley 
graph of a left-cancellative magma if and only if it is deterministic, simple, 
source-complete: for any label of the graph and from any vertex, there is at
least one edge. This equivalence does not require the axiom of choice for 
finitely labeled graphs, and in this case, these graphs are also the 
generalized Cayley graphs of left-quasigroups. Moreover, a finitely labeled 
graph is a generalized Cayley graph of a quasigroup if and only if it is also 
co-deterministic and target-complete: for any label of the graph and to any 
vertex, there is at least one edge. We also characterize all the generalized 
Cayley graphs of left-quasigroups, and of quasigroups. 
Finally, a graph is a generalized Cayley graph of a group if anf only if it is 
simple, symmetric, deterministic and co-deterministic.

\section{Directed labeled graphs}

{\indent}We consider directed labeled graphs without isolated vertex. 
We recall some basic concepts such as determinism, completeness and symmetry. 
We introduce the notions of accessible-isomorphic vertices and arc-symmetric 
graph.\\[-0.5em]

Let \,$A$ \,be an arbitrary (finite or infinite) set. 
A directed \,$A$-{\it graph} \,$(V,G)$ \,is defined by a set \,$V$ \,of 
{\it vertices} \,and a subset \,$G \,\subseteq \,V{\croix}A{\croix}V$ \,of 
{\it edges}. 
Any edge \,$(s,a,t) \in G$ \,is from the {\it source} \,$s$ \,to the 
{\it target} \,$t$ \,with {\it label} \,$a$, and is also written by the 
{\it transition} \,$s\ \fleche{a}_G\ t$ \,or directly \,$s\ \fleche{a}\ t$ \,if 
\,$G$ \,is clear from the context. 
The sources and targets of edges form the set \,$V_G$ \,of 
{\it non-isolated vertices} \,of \,$G$ \,and we denote by \,$A_G$ \,the set of 
edge labels:\\[0.25em]
\hspace*{2em}$V_G\ =\ \{\ s\ |\ \exists\ a,t\ \ (s\ \fleche{a}\ t \,\vee 
\,t\ \fleche{a}\ s)\ \}$ \ \ \ and \ \ \ 
$A_G\ =\ \{\ a\ |\ \exists\ s,t\ \ (s\ \fleche{a}\ t)\ \}$.\\[0.25em]
Thus \,$V - V_G$ \,is the set of {\it isolated vertices}. 
From now on, we assume that any graph \,$(V,G)$ \,is without isolated vertex 
\,({\it i.e.} $V = V_G$), hence the graph can be identified with its edge 
set~\,$G$. We also exclude the empty graph \,$\emptyset$\,: every graph is a 
non-empty set of labeled edges. 
For instance \,$\Upsilon\ =\ \{\ s\ \fleche{n}\ s+n\ |\ s \in \reel \,\wedge 
\,n \in \relatif\ \}$ \,is a graph of vertex set \,$\reel$ \,and of label set
\,$\relatif$. 
As any graph \,$G$ \,is a set, there are no two edges with the same source,
target and label. We say that a graph is {\it simple} \,if there are no two 
edges with the same source and target: 
\,$(s\ \fleche{a}\ t \,\wedge \,s\ \fleche{b}\ t)\ \Longrightarrow\ a=b$. 
We say that \,$G$ \,is {\it finitely labeled} \,if \,$A_G$ \,is finite. 
We denote by \,$G^{-1}\ =\ \{\ (t,a,s)\ |\ (s,a,t) \in G\ \}$ \,the 
{\it inverse} \,of \,$G$. 
A graph is {\it deterministic} \,if there are no two edges with the same source 
and label: \,$(r\ \fleche{a}\ s \,\wedge \,r\ \fleche{a}\ t)\ \Longrightarrow\ 
s=t$. A graph is {\it co-deterministic} \,if its inverse is deterministic:
there are no two edges with the same target and label: 
\,$(s\ \fleche{a}\ r \,\wedge \,t\ \fleche{a}\ r)\ \Longrightarrow\ s=t$. 
For instance, the graph \,$\Upsilon$ \,is simple, not finitely labeled, 
deterministic and co-deterministic. 
A graph \,$G$ \,is {\it complete} \,if there is an edge between any couple of 
vertices: $\forall\ s,t \in V_G\ \ \exists\ a \in A_G \ (s\ \fleche{a}_G\ t)$. 
A graph \,$G$ \,is {\it source-complete} \,if for all vertex \,$s$
\,and label \,$a$, there is an \,$a$-edge from \,$s$\,: $\forall\ s \in V_G\ \ 
\forall\ a \in A_G\ \ \exists\ t \ (s\ \fleche{a}_G\ t)$. 
A graph is {\it target-complete} \,if its inverse is source-complete: 
\,$\forall\ t \in V_G\ \ \forall\ a \in A_G\ \ \exists\ s \ 
(s\ \fleche{a}_G\ t)$. 
For instance, \,$\Upsilon$ \,is source-complete, target-complete but not 
complete. Another example is given by the graph 
\,${\rm Even} \ = \ \{(p,a,q)\ ,\ (p,b,p)\ ,\ (q,a,p)\ ,\ (q,b,q)\}$
\,represented as follows:
\begin{center}
\includegraphics{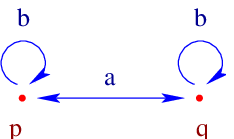}\\%magnitude 0.3  
\end{center}
It is simple, deterministic, co-deterministic, complete, source-complete and 
target-complete.\\
The {\it vertex-restriction} \,$G_{|P}$ \,of \,$G$ \,to a set \,$P$ \,is the 
induced subgraph of \,$G$ \,by \,$P \cap V_G$\,:\\[0.25em]
\hspace*{12em}$G_{|P}\ =\ \{\ (s,a,t) \in G\ |\ s,t \in P\ \}$.\\[0.25em]
The {\it label-restriction} \,$G^{|P}$ \,of \,$G$ \,to a set \,$P$ \,is the 
subset of all its edges labeled in \,$P$\,:\\[0.25em]
\hspace*{12em}$G^{|P}\ =\ \{\ (s,a,t) \in G\ |\ a \in P\ \}$.\\[0.25em]
Let \,$\fleche{}_G$ \,be the unlabeled edge relation \,{\it i.e.}
\,$s\ \fleche{}_G\ t$ \,if \,$s\ \fleche{a}_G\ t$ \,for some \,$a \in A$. 
We denote by \,$\fleche{}_G(s) \,= \,\{\ t\ |\ s\ \fleche{}_G\ t\ \}$ \,the set 
of {\it successors} \,of \,$s \in V_G$\,. 
We write \,$s\ \nofleche{}{}\,\!_G\ t$ \,if there is no edge in \,$G$ \,from 
\,$s$ \,to \,$t$ \,{\it i.e.} 
\,$G \,\cap \,\{s\}{\croix}A{\croix}\{t\} \,= \,\emptyset$. 
The {\it accessibility} \,relation 
\,$\fleche{}^*_G \,= \,\bigcup_{n \geq 0}\fleche{}^n_G$ \,is the reflexive and 
transitive closure under composition of \,$\fleche{}_G$\,. 
A graph \,$G$ \,is {\it accessible} \,from \,$P \subseteq V_G$ \,if for any
\,$s \in V_G$\,, there is \,$r \in P$ \,such that \,$r\ \fleche{}^*_G\ s$. 
We denote by \,$G_{{\downarrow}P}$ \,the induced subgraph of \,$G$ \,to the 
vertices accessible from \,$P$ \,which is the greatest subgraph of \,$G$ 
\,accessible from \,$P$. 
For instance \,$\Upsilon_{{\downarrow}\{0\}}\ =\ 
\{\ m\ \fleche{n}\ m+n\ |\ m,n \in \relatif\ \}$ \,is a complete subgraph of 
\,$\Upsilon$. 
A {\it root} \,$r$ \,is a vertex from which \,$G$ \,is accessible \,{\it i.e.} 
\,$G_{{\downarrow}\{r\}}$ \,also denoted by \,$G_{{\downarrow}r}$ \,is equal to 
\,$G$. 
A graph \,$G$ \,is {\it strongly connected} \,if every vertex is a root: 
\,$s\ \fleche{}^*_G\ t$ \ \ for all \,$s,t \in V_G$\,. 
A graph \,$G$ \,is {\it co-accessible} \,from \,$P \subseteq V_G$ \,if 
\,$G^{-1}$ \,is accessible from \,$P$. 
We denote by \,$d_G(s,t) \,= 
\,{\rm min}\{\ n\ |\ s\ \fleche{}^n_{G\,\cup\,G^{-1}}\ t\ \}$ \,the
{\it distance} \,between \,$s,t \in V_G$ \,with \,min$(\emptyset) = \omega$. 
A graph \,$G$ \,is {\it connected} \,if \,$G \,\cup \,G^{-1}$ \,is strongly 
connected \,{\it i.e.} \,$d_G(s,t) \in \entier$ \ \ for any \,$s,t \in V_G$\,. 
Recall that a {\it connected component} \,of a graph \,$G$ \,is a maximal 
connected subset of \,$G$\,; \,we denote by \,${\rm Comp}(G)$ \,the set of 
connected components of \,$G$. 
A {\it representative set} \,of \,Comp$(G)$ \,is a vertex subset 
\,$P \subseteq V_G$ \,having exactly one vertex in each connected component: 
\,$|P \,\cap \,V_C| = 1$ \,for any \,$C \in {\rm Comp}(G)$\,; \,it induces the 
{\it canonical mapping} \,$\pi_P : V_G\ \fleche{}\ P$ \,associating with each 
vertex \,$s$ \,the vertex of \,$P$ \,in the same connected component: 
\,$s\ \fleche{}^*_{G\,\cup\,G^{-1}}\ \pi_P(s)$ \,for any \,$s \in V_G$\,. 
For instance, \,$[0,1[$ \,is a representative set of \,${\rm Comp}(\Upsilon)$ 
\,and its canonical mapping is defined by 
\,$\pi_{[0,1[}(x) \,= \,x - {\lfloor}x{\rfloor}$ \,for any \,$x \in \reel$.\\
A {\it path} \,$(s_0,a_1,s_1,\ldots,a_n,s_n)$ \,of {\it length} \,$n \geq 0$
\,in a graph \,$G$ \,is a sequence 
\,$s_0\ \fleche{a_1}\ s_1 \ldots \fleche{a_n}\ s_n$ \,of \,$n$ \,consecutive 
edges, and we write \,$s_0\ \fleche{a_1{\ldots}a_n}\ s_n$ \,for indicating the 
source \,$s_0$\,, \,the target \,$s_n$ \,and the label word 
\,$a_1{\ldots}a_n \in A_G^*$ \,of the path where \,$A_G^*$ \,is the set of 
words over \,$A_G$ \,(the free monoid generated by \,$A_G$) \,and 
\,$\varepsilon$ \,is the empty word (the identity element).\\
Recall that a {\it morphism} \,from a graph \,$G$ \,into a graph \,$H$ \,is a 
mapping \,$h$ \,from \,$V_G$ \,into \,$V_H$ \,such that 
\,$s\ \fleche{a}_G\ t \ \Longrightarrow \ h(s)\ \fleche{a}_H\ h(t)$. 
If, in addition \,$h$ \,is bijective and \,$h^{-1}$ \,is a morphism, $h$ \,is 
called an {\it isomorphism} \,from \,$G$ \,to \,$H$\,; \,we write 
\,$G \,\equiv_h \,H$ \,or directly \,$G \,\equiv \,H$ \,if we do not specify 
an isomorphism, and we say that \,$G$ \,and \,$H$ \,are {\it isomorphic}. 
An {\it automorphism} \,of \,$G$ \,is an isomorphism from \,$G$ \,to \,$G$. 
Two vertices \,$s,t$ \,of a graph \,$G$ \,are {\it isomorphic} 
\,and we write \,$s \,\simeq_G t$ \,if \,$t = h(s)$ \,for some automorphism 
\,$h$ \,of \,$G$.\\
A graph \,$G$ \,is {\it symmetric} (or vertex-transitive) \,if all its 
vertices are isomorphic: \,$s \,\simeq_G t$ \,for every \,$s,t \in V_G$\,. 
For instance, the previous graphs \,$\Upsilon$ \,and \,Even \,are symmetric.\\
Two vertices \,$s,t$ \,of a graph \,$G$ \,are {\it accessible-isomorphic} 
\,and we write \,$s \downarrow_G t$ \,if \,$t = h(s)$ \,for some isomorphism 
\,$h$ \,from \,$G_{{\downarrow}s}$ \,to \,$G_{{\downarrow}t}$\,. 
A graph \,$G$ \,is {\it arc-symmetric} \,if all its vertices are 
accessible-isomorphic: \,$s \downarrow_G t$ \,for every \,$s,t \in V_G$\,.
\begin{fact}\label{AccessibleTransitive}
Any symmetric graph is arc-symmetric which is source-complete.
\end{fact}
For instance \,$\Upsilon^{|\{1\}}_{|\smallentier}\ =\ 
\{\ n\ \fleche{1}\ n+1\ |\ n \in \entier\ \}$ \,is arc-symmetric but not 
symmetric. 
On the other hand \,$\Upsilon^{|\{-1\}}_{|\smallentier}\ =\ 
\{\ n\ \fleche{-1}\ n-1\ |\ n \in \entier\ \}$ \,is not arc-symmetric: two
distinct vertices are not accessible-isomorphic.

\section{Cayley graphs of left-cancellative and cancellative monoids}
\label{CayMon}

We present graph-theoretic characterizations for the Cayley graphs of 
left-cancellative monoids (Theorem~\ref{LeftMonoid}), of cancellative monoids 
(Theorem~\ref{Monoid}), of cancellative semigroups (Theorem~\ref{CanSemigroup}).
\\[-0.5em]

A {\it magma} \,(or groupoid) is a set \,$M$ \,equipped with a binary 
operation \,$\cdot\,: M{\croix}M\ \fleche{}\ M$ \,that sends any two elements 
\,$p,q \in M$ \,to the element \,$p \cdot q$.\\
Given a subset \,$Q \subseteq M$ \,and an injective mapping 
\,$\inter{\ }\,: Q\ \fleche{}\ A$, we define the graph\\[0.25em]
\hspace*{10em}
${\cal C}\inter{M,Q}\ =\ \{\ p\ \fleche{\interFootnote{q}}\ p \cdot q\ |\ 
p \in M \,\wedge \,q \in Q\ \}$\\[0.25em]
which is called a {\it generalized Cayley graph} \,of \,$M$. 
It is of  vertex set \,$M$ \,and of label set 
\,$\inter{Q} \,= \,\{\ \inter{q}\ |\ q \in Q\ \}$. 
We denote \,${\cal C}\inter{M,Q}$ \,by \,${\cal C}(M,Q)$ \,when 
\,$\inter{\ }$ \,is the identity. 
For instance \,$\Upsilon \,= \,{\cal C}(\reel,\relatif)$ \,for the magma 
\,$(\reel,+)$. 
We also write \,${\cal C}\inter{M}$ \,instead of \,${\cal C}\inter{M,M}$ \,and 
\,${\cal C}(M) \,= \,{\cal C}(M,M) \,= \,\{\ p\ \fleche{q}\ p \cdot q\ |\ 
p,q \in M\ \}$.
\begin{fact}\label{Magma}
Any generalized Cayley graph is deterministic and source-complete.
\end{fact}
For instance taking the magma \,$(\relatif,-)$ \,and
\,$\inter{-1} = a$, ${\cal C}\inter{\relatif,\{-1\}}\ =\ 
\{\ n\ \fleche{a}\ n+1\ |\ n \in \relatif\ \}$. 
By adding \,$\inter{1} = b$, 
\,${\cal C}\inter{\relatif,\{1,-1\}} \,=
\,\{\ n\ \fleche{a}\ n+1\ |\ n \in \relatif\ \} \,\cup
\,\{\ n\ \fleche{b}\ n-1\ |\ n \in \relatif\ \}$.\\
We say that a magma \,$(M,\cdot)$ \,is {\it left-cancellative} \,if 
\,$r \cdot p = r \cdot q \ \Longrightarrow \ p = q$ \,for any \,$p,q,r \in M$.\\
Similarly \,$(M,\cdot)$ \,is {\it right-cancellative} \,if 
\,$p \cdot r = q \cdot r \ \Longrightarrow \ p = q$ \,for any \,$p,q,r \in M$.\\
A magma is {\it cancellative} \,if it is both left-cancellative and 
right-cancellative.
\begin{fact}\label{CancelMagma}
Any generalized Cayley graph of a left-cancellative magma is simple.\\
\hspace*{4.8em}Any generalized Cayley graph of a right-cancellative magma is 
co-deterministic.
\end{fact}
Recall also that \,$(M,\cdot)$ \,is a {\it semigroup} \,if \,$\cdot$ \,is 
associative: \,$(p \cdot q) \cdot r \,= \,p \cdot (q \cdot r)$ \,for any 
\,$p,q,r \in M$. 
A {\it monoid} \,$(M,\cdot)$ \,is a semigroup with an {\it identity} element 
\,$1$\,: \,$1{\cdot}p \,= \,p{\cdot}1 \,= \,p$ \,for all \,$p \in M$. 
The {\it submonoid generated} \,by \,$Q \subseteq M$ \,is the least submonoid 
\,$Q^* \,= \,\{\ q_1{\cdot}\ldots{\cdot}q_n \mid n \geq 0 \,\wedge 
\,q_1,\ldots,q_n \in Q\ \}$ \,containing \,$Q$.\\
When a monoid is left-cancellative, its generalized Cayley graphs are 
arc-symmetric.
\begin{proposition}\label{Semigroup}
Any generalized Cayley graph of a left-cancellative monoid is 
arc-symmetric.
\end{proposition}
\proof\mbox{}\\
Let \,$G \,= \,{\cal C}\inter{M,Q}$ \,for some left-cancellative monoid 
\,$(M,\cdot)$ \,and some \,$Q \subseteq M$.\\
Let \,$r \in M$. We have to check that \,$1 \downarrow_G r$.\\
By induction on \,$n \geq 0$ \,and for any \,$q_1,\ldots,q_n \in Q$ \,and 
\,$s \in M$, we have\\[0.25em]
\hspace*{10em}$r\ \fleche{\interInd{q_1}\ldots\interInd{q_n}}_G\ s \ \ 
\Longleftrightarrow \ \ s \,= \,({\ldots}(r{\cdot}q_1){\ldots}){\cdot}q_n$\,.
\\[0.25em]
As \,$\cdot$ \,is associative, we get $V_{G_{{\downarrow}r}} \,= 
\,\{\ s \mid r\ \fleche{}^*_G\ s\ \} \,= \,r{\cdot}Q^*$. 
In particular \,$V_{G_{{\downarrow}1}} \,= \,Q^*$.\\
We consider the mapping \,$f_r : \,M\ \fleche{}\ M$ \,defined by 
\,$f_r(p) = r{\cdot}p$ \,for any \,$p \in M$.\\
As \,$\cdot$ \,is left-cancellative, \,$f_r$ \,is injective.\\
Furthermore \,$f_r$ \,is an isomorphism on its image: for any \,$p,q,p' \in M$,
\\[0.25em]
\hspace*{10em}$p\ \fleche{\interInd{q}}_G\ p' \ \ \Longleftrightarrow \ \ 
f_r(p)\ \fleche{\interInd{q}}_G\ f_r(p')$.\\[0.25em]
The associativity of \,$\cdot$ \,gives the necessary condition.\\
The associativity and the left-cancellative property of  \,$\cdot$ \,gives the 
sufficient condition.\\
Thus \,$f_r$ \,restricted to \,$Q^*$ \,is an isomorphism from 
\,$G_{{\downarrow}1}$ \,to \,$G_{{\downarrow}r}$ \,hence 
\,$1 \downarrow_G \,r$.
\qed\\[1em]
We can not generalize Proposition~\ref{Semigroup} to the left-cancellative 
semigroups. For instance the semigroup \,$M = \{a,b\}$ \,with \,$x{\cdot}y = y$ 
\,for any \,$x,y \in M$ \,is left-cancellative but the graph \,${\cal C}(M)$ 
\,represented below is not arc-symmetric.
\begin{center}
\includegraphics{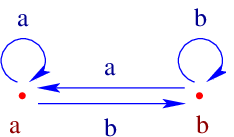}\\%magnitude 0.3
\end{center}
A {\it monoid Cayley graph} \,is a generalized Cayley graph 
\,${\cal C}\inter{M,Q}$ \,of a monoid \,$M$ \,generated by \,$Q$ \,which means
that the identity element \,$1$ \,is a root of \,${\cal C}\inter{M,Q}$. 
\begin{fact}\label{RootCayley}
A monoid \,$M$ \,is generated by \,$Q$ \ \ $\Longleftrightarrow$ \ \ 
$1$ \,is a root of \,${\cal C}\inter{M,Q}$.
\end{fact}
Under additional simple conditions, let us establish the converse of 
Proposition~\ref{Semigroup}.\\
For any graph \,$G$ \,and any vertex \,$r$, we introduce the 
{\it path-relation} \,${\rm Path}_G(r)$ \,as the ternary relation on \,$V_G$ 
\,defined by\\[0.25em]
\hspace*{3em}$(s,t,x) \in {\rm Path}_G(r)$ \ if there exists \,$u \in A_G^*$ 
\,such that \,$r\ \fleche{u}_G\ t$ \,and \,$s\ \fleche{u}_G\ x$.\\[0.25em]
If for any \,$s,t \in V_G$ \,there exists a unique \,$x$ \,such that 
\,$(s,t,x) \in {\rm Path}_G(r)$, we denote by 
\,$\ast_r : V_G{\croix}V_G\ \fleche{}\ V_G$ \,the binary {\it path-operation} 
\,on \,$V_G$ \,defined by \,$(s,t,s \ast_r t) \in {\rm Path}_G(r)$ \,for any 
\,$s,t \in V_G$\,; we also write \,$_G\ast_r$ \,when we need to specify \,$G$. 
Let us give conditions so that this path-operation exists and is associative 
and left-cancellative.
\begin{proposition}\label{PathProduct}
Let \,$r$ \,be a root of a deterministic and arc-symmetric graph \,$G$.
\\
\hspace*{0.5em}Then \,$(V_G,\ast_r)$ \,is a left-cancellative monoid of 
identity \,$r$ \,and generated by \ $\fleche{}_G(r)$.\\
\hspace*{0.5em}If \,$G$ \,is co-deterministic \,then \ $\ast_r$ \,is 
cancellative.\\
\hspace*{0.5em}If \,$G$ \,is simple \,then \ 
$G \,= \,{\cal C}\inter{V_G,\fleche{}_G(r)}$ \ with \ $\inter{s} \,= \,a$ \
for any \,$r\ \fleche{a}_G\ s$.
\end{proposition}
\proof\mbox{}\\
{\bf i)} Let \,$s,t \in V_G$\,. \,Let us check that there is a unique \,$x$ 
\,such that \,$(s,t,x) \in {\rm Path}_G(r)$.\\
As \,$r$ \,is a root, there exists \,$u$ \,such that \,$r\ \fleche{u}_G\ t$.\\
As \,$G$ \,is source-complete, there exists \,$x$ \,such that 
\,$s\ \fleche{u}_G\ x$. Hence \,$(s,t,x) \in {\rm Path}_G(r)$.\\
Let \,$(s,t,y) \in {\rm Path}_G(r)$. There exists \,$v \in A_G^*$ \,such that 
\,$r\ \fleche{v}_G\ t$ \,and \,$s\ \fleche{v}_G\ y$.\\
As \,$G$ \,is arc-symmetric, we have \,$r \downarrow_G s$ \,and as \,$G$ \,is 
deterministic, we get \,$s\ \fleche{v}_G\ x$.\\
As \,$G$ \,is deterministic, it follows that \,$x \,= \,y$.\\
Thus \,$\ast_r$ \,exists and is denoted by \,$\cdot$ \,in the rest of this 
proof.\\
Let us show that $(V_G,\cdot)$ \,is a left-cancellative monoid.\\[0.5em]
{\bf ii)} Let us show that \,$\cdot$ \,is associative.\\
Let \,$x,y,z \in V_G$\,. 
\,We have to check that \,$(x{\cdot}y){\cdot}z \,= \,x{\cdot}(y{\cdot}z)$.\\
As \,$r$ \,is a root, there exists \,$v,w \in A_G^*$ \,such that 
\,$r\ \fleche{v}\ y$ \,and \,$r\ \fleche{w}\ z$.\\
By (i), \,$x\ \fleche{v}\ x{\cdot}y\ \fleche{w}\ (x{\cdot}y){\cdot}z$ \ and \ 
$y\ \fleche{w}\ y{\cdot}z$\\
So \ $r\ \fleche{vw}\ y{\cdot}z$ \ \ hence \ \ 
$x\ \fleche{vw}\ x{\cdot}(y{\cdot}z)$.\\
As \,$G$ \,is deterministic, we get
\,$(x{\cdot}y){\cdot}z \,= \,x{\cdot}(y{\cdot}z)$.\\[0.5em]
{\bf iii)} Let us check that \,$r$ \,is an identity element.\\
Let \,$s \in V_G$\,. \,As \,$r\ \fleche{\varepsilon}\ r$, we get 
\,$s\ \fleche{\varepsilon}\ s{\cdot}r$ \ {\it i.e.} \ $s{\cdot}r \,= \,s$.\\
For \,$r\ \fleche{u}\ s$, we have \,$r\ \fleche{u}\ r{\cdot}s$. 
As \,$G$ \,is deterministic, we get \,$r{\cdot}s \,= \,s$.\\[0.5em]
{\bf iv)} Let us check that \,$\cdot$ \,is left-cancellative. 
Let \,$s,t,t' \in V_G$ \,such that \,$s{\cdot}t \,= \,s{\cdot}t'$.\\
There exists \,$u,v \in A_G^*$ \,such that \,$r\ \fleche{u}\ t$ \,and 
\,$r\ \fleche{v}\ t'$.\\
So \ $s\ \fleche{u}\ s{\cdot}t$ \ and \ $s\ \fleche{v}\ s{\cdot}t'$. 
As \,$s{\cdot}t \,= \,s{\cdot}t'$ \,and \,$r \downarrow_G s$, we get 
\,$r\ \fleche{v}\ t$.\\
As \,$G$ \,is deterministic, we have \,$t = t'$.\\[0.5em]
{\bf v)} \,Let us check that \,$Q \,= \,\fleche{}_G(r)$ \,is a generating 
subset of \,$V_G$\,. Let \,$s \in V_G$\,.\\
There exists $n \geq 0$, $a_1,\ldots,a_n \in A_G$ and 
$s_0,\ldots,s_n$ such that 
$r = s_0\ \fleche{a_1}\ s_1{\ldots}s_{n-1}\ \fleche{a_1}\ s_n~=~s$.\\
By Fact~\ref{AccessibleTransitive}, there exists \,$r_1,\ldots,r_n$ \,such 
that \,$r\ \fleche{a_1}\ r_1,\ldots,r\ \fleche{a_n}\ r_n$\,.\\
For every \,$1 \leq i \leq n$, $s_i \,= \,s_{i-1}{\cdot}r_i$ \,hence 
\,$s \,= \,r{\cdot}r_1\cdot\ldots{\cdot}r_n \,= \,r_1\cdot\ldots{\cdot}r_n 
\,\in \,Q^*$.\\[0.5em]
{\bf vi)} \,Assume that \,$G$ \,is co-deterministic. 
Let us check that \,$\cdot$ \,is right-cancellative.\\
Let \,$s,s',t \in V_G$ \,such that \,$s{\cdot}t \,= \,s'{\cdot}t$.\\
There exists \,$u \in A_G^*$ \,such that \,$r\ \fleche{u}\ t$. 
So \ $s\ \fleche{u}\ s{\cdot}t$ \ and \ 
$s'\ \fleche{u}\ s'{\cdot}t \,= \,s{\cdot}t$.\\
As \,$G$ \,is co-deterministic, we get \,$s = s'$.\\[0.5em]
{\bf vii)} \,Assume that \,$G$ \,is simple.
Let \,$Q\ =\ \{\ s\ |\ r\ \fleche{}_G\ s\ \}$.\\
As \,$G$ \,is simple and deterministic, we define the following injection 
\,$\inter{\ }$ \,from \,$Q$ \,into \,$A_G$ \,by\\[0.25em]
\hspace*{12em}
$\inter{s} \,= \,a$ \ \ for \,$r\ \fleche{a}_G\ s$.\\
Let \,$K = {\cal C}\inter{V_G,Q}$. Let us show that \,$G \,= \,K$.\\[0.25em]
$\subseteq$\,: \,Let \,$s\ \fleche{a}_G\ t$. 
As \,$r \downarrow_G s$, \,there exists \,$r'$ \,such that 
\,$r\ \fleche{a}_G\ r'$.\\
So \,$s\ \fleche{a}_G\ s{\cdot}r'$. 
As \,$G$ \,is deterministic, $s{\cdot}r' \,= \,t$.\\
Furthermore \,$r' \in Q$ \,and \,$\inter{r'} \,= \,a$.
So \,$s\ \fleche{a}_K\ s{\cdot}r' \,= \,t$.\\[0.25em]
$\supseteq$\,: \,Let \,$s\ \fleche{a}_K\ t$.
So \,$a = \inter{r'}$ \,for some \,$r' \in Q$.\\
Thus \,$t \,= \,s{\cdot}r'$ \,and \,$r\ \fleche{a}_G\ r'$. 
So \,$s\ \fleche{a}_G\ s{\cdot}r' \,= \,t$.
\qed\\[1em]
For instance let us consider a graph \,$G$ \,of the following representation:
\begin{center}
\includegraphics{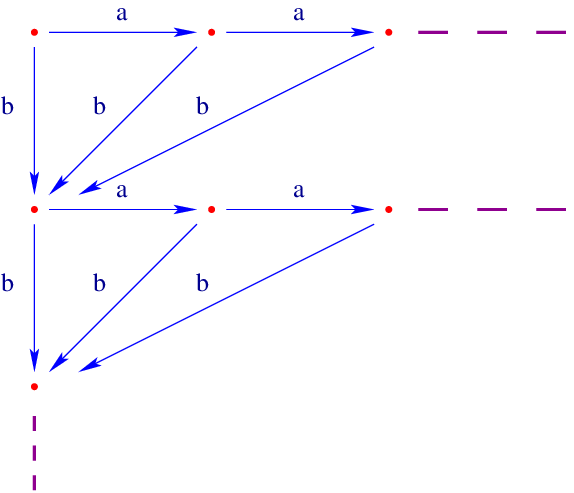}\\%magnitude 0.25
\end{center}
It is a {\it skeleton} of the graph of \,$\omega^2$ \,where \,$a$ \,is the 
successor and \,$b$ \,goes to the next limit ordinal: \,$(V_G,\fleche{}^*_G)$ 
\,is isomorphic to \,$(\omega^2,\leq)$. 
By Proposition~\ref{PathProduct}, it is a Cayley graph of a left-cancellative 
monoid. Precisely to each word \,$u \in b^*a^*$, we associate the unique vertex 
\,$\InfSup{u} \in V_G$ \,accessible from the root by the path labeled by 
\,$u$. Thus\\[0.25em]
\hspace*{1.5em}$G \,= \,\{\ \InfSup{b^ma^n}\ \fleche{a}\ \InfSup{b^ma^{n+1}}\ |
\ m,n \geq 0\ \} \,\cup 
\,\{\ \InfSup{b^ma^n}\ \fleche{b}\ \InfSup{b^{m+1}}\ |\ m,n \geq 0\ \}$.
\\[0.25em]
By Proposition~\ref{PathProduct}, $(V_G,\ast_{\InfSupInd{\varepsilon}})$ \,is a 
left-cancellative monoid where for any \,$m,n,p,q \geq 0$,\\[0.25em]
\hspace*{10em}{$\InfSup{b^ma^n} \,\ast_{\InfSupInd{\varepsilon}} 
\,\InfSup{b^pa^q} \ = \ \left\{\begin{tabular}{ll}
$\InfSup{b^ma^{n+q}}$ & if \ $p = 0$\\[0.25em]
$\InfSup{b^{m+p}a^q}$ & if \ $p \neq 0$
\end{tabular}\right.$}\\[0.25em]
and we have \,$G \,= \,{\cal C}\inter{V_G,\{\InfSup{a},\InfSup{b}\}}$ \,with 
\,$\inter{\InfSup{a}} = a$ \,and \,$\inter{\InfSup{b}} = b$.\\[0.25em]
Propositions~\ref{Semigroup} and \ref{PathProduct} give a graph-theoretic 
characterization of the Cayley graphs of left-cancellative monoids.
\begin{theorem}\label{LeftMonoid}
A graph is a Cayley graph of a left-cancellative monoid if and only if\\
\hspace*{7em}it is rooted, simple, deterministic and arc-symmetric.
\end{theorem}
\proof\mbox{}\\
We obtain the necessary condition by Proposition~\ref{Semigroup} with
Facts~\ref{Magma},\ref{CancelMagma},\ref{RootCayley}.\\
The sufficient condition is given by Proposition~\ref{PathProduct}.
\qed\\[1em]
We can restrict Theorem~\ref{LeftMonoid} to cancellative monoids.
\begin{theorem}\label{Monoid}
A graph is a Cayley graph of a cancellative monoid if and only if\\
\hspace*{7em}it is rooted, simple, deterministic, co-deterministic, 
arc-symmetric.
\end{theorem}
\proof\mbox{}\\
$\Longrightarrow$\,: \,By Theorem~\ref{LeftMonoid} and Fact~\ref{CancelMagma}.\\
$\Longleftarrow$\,: \,By Proposition~\ref{PathProduct}.
\qed\\[1em]
The previous graph is not co-deterministic hence, by Theorem~\ref{Monoid} 
or Fact~\ref{CancelMagma}, is not a Cayley graph of a cancellative monoid.
On the other hand and according to Proposition~\ref{PathProduct}, a 
{\it quater-grid} \,$G$ \,of the following representation:
\begin{center}
\includegraphics{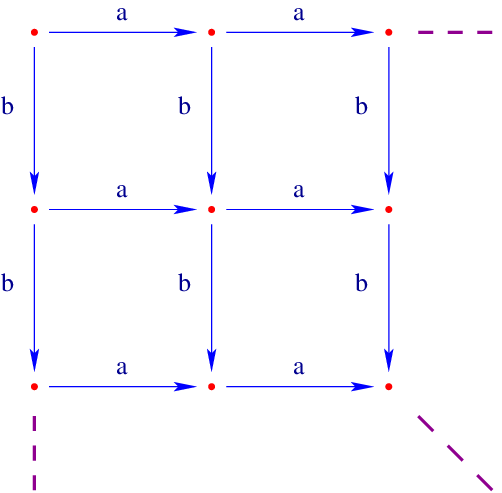}\\%magnitude 0.25
\end{center}
is a Cayley graph of a cancellative monoid. Precisely and as for the previous 
graph, we associate to each word \,$u \in b^*a^*$ \,the unique vertex 
\,$\InfSup{u}$ \,accessible from the root by the path labeled by \,$u$. 
By Proposition~\ref{PathProduct}, $(V_G,\ast_{\InfSupInd{\varepsilon}})$ \,is a 
cancellative monoid where\\[0.25em]
\hspace*{6em}$\InfSup{b^ma^n} \,\ast_{\InfSupInd{\varepsilon}} \,\InfSup{b^pa^q} 
\ = \ \InfSup{b^{m+p}a^{n+q}}$ \ \ for any \,$m,n,p,q \geq 0$.\\[0.25em]
and we have \,$G \,= \,{\cal C}\inter{V_G,\{\InfSup{a},\InfSup{b}\}}$ \,with 
\,$\inter{\InfSup{a}} = a$ \,and \,$\inter{\InfSup{b}} = b$.\\[0.5em]
Recall that a Cayley graph of a semigroup \,$M$ \,is a generalized Cayley 
graph \,${\cal C}\inter{M,Q}$ \,such that \,$M \,= \,Q^+$ \,whose 
\,$Q^+ \,= \,\{\ q_1{\cdot}\ldots{\cdot}q_n \mid n > 0 \,\wedge 
\,q_1,\ldots,q_n \in Q\ \}$ \,is the {\it subsemigroup generated} \,by \,$Q$. 
Theorem~\ref{Monoid} can be easily extended into a characterization of the 
Cayley graphs of cancellative semigroups. Indeed, a semigroup without an 
identity is turned into a monoid by just adding an identity. 
Precisely a {\it monoid-completion} \,$\overline{M}$ \,of a semigroup \,$M$ 
\,is defined by \,$\overline{M} \,= \,M$ \,if \,$M$ \,has an identity element, 
otherwise \,$\overline{M} \,= \,M \,\cup \,\{1\}$ \,whose \,$1$ \,is an 
identity element of \,$\overline{M}$\,: \,$p{\cdot}1 \,= \,1{\cdot}p \,= \,p$ 
\,for any \,$p \in \overline{M}$. This natural completion does not preserve the 
left-cancellative property but it preserves the cancellative property.
\begin{lemma}\label{Completion}
Any monoid-completion of a cancellative semigroup is a cancellative monoid.
\end{lemma}
\proof\mbox{}\\
Let \,$\overline{M} \,= \,M \,\cup \,\{1\}$ \,be a monoid-completion of a 
cancellative semigroup \,$M$ \,without an identity element.\\[0.25em]
{\bf i)} Suppose there are \,$m,e \in M$ \,such that \,$m{\cdot}e \,= \,m$.\\
In this case, let us check that \,$e$ \,is an identity element.\\
We have \,$m{\cdot}(e{\cdot}e) \,= \,(m{\cdot}e){\cdot}e \,= \,m{\cdot}e$. 
As \,$\cdot$ \,is left-cancellative, we get \,$e{\cdot}e \,= \,e$.\\
Let \,$n \in M$. So \,$(n{\cdot}e){\cdot}e \,= \,n{\cdot}(e{\cdot}e) \,= 
\,n{\cdot}e$. 
As \,$\cdot$ \,is right-cancellative, we get \,$n{\cdot}e \,= \,n$.\\
Finally \,$e{\cdot}(e{\cdot}n) \,= \,(e{\cdot}e){\cdot}n \,= \,e{\cdot}n$. 
As \,$\cdot$ \,is left-cancellative, we get \,$e{\cdot}n \,= \,n$.\\[0.25em]
{\bf ii)} By hypothesis \,$M$ \,has no identity element. By (i), there are no 
\,$m,e \in M$ \,such that \,$m{\cdot}e \,= \,m$. 
Let us show that \,$\overline{M}$ \,is left-cancellative.\\
Let \,$m{\cdot}p \,= \,m{\cdot}q$ \,for some \,$m,p,q \in \overline{M}$. 
Let us check that \,$p = q$.\\
As \,$M$ \,is left-cancellative, we only have to consider the case where 
\,$1 \in \{m,p,q\}$.\\
If \,$m = 1$ \,then \,$p \,= \,1{\cdot}p \,= \,1{\cdot}q \,= \,q$.\\
Otherwise \,$m \in M$ \,and \,$1 \in \{p,q\}$. By (i), we get \,$p = q = 1$.
\\[0.25em]
{\bf iii)} Similarly there are no \,$m,e \in M$ \,such that 
\,$e{\cdot}m \,= \,m$ \,hence \,$\overline{M}$ \,remains also 
right-cancellative.
\qed\\[1em]
Let us translate the monoid-completion of cancellative semigroups into their 
Cayley graphs. 
A {\it root-completion} \,of a graph \,$G$ \,is a graph \,$\overline{G}$ 
\,defined by \,$\overline{G} \,= \,G$ \,if \,$G$ \,is rooted, otherwise 
\,$G \,\subset \,\overline{G} \,\subseteq 
\,G \,\cup \,\{r\}{\croix}A_G{\croix}V_G$ \,and \,$r$ \,is the root of 
\,$\overline{G}$\,; \,we say that \,$G$ \,is {\it rootable} 
\,into \,$\overline{G}$. For instance the following non connected graph:
\begin{center}
\includegraphics{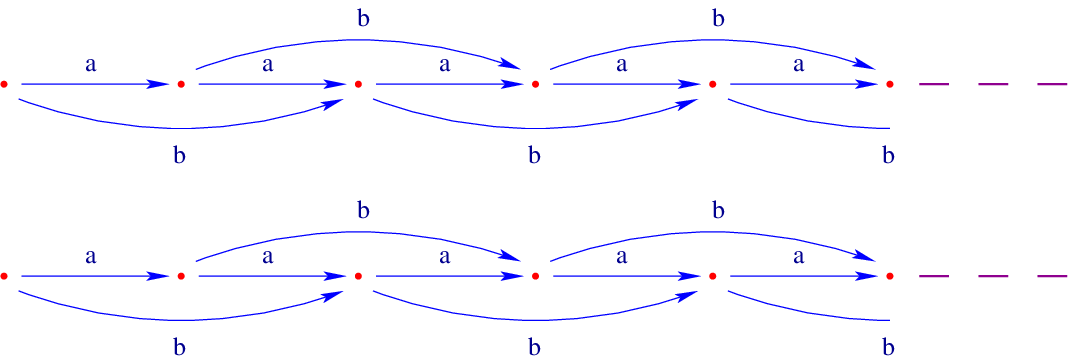}\\%magnitude 0.3
\end{center}
is arc-symmetric but is not rootable into an arc-symmetric 
graph. 
On the other hand, a graph consisting of two (isomorphic) deterministic and 
source-complete trees over \,$\{a,b\}$ \,is rootable into a deterministic 
source-complete tree over \,$\{a,b\}$. 
Finally the following graph:
\begin{center}
\includegraphics{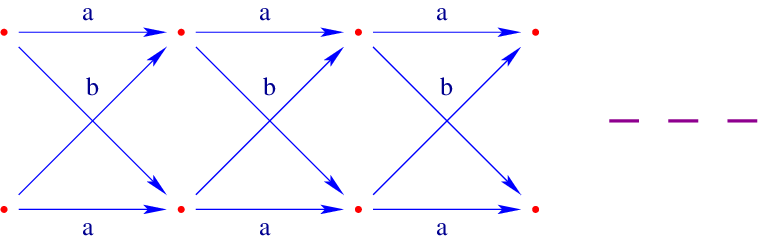}\\%magnitude 0.3
\end{center}
is also rootable into a simple, deterministic, co-deterministic, 
arc-symmetric graph. 
We can apply Theorem~\ref{Monoid}.
\begin{theorem}\label{CanSemigroup}
A graph is a Cayley graph of a cancellative semigroup if and only if\\
\hspace*{1em}it is rootable into a simple, deterministic, co-deterministic, 
arc-symmetric graph.
\end{theorem}
\proof\mbox{}\\
$\Longrightarrow$\,: \,Let \,$G \,= \,{\cal C}\inter{M,Q}$ \,for some 
cancellative semigroup \,$M$ \,and some generating subset\,$Q$ \,of \,$M$ 
\,{\it i.e.} \,$Q^+ = M$. We have the following two complementary cases.\\
{\it Case 1}\,: \,$M$ \,has an identity element. 
By Theorem~\ref{Monoid}, $G$ \,is rooted, simple, arc-symmetric, 
deterministic and co-deterministic.
As \,$G$ \,has a root, it is rootable into itself.\\
{\it Case 2}\,: \,$M$ \,is not a monoid. 
Let \,$\overline{M} \,= \,M \,\cup \,\{1\}$ \,be a monoid-completion 
of \,$M$.\\
By Lemma~\ref{Completion}, ,$\overline{M}$ \,remains cancellative. 
Furthermore \,$Q^* \,= \,\overline{M}$. Let\\[0.25em]
\hspace*{9em}$\overline{G} \ = \ {\cal C}\inter{\overline{M},Q} \ = \ 
G \,\cup \,\{\ 1\ \fleche{\interInd{q}}\ q\ |\ q \in Q\ \}$.\\[0.25em]
By Theorem~\ref{Monoid}, 
\,$\overline{G}$ \,is rooted, simple, arc-symmetric, deterministic and
co-deterministic.
Moreover \,$G$ \,is rootable into \,$\overline{G}$.\\[0.25em]
$\Longleftarrow$\,: \,Let a graph \,$G$ \,rootable into a simple, 
deterministic, co-deterministic, arc-symmetric graph \,$\overline{G}$. 
We have the following two complementary cases.\\
{\it Case 1}\,: \,$G$ \,is rooted. By Theorem~\ref{Monoid} 
(or Proposition~\ref{PathProduct}), $G$ \,is a Cayley graph of a cancellative 
monoid.\\
{\it Case 2}\,: \,$G$ \,has no root. 
Let \,$r$ \,be the root of \,$\overline{G}$ \,and 
\,$Q = \fleche{}_{\overline{G}}(r)$.\\
So \,$Q \subseteq V_G$ \,and \,$V_G \,= \,V_{\overline{G}}-\{r\}$.\\
By Proposition~\ref{PathProduct}, 
\,$\overline{G} \,= \,{\cal C}\inter{V_{\overline{G}},Q}$ \,for the associative 
and cancellative path-operation \,$\ast_r$ \,on \,$V_{\overline{G}}$ \,of 
identity element \,$r$ \,with \,$V_{\overline{G}}$ \,generated by \,$Q$.\\
As \,$r$ \,is not the target of an edge of \,$\overline{G}$ \,and by 
definition, $\ast_r$ \,remains an internal operation on \,$V_G$ \,{\it i.e.} 
\,$p \ast_r q \neq r$ \,for any \,$p,q \in V_G$\,.\\
Finally \,$G \,= \,{\cal C}\inter{V_G\,,\,Q}$ \,and \,$(V_G\,,\,\ast_r)$ 
\,is a cancellative semigroup.
\qed\\[1em]
For instance by Theorem~\ref{CanSemigroup}, the previous graph is a Cayley 
graph of a cancellative semigroup. It is isomorphic to\\[0.25em]
\hspace*{3em}\begin{tabular}{rclcl}
$G$ & $=$ & $\{\ n\ \fleche{a}\ n+1\ |\ n > 0\ \}$ & $\cup$ & 
$\{\ n\ \fleche{a}\ n-1\ |\ n < 0\ \}$\\[0.25em]
 & $\cup$ & $\{\ n\ \fleche{b}\ -n-1\ |\ n > 0\ \}$ & $\cup$ & 
$\{\ n\ \fleche{b}\ -n+1\ |\ n < 0\ \}$.
\end{tabular}\\[0.25em]
We have \,$G \,= 
\,{\cal C}\inter{\relatif-\{0\}\,,\,\{-1,1\}}$ \,with \,$\inter{1} = a$ \,and 
\,$\inter{-1} = b$ \,for the following associative and cancellative 
path-operation \,$\ast$ \,defined by\\[0.25em]
\hspace*{6em}$m \ast n\ =\ sign(m{\croix}n)\,(|m|+|n|)$ \ for any 
\,$m,n \in \relatif-\{0\}$.\\[0.25em]
We can now restrict Theorem~\ref{Monoid} to the Cayley graphs of groups.

\section{Cayley graphs of groups}

We present a graph-theoretic characterization for the Cayley graphs of groups: 
they are the deterministic, co-deterministic, symmetric, simple and 
connected graphs (Theorem~\ref{CayleyGroup}). 
By removing the connectivity condition and under the assumption of the axiom 
of choice, we get a characterization for the generalized Cayley graphs of 
groups (Theorem~\ref{MainFour}).\\[-0.5em]

Recall that a {\it group} \,$(M,\cdot)$ \,is a monoid whose each element 
\,$p \in M$ \,has an inverse \,$p^{-1}$\,: 
\,$p{\cdot}p^{-1} \,= \,1 \,= \,p^{-1}{\cdot}p$. 
So \,${\cal C}(M)$ \,is strongly connected hence by Proposition~\ref{Semigroup} 
is symmetric. 
\begin{fact}\label{Group}
Any generalized Cayley graph of a group is symmetric.
\end{fact}
\proof\mbox{}\\
Let \,$(M,\cdot)$ \,be a group and \,$\inter{\ }\,: M\ \fleche{}\ A$ \,be an 
injective mapping.\\
By Proposition~\ref{Semigroup}, \,${\cal C}\inter{M}$ \,is 
arc-symmetric.\\
As \,${\cal C}\inter{M}$ \,is strongly connected, ${\cal C}\inter{M}$ \,is
symmetric.\\
For any \,$Q \subseteq M$, 
\,${\cal C}\inter{M,Q} \,= \,{\cal C}\inter{M}^{|\interInd{Q}}$ \,remains 
symmetric.
\qed\\[1em]
We start by considering the monoid Cayley graphs of a group \,$M$ \,which are 
the generalized Cayley graph \,${\cal C}\inter{M,Q}$ \,with \,$Q^* = M$.
\begin{fact}\label{GroupConnex}
Any monoid Cayley graph of a group is strongly connected.
\end{fact}
\proof\mbox{}\\
Let \,$G \,= \,{\cal C}\inter{M,Q}$ \,for some group \,$M$ \,and some 
\,$Q \subseteq M$ \,with \,$Q^* = M$.\\
Let \,$p \in M$. 
We have to check that \,$1\ \fleche{}^*_G\ p\ \fleche{}^*_G\ 1$.\\
There exists \,$n \geq 0$ \,and \,$q_1,\ldots,q_n \in Q$ \,such that 
\,$p \,= \,q_1{\cdot}\ldots{\cdot}q_n$\,. 
So \,$1\ \fleche{\interInd{q_1}{\ldots}\interInd{q_n}}_G\ p$.\\
For any \,$1 \leq i \leq n$, we have 
\,$q_i^{-1} \,= \,q_{i,1}{\cdot}\ldots{\cdot}q_{i,m_i}$ \,for some 
\,$m_i \geq 0$ \,and \,$q_{i,1},\ldots,q_{i,m_i} \in Q$.\\
Thus \,$p\ \fleche{u}_G\ 1$ \,for \,$u \,=
\,\inter{q_{n,1}}{\ldots}\inter{q_{n,m_n}}\,\ldots\,\inter{q_{1,1}}{\ldots}\inter{q_{1,m_1}}$.
\qed\\[1em]
Let us complete Proposition~\ref{PathProduct} in the case where the graph is 
symmetric. In this case, the path-operation is also invertible.
\begin{proposition}\label{PathProductGroup}
For any root \,$r$ \,of a deterministic and symmetric graph \,$G$,\\
\hspace*{8.5em}$(V_G,\ast_r)$ \,is a group.
\end{proposition}
\proof\mbox{}\\
It suffices to complete the proof of Proposition~\ref{PathProduct} when 
\,$G$ \,is in addition symmetric.\\
Let \,$s \in V_G$\,. Let us show that \,$s$ \,has an inverse.\\
There exists \,$u \in A_G^*$ \,such that \,$r\ \fleche{u}\ s$.\\
As \,$r \,\simeq_G s$, \,$s$ \,is also a root hence there exists \,$v$ \,such 
that \,$s\ \fleche{v}\ r$.\\
Let \,$\overline{s}$ \,be the vertex such that \,$r\ \fleche{v}\ \overline{s}$. 
So\\[0.25em]
\hspace*{10em}$\overline{s}\ \fleche{u}\ \overline{s}{\cdot}s$ \ \ and \ \ 
$s\ \fleche{v}\ s{\cdot}\overline{s}$.\\[0.25em]
As \,$G$ \,is deterministic, we get \,$s{\cdot}\overline{s} \,= \,r$.\\
As \,$r \,\simeq_G s$ \,and \,$s\ \fleche{vu}\ s$, we get \,$r\ \fleche{vu}\ r$.
\\
As \,$G$ \,is deterministic, we get \,$\overline{s}\ \fleche{u}\ r$ \,hence 
\,$\overline{s}{\cdot}s \,= \,r$.
\qed\\[1em]
We describe the monoid Cayley graphs of groups from the 
characterization of the Cayley graphs of left-cancellative monoids 
(Theorem~\ref{LeftMonoid}) just by replacing the arc-symmetry by the symmetry.
\begin{theorem}\label{RestrictedGroup}
A graph is a monoid Cayley graph of a group if and only if\\
\hspace*{7em}it is rooted, simple, deterministic and symmetric.
\end{theorem}
\proof\mbox{}\\
$\Longrightarrow$\,: \,By Theorem~\ref{LeftMonoid} and Fact~\ref{Group}.
\\[0.25em]
$\Longleftarrow$\,: \,\,By Propositions~\ref{PathProduct} and 
\ref{PathProductGroup}.
\qed\\[1em]
For instance by Theorem~\ref{RestrictedGroup}, a graph of the following 
representation:
\begin{center}
\includegraphics{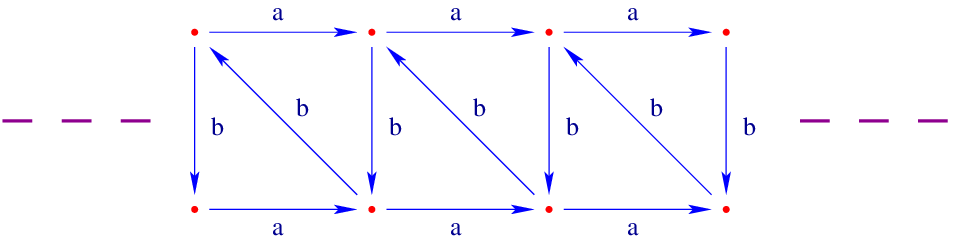}\\%magnitude 0.3
\end{center}
is a monoid Cayley graph of a group: it is isomorphic to 
\,${\cal C}\inter{\relatif,\{2,-1\}}$ \,for the group \,$(\relatif,+)$ \,with 
\,$\inter{2} = a$ \,and \,$\inter{-1} = b$.\\
From Fact~\ref{GroupConnex}, we can replace in Theorem~\ref{RestrictedGroup} 
the rooted condition by the fact to be strongly connected. 
By Fact~\ref{CancelMagma}, we can also add the co-determinism condition.
\begin{corollary}\label{RestrictedGroupBis}
Any rooted, simple, deterministic and symmetric graph is 
strongly connected and co-deterministic.
\end{corollary}
We can now consider a {\it group Cayley graph} \,as a generalized Cayley graph 
\,${\cal C}\inter{M,Q}$ \,such that \,$M$ \,is a group equal to the 
{\it subgroup generated} by \,$Q$ \,which is the least subgroup 
\,$(Q \,\cup \,Q^{-1})^*$ \,containing \,$Q$ \,where 
\,$Q^{-1} \,= \,\{\ q^{-1}\ |\ q \in Q\ \}$ \,is the set of inverses of the 
elements in \,$Q$.\\
For instance, the \,$a$-{\it line} 
\,$\{\ n\ \fleche{a}\ n+1\ |\ n \in \relatif\ \}$ \,is the Cayley graph 
\,${\cal C}\inter{\relatif,\{1\}}$ \,of the group 
\,$(\relatif,+)$ \,with \,$\inter{1} = a$. 
This unrooted graph is not a monoid Cayley graph.\\
Let us generalize Theorem~\ref{RestrictedGroup} to these well-known Cayley 
graphs. 
We need to be able to circulate in a graph in the direct and inverse direction 
of the arrows. 
Let \,$G$ \,be a graph and let \,$\haut{\mbox{---}} : A_G\ \fleche{}\ A - A_G$ 
\,be an injective mapping of image 
\,$\overline{A_G} \,= \,\{\ \overline{a}\ |\ a \in A_G\ \}$.
A {\it chain} \,$s\ \fleche{u}_G\ t$ \,is a path labeled by 
\,$u \in (A_G \,\cup \,\overline{A_G})^*$ \,where for any \,$a \in A_G$\,, we 
have \,$s\ \fleche{\overline{a}}_G\ t$ \,for \,$t\ \fleche{a}_G\ s$. 
Given a vertex \,$r$, the path-relation \,${\rm Path}_G(r)$ \,is extended into 
the {\it chain-relation} \,${\rm Chain}_G(r)$ \,as the ternary relation on 
\,$V_G$ \,defined by\\[0.25em]
\hspace*{0.5em}$(s,t,x) \in {\rm Chain}_G(r)$ \ if there exists 
\,$u \in (A_G \,\cup \,\overline{A_G})^*$ \,such that \,$r\ \fleche{u}_G\ t$ 
\,and \,$s\ \fleche{u}_G\ x$.\\[0.25em]
Thus \,${\rm Path}_G(r) \,\subseteq \,{\rm Chain}_G(r) \,= 
\,{\rm Path}_{\overline{G}}(r)$ \,for \,$\overline{G}\ =\ 
G \,\cup \,\{\ t\ \fleche{\overline{a}}\ s\ |\ s\ \fleche{a}_G\ t\ \}$.\\
If for any \,$s,t \in V_G$ \,there exists a unique \,$x$ \,such that 
\,$(s,t,x) \in {\rm Chain}_G(r)$, we denote by 
\,$\overline{\ast}_r : V_G{\croix}V_G\ \fleche{}\ V_G$ \,the binary 
{\it chain-operation} \,on \,$V_G$ \,defined by 
\,$(s,t,s\ \overline{\ast}_r \,t) \in {\rm Chain}_G(r)$ \,for any 
\,$s,t \in V_G$\,; we also write \,$_G\overline{\ast}_r$ \,when we need to 
specify \,$G$.\\
Let us adapt Propositions~\ref{PathProduct} and \ref{PathProductGroup} to this 
chain-operation.
\begin{proposition}\label{ChainProduct}
Let \,$r$ \,be a vertex of a connected, symmetric, deterministic and\\
\hspace*{0.5em}co-deterministic graph \,$G$.\\
\hspace*{0.5em}Then \,$(V_G,\overline{\ast}_r)$ \,is a group of identity \,$r$ 
\,generated by \ $\fleche{}_G(r)$.\\
\hspace*{0.5em}If \,$G$ \,is simple \,then \ 
$G \,= \,{\cal C}\inter{V_G,\fleche{}_G(r)}$ \ with \ $\inter{s} \,= \,a$ \
for any \,$r\ \fleche{a}_G\ s$.
\end{proposition}
\proof\mbox{}\\
{\bf i)} The graph \,$\overline{G}$ \,remains symmetric. 
As \,$G$ \,is deterministic and co-deterministic, $\overline{G}$ \,is 
deterministic. 
As \,$G$ \,is connected, $\overline{G}$ \,is strongly connected.
By applying Propositions~\ref{PathProduct} and~\ref{PathProductGroup} to 
\,$\overline{G}$, we get that \,$(V_{\overline{G}}\,,\ast_r)$ \,is a group of 
identity \,$r$ \,and \,$V_G \,= \,V_{\overline{G}} \,= 
\,(\fleche{}_{\overline{G}}(r))^*$ \ with \ $\fleche{}_{\overline{G}}(r) \ \,= \ 
\,\fleche{}_G(r) \,\cup \,\fleche{}_{G^{-1}}(r)$.\\[0.25em]
{\bf ii)} Let us check that 
\,$\fleche{}_{G^{-1}}(r) \,= \,(\fleche{}_G(r))^{-1}$.\\
$\subseteq$ \,: \,Let \,$s \in \fleche{}_{G^{-1}}(r)$ \,{\it i.e.} 
\,$r\ \fleche{}_{G^{-1}}\ s$. 
So \,$s\ \fleche{a}_G\ r$ \,for some \,$a \in A$.\\
As \,$G$ \,is source-complete, there exists \,$t$ \,such that 
\,$r\ \fleche{a}_G\ t$. Thus \,$s\ \fleche{a}_G\ s \ast_r t$.\\
As \,$G$ \,is deterministic, $s \ast_r t \,= \,r$ \,hence 
\,$s \,= \,t^{-1} \,\in \,(\fleche{}_G(r))^{-1}$.\\[0.25em]
$\supseteq$ \,: \,Let \,$s \in (\fleche{}_G(r))^{-1}$ \,{\it i.e.} 
$r\ \fleche{a}_G\ s^{-1}$ \,for some \,$a \in A$.\\
So \,$s\ \fleche{a}_G\ s \ast_r s^{-1} \,= \, r$ \,{\it i.e.} 
\,$s \in \fleche{}_{G^{-1}}(r)$.\\[0.25em]
{\bf iii)} Suppose that in addition \,$G$ \,is simple. 
Note that \,$\overline{G}$ \,can be not simple. 
Thus we define the graph\\[0.25em]
\hspace*{10em}$\widehat{G} \ = \ G \,\cup \,\{\ t\ \fleche{\overline{a}}\ s\ |\ 
s\ \fleche{a}_G\ t\ \nofleche{}{}\,\!_G\ s\ \}$.\\[0.25em]
Let us show that \ $_{\widehat{G}}\ast_r \ = \ _G\overline{\ast}_r$\,.\\
As \,$\widehat{G} \,\subseteq \,\overline{G}$, we get \ 
\,$_{\widehat{G}}\ast_r \ \subseteq \ _{\overline{G}}\ast_r \ = \ 
_{G}\overline{\ast}_r$\,. 
Let us show the inverse inclusion.\\
We consider the mapping \,$\pi : A_G \,\cup \,\overline{A_G}\ \fleche{}\ 
A_{\widehat{G}}$ \,defined for any \,$a \in A_{\widehat{G}}$ \,by \,$\pi(a) = a$, 
and for any \,$\overline{a} \in \overline{A_G} - A_{\widehat{G}}$\,, 
$\pi(\overline{a})$ \,is the unique letter in \,$A_{\widehat{G}}$ \,such that 
\,$t\ \fleche{\pi(\overline{a})}_G\ s$ \,for any \,$s\ \fleche{a}_G\ t$.\\
This makes sense because \,$G$ \,is deterministic, co-deterministic and 
symmetric. Thus\\[0.25em]
\hspace*{10em}$s\ \fleche{a}_G\ t \ \ \Longrightarrow \ \ 
s\ \fleche{\pi(a)}_{\widehat{G}}\ t$ \ for any
\,$a \in A_G \,\cup \,\overline{A_G}$\,.\\
By extending \,$\pi$ \,by morphism on \,$(A_G \,\cup \,\overline{A_G})^*$, we 
get\\[0.25em]
\hspace*{10em}$s\ \fleche{u}_G\ t \ \ \Longrightarrow \ \ 
s\ \fleche{\pi(u)}_{\widehat{G}}\ t$ \ for any
\,$u \in (A_G \,\cup \,\overline{A_G})^*$\,.\\
This implies that \,$_{G}\overline{\ast}_r \ \subseteq \ _{\widehat{G}}\ast_r$\,.
\\[0.25em]
{\bf iv)} Having \,$G$ \,simple, symmetric, deterministic and 
co-deterministic, $\widehat{G}$ \,remains simple, symmetric, 
deterministic, and is strongly connected.\\
By Proposition~\ref{PathProduct}, 
\,$\widehat{G} \,= \,{\cal C}\inter{V_G,\fleche{}_{\widehat{G}}(r)}$ \ with \ 
$\inter{s} \,= \,a$ \ for any \,$r\ \fleche{a}_{\widehat{G}}\ s$.\\
Precisely for any \,$s \in \fleche{}_{\widehat{G}}(r)$, we have\\[0.25em]
\hspace*{10em}{$\inter{s} \ = \ \left\{\begin{tabular}{ll}
$a$ & if \ $r\ \fleche{a}_G\ s$\\[0.25em]
$\overline{a}$ & if \ $s\ \fleche{a}_G\ r\ \nofleche{}{}\,\!_G\ s$.
\end{tabular}\right.$}\\[0.25em]
Thus \,$G \,= \,\widehat{G}^{|A_G} \,= \,{\cal C}\inter{V_G,\fleche{}_G(r)}$.
\qed\\[1em]
For instance by Proposition~\ref{ChainProduct}, a graph of the following 
representation:
\begin{center}
\includegraphics{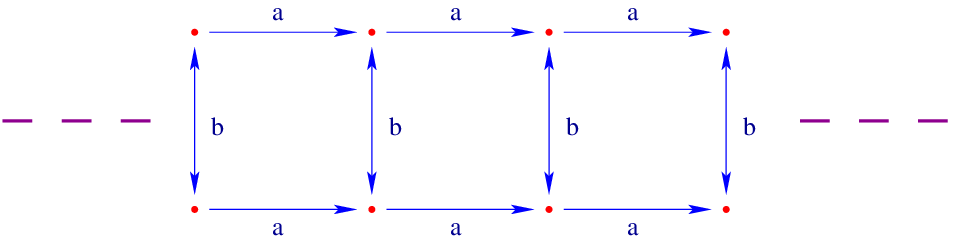}\\%magnitude 0.3
\end{center}
is a Cayley graph of a group: it is isomorphic to 
\,${\cal C}\inter{\relatif{\croix}\{0,1\},\{(1,0),(0,1)\}}$ \,with 
\,$\inter{(1,0)} = a$, \,$\inter{(0,1)} = b$ \,and for the chain-operation 
\,$(m,i) \ \overline{\ast}_{(0,0)} \ (n,j) = (m+n,i+j \,({\rm mod} 2))$. 
It is also isomorphic to the Cayley graph of the group of finite presentation 
\,$<\!a,b\ |\ ab = ba,b^2=1\!>$.\\[0.25em]
Let us adapt Theorem~\ref{RestrictedGroup} to simply describe the 
Cayley graphs of groups.
\begin{theorem}\label{CayleyGroup}
A graph is a Cayley graph of a group if and only if\\
\hspace*{7em}it is connected, simple, deterministic, co-deterministic and 
symmetric.
\end{theorem}
\proof\mbox{}\\
$\Longrightarrow$\,: \,Let \,$G \,= \,{\cal C}\inter{M,Q}$ \,for some group 
\,$M$ \,and \,$Q \subseteq M$ \,with \,$(Q \,\cup \,Q^{-1})^* \,= \,M$.\\
By Facts~\ref{Magma}, \ref{CancelMagma}, \ref{Group}, \,$G$ \,is simple, 
symmetric, deterministic and co-deterministic.\\
By Fact~\ref{GroupConnex}, the monoid Cayley graph 
\,${\cal C}\inter{M,Q \cup Q^{-1}}$ \,is strongly connected.\\
Thus \,$G \,= \,{\cal C}\inter{M,Q \cup Q^{-1}}^{|\interInd{Q}}$ \,is connected.
\\[0.25em]
$\Longleftarrow$\,: \,By Proposition~\ref{ChainProduct}.
\qed\\[1em]
Theorems~\ref{RestrictedGroup} and \ref{CayleyGroup} give respectively a 
characterization of the monoid Cayley graphs of groups, and the group Cayley 
graphs. We can now deduce a characterization of the generalized Cayley graphs 
of groups. First, let us apply Corollary~\ref{RestrictedGroupBis} and 
Theorem~\ref{CayleyGroup}.
\begin{corollary}\label{Component}
The connected (resp. strongly connected) components of generalized Cayley
graphs of groups are the (resp. monoid) Cayley graphs of groups.
\end{corollary}
Let us extend Proposition~\ref{ChainProduct} for non connected graphs.\\
Let a magma \,$(P,\cdot)$ \,for \,$P$ \,a representative set of \,Comp$(G)$.\\ 
We define the {\it extended chain-relation} \,${\rm Chain}_G(P)$ \,as the 
ternary relation on \,$V_G$ \,defined~by\\[0.25em]
\hspace*{3em}$(s,t,x) \in {\rm Chain}_G(P)$ \ if there exists 
\,$u,v \in (A_G \,\cup \,\overline{A_G})^*$ \,such that\\[0.25em]
\hspace*{9em}$\pi_P(s)\ \fleche{u}_G\ s$ \ and \ $\pi_P(t)\ \fleche{v}_G\ t$ 
\ and \ $\pi_P(s)\cdot\pi_P(t)\ \fleche{uv}_G\ x$.\\[0.25em]
For any connected and deterministic graph \,$G$ \,and any vertex \,$r$, 
${\rm Chain}_G(\{r\}) \,= \,{\rm Chain}_G(r)$.\\
If for any \,$s,t \in V_G$ \,there exists a unique \,$x$ \,such that 
\,$(s,t,x) \in {\rm Chain}_G(P)$, we denote by 
\,$\overline{\ast}_P : V_G{\croix}V_G\ \fleche{}\ V_G$ \,the binary 
{\it extended chain-operation} \,on \,$V_G$ \,defined for any \,$s,t \in V_G$ 
\,by \,$(s,t,s \,\overline{\ast}_P \,t) \in {\rm Chain}_G(P)$\,; 
we also write \,$_G\overline{\ast}_P$ \,when we need to specify \,$G$.\\
We can extend Proposition~\ref{ChainProduct}.
\begin{proposition}\label{ExtendedChainProduct}
Let \,$G$ \,be a symmetric, deterministic and co-deterministic graph.\\
\hspace*{0.5em}Let a group on a representative set \,$P$ \,of 
\,${\rm Comp}(G)$ \,generated by \,$P_0$ \,and of identity \,$r$.\\
\hspace*{0.5em}Then \,$(V_G,\overline{\ast}_P)$ \,is a group of identity \,$r$ 
\,generated by \,$P_0 \,\cup \,\fleche{}_G(r)$.\\
\hspace*{0.5em}If \,$G$ \,is simple \,then \ 
$G \,= \,{\cal C}\inter{V_G,\fleche{}_G(r)}$ \ with \ $\inter{s} \,= \,a$ \
for any \,$r\ \fleche{a}_G\ s$.
\end{proposition}
\proof\mbox{}\\
{\bf i)} Let \,$(P,\cdot)$ \,be a group generated by \,$P_0$ \,of identity 
\,$r$. 
Let \,$C \in {\rm Comp}(G)$ \,with \,$r \in V_C$\,.\\
By Proposition~\ref{ChainProduct}, \,$(V_C,\overline{\ast}_r)$ \,is a group of 
identity \,$r$ \,and is generated by \,$\fleche{}_G(r)$.\\
We take the group product \,$(P{\croix}V_C,\cdot)$ \,with 
\,$(p,x) \cdot (q,y) \,= \,(p\,{\cdot}\,q,x \ \overline{\ast}_r \,y)$ \,for 
any \,$p,q \in P$ \,and \,$x,y \in V_C$\,. 
This group is of identity \,$(r,r)$ \,and is generated by 
\,$P_0{\croix}\{r\} \,\cup \,\{r\}{\croix}\fleche{}_G(r)$.\\
As \,$G$ \,is symmetric, deterministic and co-deterministic, we can define 
the mapping\\
\hspace*{0.5em}$f : P{\croix}V_C\ \fleche{}\ V_G$ \,such that 
\,$r\ \fleche{u}_G\ x \ \ \Longrightarrow \ \ p\ \fleche{u}_G\ f(p,x)$ \,for 
(any) \,$u \in (A_G \,\cup \,\overline{A_G})^*$.\\
Thus \,$f$ \,is a bijection hence \,$(V_G,\cdot)$ \,is a group where 
\,$f(p,x) \cdot f(q,y) \,= \,f(p\,{\cdot}\,q,x \ \overline{\ast}_r \,y)$ \,for 
any \,$p,q \in P$ \,and \,$x,y \in V_C$\,. 
This group \,$(V_G,\cdot)$ \,is of identity \,$f(r,r) = r$ \,and is generated 
by \,$f(P_0{\croix}\{r\}) \,\cup \,f(\{r\}{\croix}\fleche{}_G(r))\ =\ 
P_0 \,\cup \,\fleche{}_G(r)$.\\[0.25em]
{\bf ii)} Let us show that the operation \,$\cdot$ \,on \,$V_G$ \,is equal to 
\,$\overline{\ast}_P$\,.\\
Let \,$p,q \in P$ \,and \,$x,y \in V_C$\,. 
We have to check that 
\,$f(p,x)\ \overline{\ast}_P \,f(q,y) \,= \,f(p,x) \cdot f(q,y)$.\\
Let \,$u,v \in (A_G \,\cup \,\overline{A_G})^*$ \,such that 
\,$r\ \fleche{u}_G\ x$ \,and \,$r\ \fleche{v}_G\ y$.\\
By definition of \,$\overline{\ast}_P$\,, we have 
\,$x\ \fleche{v}_G\ x\ \overline{\ast}_r \,y$ \,hence 
\,$r\ \fleche{uv}_G\ x\ \overline{\ast}_r \,y$.\\
By definition of \,$f$, we get
\,$p \cdot q\ \fleche{uv}_G\ f(p \cdot q,x\ \overline{\ast}_r \,y)$, 
\,$p\ \fleche{u}_G\ f(p,x)$ \,and \,$q\ \fleche{v}_G\ f(q,y)$.\\
By definition of \,$\overline{\ast}_P$\,, we have 
\,$p \cdot q\ \fleche{uv}_G\ f(p,x)\ \overline{\ast}_P \,f(q,y)$.\\
As \,$G$ \,is deterministic, we get \,$f(p,x)\ \overline{\ast}_P \,f(q,y) \,= 
\,f(p \cdot q,x\ \overline{\ast}_r \,y) \,= \,f(p,x) \cdot f(q,y)$.\\[0.25em]
{\bf iii)} Suppose that in addition \,$G$ \,is simple. 
Let \,$\inter{y} \,= \,a$ \ for any \,$r\ \fleche{a}_G\ y$. Thus\\
\hspace*{3em}\begin{tabular}{rcl}
${\cal C}\inter{V_G,\fleche{}_G(r)}$ & $=$ & 
$\{\ s\ \fleche{a}\ s \cdot y\ |\ s \in V_G \,\wedge \,r\ \fleche{a}\ y\ \}$
\\[0.25em]
 & $=$ & $\{\ s\ \fleche{a}\ s\ \overline{\ast}_P \,y\ |\ s \in V_G \,\wedge 
\,r\ \fleche{a}\ y\ \}\ =\ G$.
\end{tabular}
\qed\\[1em]
In ZF set theory, the axiom of choice is equivalent to the property that any 
non-empty set has a group structure \cite{HK}. 
Under the assumption of the axiom of choice, we can characterize the 
generalized Cayley graphs of groups.
\begin{theorem}\label{MainFour}
In ZFC set theory, a graph is a generalized Cayley graph of a group if and 
only if it is simple, deterministic, co-deterministic, symmetric.
\end{theorem}
\proof\mbox{}\\
By Facts~\ref{Magma}, \ref{CancelMagma}, \ref{Group}, any generalized Cayley 
graph is symmetric, deterministic and co-deterministic.\\
Conversely let \,$G$ \,be a simple, deterministic, co-deterministic, symmetric 
graph.\\
Using ZFC set theory, there exists a representative set \,$P$ \,of 
\,${\rm comp}(G)$ \,and a binary operation \,$\cdot$ \,such that \,$(P,\cdot)$ 
\,is a group. 
By Proposition~\ref{ExtendedChainProduct}, \,$(V_G,\overline{\ast}_P)$ \,is a 
group and \,$G \,= \,{\cal C}\inter{V_G,\fleche{}_G(r)}$ \,with 
\,$\inter{s} \,= \,a$ \ for any \,$r\ \fleche{a}_G\ s$.
\qed\\[1em]
{\noindent}For instance let us consider the following graph:\\[0.25em]
\hspace*{6em}\begin{tabular}{rcl}
$G$ & $=$ & $\{\ (m,i,p)\ \fleche{a}\ (m+1,i,p)\ |\ m,p \in \relatif 
\,\wedge \,i \in \{0,1\}\ \}$\\[0.25em]
 & $\cup$ & $\{\ (m,i,p)\ \fleche{b}\ (m,1-i,p)\ |\ m,p \in \relatif 
\,\wedge \,i \in \{0,1\}\ \}$
\end{tabular}\\[0.25em]
of representation the countable repetition of the preceding one.\\
By Theorem~\ref{MainFour}, $G \,= 
\,{\cal C}\inter{\relatif{\croix}\{0,1\}{\croix}\relatif,\{(1,0,0),(0,1,0)\}}$ 
\,with \,$\inter{(1,0,0)} = a$, \,$\inter{(0,1,0)} = b$ \,and for the group 
operation \,$(m,i,p) \cdot (n,j,q) \,= \,(m+n,i+j \,({\rm mod} 2),p+q)$.
\\[0.25em]
Let us summarize the characterizations obtained for the Cayley 
graphs of monoids.\\[-0.5em]\mbox{}
\begin{center}
\includegraphics{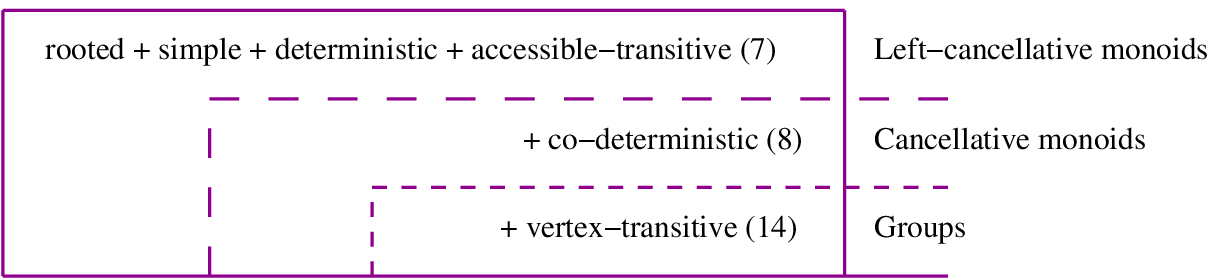}\\%magnitude 0.25
\end{center}%\mbox{}\\
By relaxing the condition of being rooted by that of connectivity, we have 
obtained a graph-theoretic characterization for the Cayley graphs of groups 
(Theorem~\ref{CayleyGroup}).

\section{Generalized Cayley graphs of left-cancellative magmas}

Under ZFC set theory, we will give a fully graph-theoretic characterization 
for generalized Cayley graphs of left-cancellative magmas 
(Theorem~\ref{LeftCancel}), and then when they have an identity
(Theorem~\ref{UnitCancel}).\\[-0.5em]

Recall that an element \,$e$ \,of a magma $(M,\cdot)$ \,is a 
{\it left identity} (resp. {\it right identity}) \,if \,$e{\cdot}p = p$ 
(resp. \,$p{\cdot}e = p$) for any \,$p \in M$. 
If \,$M$ \,has a left identity \,$e$ \,and a right identity \,$e'$ \,then 
\,$e = e'$ \,which is an {\it identity} \,(or neutral element) of \,$M$.
\\
In order to characterize the Cayley graphs of left-cancellative magmas with or 
without an identity, we need to restrict the path-relation. 
For any graph \,$G$ \,and any vertex \,$r$, we define the {\it edge-relation} 
\,${\rm Edge}_G(r)$ \,as the ternary relation on \,$V_G$ \,defined by\\[0.25em]
\hspace*{3em}$(s,t,x) \in {\rm Edge}_G(r)$ \ if there exists \,$a \in A_G$ 
\,such that \,$r\ \fleche{a}_G\ t$ \,and \,$s\ \fleche{a}_G\ x$.\\[0.25em]
So \,${\rm Edge}_G(r) \,\subseteq \,{\rm Path}_G(r)$. 
If for any \,$s,t \in V_G$ \,there exists a unique \,$x$ \,such that 
\,$(s,t,x) \in {\rm Edge}_G(r)$, we denote by 
\,$\croix_r : V_G{\croix}V_G\ \fleche{}\ V_G$ \,the binary {\it edge-operation} 
\,on \,$V_G$ \,defined by \,$(s,t,s \,\croix_r \,t) \in {\rm Edge}_G(r)$ \,for 
any \,$s,t \in V_G$\,; we also write \,$_G\croix_r$ \,when we need to specify 
\,$G$.\\
Let us give conditions for the existence of this edge-operation. 
We need to introduce two basic graph notions. 
We say that a vertex \,$r$ \,of a graph \,$G$ \,is an \,$1${\it -root} \,if 
\,$r\ \fleche{}_G\ s$ \,for any vertex \,$s$ \,of \,$G$. 
Thus a graph is complete if and only if all its vertices are $1$-roots.
\begin{fact}\label{1Root}
Any left identity of a magma \,$M$ \,is an 1-root of \,${\cal C}(M)$.
\end{fact}
Moreover we say that a graph is {\it loop-complete} \,if one vertex has an 
\,$a$-loop then all the vertices have an \,$a$-loop:\\[0.25em]
\hspace*{6em}\,$\exists\ r \in V_G\ (r\ \fleche{a}_G\ r) \ \ \Longrightarrow \ 
\ \forall\ s \in V_G\ (s\ \fleche{a}_G\ s)$.
\begin{fact}\label{LoopComp}
Any generalized Cayley graph of a left-cancellative magma with a right 
identity\\
\hspace*{1.25em}is loop-complete.
\end{fact}
\proof\mbox{}\\
Let \,$M$ \,be a left-cancellative magma with a right-identity \,$e$.\\
Let \,$G \,= \,{\cal C}\inter{M,Q}$ \,be a generalized Cayley graph of \,$M$.
\\[0.25em]
Let \,$p\ \fleche{\interInd{q}}_G\ p$ \,for some \,$p \in M$ \,and \,$q \in Q$. 
So \,$p{\cdot}q = p = p{\cdot}e$.\\
As \,$M$ \,is left-cancellative, $q = e$. 
Thus \,$r\ \fleche{\interInd{q}}_G\ r{\cdot}q \,= \,r{\cdot}e \,= \,r$ \,for 
any \,$r \in M$.
\qed\\[1em]
Let us adapt Propositions~\ref{PathProduct} to the edge-operation.
\begin{proposition}\label{EdgeProduct}
Let \,$r$ \,be an $1$-root of a deterministic source-complete simple graph 
\,$G$.\\
\hspace*{0.5em}Then \,$(V_G,\croix_r)$ \,is a left-cancellative magma of 
left-identity \,$r$ \,and\\
\hspace*{10em}$G \,= \,{\cal C}\inter{V_G}$ \ with \ $\inter{s} \,= \,a$ \ for 
any \,$r\ \fleche{a}_G\ s$.\\
\hspace*{0.5em}If \,$G$ \,is loop-complete then \,$r$ \,is an identity.
\end{proposition}
\proof\mbox{}\\
{\bf i)} Let \,$s,t \in V_G$\,. 
\,Let us check that there is a unique \,$x$ \,such that 
\,$(s,t,x) \in {\rm Edge}_G(r)$.\\
As \,$r$ \,is an \,$1$-root and \,$G$ \,is simple, there exists a unique 
\,$a \in A_G$ \,such that \,$r\ \fleche{a}_G\ t$.\\
As \,$G$ \,is source-complete and deterministic, there exists a unique vertex 
\,$x$ \,such that \,$s\ \fleche{a}_G\ x$. 
Thus \,$\croix_r$ \,exists and is denoted by \,$\cdot$ \,in the rest of this 
proof.\\[0.25em]
{\bf ii)} Let us check that \,$(V_G,\cdot)$ \,is left-cancellative. 
Assume that \,$s{\cdot}t \,= \,s{\cdot}t'$.\\
As \,$r$ \,is an \,$1$-root, there exists \,$a,a' \in A_G$ \,such that 
\,$r\ \fleche{a}_G\ t$ \,and \,$r\ \fleche{a'}_G\ t'$.\\
By definition of \,$\cdot$ \,we get \,$s\ \fleche{a}_G\ s \cdot t$ \,and 
\,$s\ \fleche{a'}_G\ s{\cdot}t' \,= \,s{\cdot}t$.\\
As \,$G$ \,is simple, we have \,$a = a'$. 
As \,$G$ \,is deterministic, it follows that \,$t = t'$.\\[0.25em]
{\bf iii)} Let us check that \,$r$ \,is a left identity of \,$(V_G,\cdot)$.\\
Let \,$s \in V_G$\,. As \,$r$ \,is an \,$1$-root, there exists \,$a \in A_G$ 
\,such that \,$r\ \fleche{a}_G\ s$.\\
By definition of \,$\cdot$ \,we have \,$r\ \fleche{a}_G\ r{\cdot}s$. 
As \,$G$ \,is deterministic, we get \,$r{\cdot}s = s$.\\[0.25em]
{\bf iv)} As \,$r$ \,is an \,$1$-root and \,$G$ \,is simple, we can define the 
mapping \,$\inter{\ } : V_G \,\fleche{} \,A_G$ \,by\\[0.25em]
\hspace*{12em}$\inter{s} \,= \,a$ \ \ for \,$r\ \fleche{a}_G\ s$.\\[0.25em]
As \,$G$ \,is deterministic, \,$\inter{\ }$ \,is an injection. 
As \,$G$ \,is source-complete, \,$\inter{\ }$ \,is a bijection.\\
Let us show that \,$G \,= \,{\cal C}\inter{V_G}$.\\[0.25em]
$\subseteq$\,: \,Let \,$s\ \fleche{a}_G\ t$.
As \,$G$ \,is source-complete, there exists \,$r'$ \,such that 
\,$r\ \fleche{a}_G\ r'$.\\
So \,$s\ \fleche{a}_G\ s{\cdot}r'$. 
As \,$G$ \,is deterministic, $s{\cdot}r' \,= \,t$. 
We have \,$\inter{r'} \,= \,a$ \,hence 
\,$s\ \fleche{a}_{{\cal C}\interInd{V_G}}\ s{\cdot}r' \,= \,t$.\\[0.25em]
$\supseteq$\,: \,Let \,$s\ \fleche{a}_{{\cal C}\interInd{V_G}}\ t$.
There exists (a unique) \,$r' \in V_G$ \,such that \,$\inter{r'} = a$.\\
Thus \,$t \,= \,s{\cdot}r'$ \,and \,$r\ \fleche{a}_G\ r'$. 
So \,$s\ \fleche{a}_G\ s{\cdot}r' \,= \,t$.\\[0.25em]
{\bf v)} Assume that \,$G$ \,is loop-complete. 
Let us check that \,$r$ \,is also a right identity.\\
As \,$r$ \,is an \,$1$-root, there is (a unique) \,$a \in A_G$ 
\,such that \,$r\ \fleche{a}_G\ r$.\\
Let \,$s \in V_G$\,. \,As \,$G$ \,is loop-complete, we get 
\,$s\ \fleche{a}_G\ s$. 
By definition of \,$\cdot$ \,we have \,$s\ \fleche{a}\ s{\cdot}r$.\\
As \,$G$ \,is deterministic, \,$s = s{\cdot}r$. 
Thus \,$r$ \,is a right identity.
\qed\\[1em]
We get a fully graph-theoretic characterization of the Cayley graphs 
\,${\cal C}\inter{M}$ \,for any left-cancellative magma \,$M$ \,with a left 
identity.
\begin{proposition}\label{CayleyMagma}
A graph is equal to \,${\cal C}\inter{M}$ \,for some left-cancellative magma 
\,$M$ \,with a left identity \ if and only if \ it is simple, deterministic, 
source-complete and \,$1$-rooted.
\end{proposition}
\proof\mbox{}\\
$\Longrightarrow$\,: \,let \,$G \,= \,{\cal C}\inter{M}$ \,for some 
left-cancellative magma \,$(M,\cdot)$ \,with a left identity \,$r$, and some 
injective mapping \,$\inter{\ }$. 
By Facts~\ref{Magma} and \ref{CancelMagma}, \,$G$ \,is deterministic, 
source-complete and simple. 
By Fact~\ref{1Root}, $r$ \,is an \,$1$-root of \,$G$.\\[0.25em]
$\Longleftarrow$\,: \,By Proposition~\ref{EdgeProduct}.
\qed\\[1em]
Under the assumption of the axiom of choice, we can characterize the 
generalized Cayley graphs of left-cancellative magmas.
\begin{theorem}\label{LeftCancel}
In ZFC set theory, the following graphs define the same family\,{\sf :}\\
{\bf a)} the generalized Cayley graphs of left-cancellative magmas,\\
{\bf b)} the generalized Cayley graphs of left-cancellative magmas with a left
identity,\\
{\bf c)} the simple, deterministic, source-complete graphs.
\end{theorem}
\proof\mbox{}\\
{\bf b)} \,$\Longrightarrow$ \,{\bf a)}\,: \,immediate.\\
{\bf a)} \,$\Longrightarrow$ \,{\bf c)}\,: \,by Facts~\ref{Magma} and 
\ref{CancelMagma}.\\
{\bf c)} \,$\Longrightarrow$ \,{\bf b)}\,: \,let \,$G$ \,be a simple, 
deterministic and source-complete graph.\\
Assume the axiom of choice. 
Let \,$r$ \,be a vertex of \,$G$ \,with\\
\hspace*{6em}$V_G - \fleche{}_G(r) \ = 
\ \{\ s \in V_G\ |\ r\ \nofleche{}{}\,\!_G\ s\ \}$ \ of minimal cardinality.\\
For each vertex \,$s$, we take an injection \,$f_s$ \,from 
\,$V_G - \fleche{}_G(r)$ \,to \,$V_G - \fleche{}_G(s)$ \,and whose \,$f_r$ \,is 
the identity. We define the graph\\[0.25em]
\hspace*{1em}$\InfSup{G} \ = \ 
\{\ s\ \fleche{p}\ t\ |\ \exists\ a\ (s\ \fleche{a}_G\ t 
\,\wedge\ r\ \fleche{a}_G\ p)\ \} \,\cup \,\{\ s\ \fleche{t}\ f_s(t)\ |\ 
s \in V_G \,\wedge \,t \in {\rm Dom}(f_s)\ \}$.\\[0.25em]
Thus \,$\InfSup{G}$ \,remains simple, deterministic, source-complete with 
\,$V_{\InfSupInd{G}} \,= \,V_G \,= \,A_{\InfSupInd{G}}$\,.\\
Furthermore \,$r\ \fleche{s}_{\InfSupInd{G}}\ s$ \,for any \,$s \in V_G$ \,hence 
\,$r$ \,is an $1$-root of \,$\InfSup{G}$.\\
By Proposition~\ref{EdgeProduct}, $\InfSup{G} \,= \,{\cal C}(V_G)$ \,for 
\,$(V_G,\croix_r)$ \,left-cancellative, of left identity \,$r$ \,with\\[0.25em]
\hspace*{10em}$s\ \fleche{t}_{\InfSupInd{G}}\ s \,\croix_r \,t$ \ for any 
\,$s,t \in V_G$\,.\\[0.25em]
Finally \,$G \,= \,{\cal C}\inter{V_G\,,\,\fleche{}_G(r)}$ \,with 
\,$\inter{s} = a$ \,for any \,$r\ \fleche{a}_G\ s$.
\qed\\[1em]
For instance, let \,$G\ =\ \{\ m\ \fleche{0}\ 0\ |\ m \geq 0\ \} \,\cup 
\,\{\ m\ \fleche{1}\ m+1\ |\ m \geq 0\ \}$ \,represented by
\begin{center}
\includegraphics{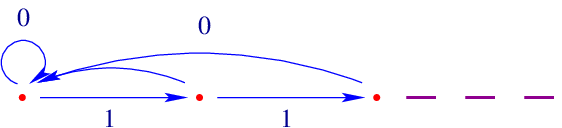}\\%magnitude 0.3
\end{center}
It is simple, deterministic, source-complete but not loop-complete.
By Theorem~\ref{LeftCancel}, this graph is a generalized Cayley graph of a 
left-cancellative magma with a left identity. Precisely we complete \,$G$ 
\,into the graph\\[0.25em]
\hspace*{6em}$\InfSup{G}\ =\
\{\ m\ \fleche{0}\ 0\ |\ m \geq 0\ \} \,\cup 
\,\{\ m\ \fleche{n}\ m+n\ |\ m \geq 0 \,\wedge \,n > 0\ \}$\\[0.25em]
having \,$0$ as \,$1$-root, and which remains simple, deterministic, 
source-complete.\\
By Proposition~\ref{EdgeProduct}, the magma \,$(\entier,{\croix}_0)$ \,with 
the edge-operation \,$\croix_0$ \,of \,$\InfSup{G}$ \,{\it i.e.}\\
\hspace*{6em}$m \,\croix_0 \,0 \,= \,0$ \ and \ $m \,\croix_0 \,n \,= \,m + n$ 
\ for any \,$m \geq 0$ \,and \,$n > 0$\\
is left-cancellative and \,$0$ \,is a left identity. 
Furthermore \,$G \,= \,{\cal C}(\entier,\{0,1\})$.\\[0.5em]
We can now characterize the generalized Cayley graphs of the left-cancellative 
magmas with an identity. 
We just have to add the loop-complete property to restrict 
Proposition~\ref{CayleyMagma} to left-cancellative magmas with an identity 
element.
\begin{proposition}\label{CayleyMagmaUnit}
A graph is equal to \,${\cal C}\inter{M}$ \,for some left-cancellative magma 
\,$M$ \,with an identity \ if and only if \ it is simple, deterministic, 
source-complete, loop-complete and \,$1$-rooted.
\end{proposition}
\proof\mbox{}\\
$\Longrightarrow$\,: \,By Proposition~\ref{CayleyMagma} and 
Fact~\ref{LoopComp}.\\[0.25em]
$\Longleftarrow$\,: \,By Proposition~\ref{EdgeProduct}.
\qed\\[1em]
We restrict Theorem~\ref{LeftCancel} to left-cancellative magmas having a 
right identity.
\begin{theorem}\label{UnitCancel}
In ZFC set theory, the following graphs define the same family\,{\sf :}\\
{\bf a)} the generalized Cayley graphs of left-cancellative magmas with a right
identity,\\
{\bf b)} the generalized Cayley graphs of left-cancellative magmas with an 
identity,\\
{\bf c)} the simple, deterministic, source-complete and loop-complete graphs.
\end{theorem}
\proof\mbox{}\\
{\bf b)} \,$\Longrightarrow$ \,{\bf a)}\,: \,immediate.\\
{\bf a)} \,$\Longrightarrow$ \,{\bf c)}\,: 
\,By Facts~\ref{Magma}, \ref{CancelMagma}, \ref{LoopComp}.\\
{\bf c)} \,$\Longrightarrow$ \,{\bf b)}\,: \,Let \,$G$ \,be a graph which is 
simple, deterministic, source and loop-complete.\\
Let us apply the construction given in the proof of Theorem~\ref{LeftCancel}. 
As \,$G$ \,is loop-complete and if \,$r\ \nofleche{}{}\,\!_G\ r$, we can add 
the condition that \,$f_s(r) = s$ \,for any \,$s \in V_G$\,. 
Thus \,$\InfSup{G}$ \,remains loop-complete and by 
Proposition~\ref{EdgeProduct}, $r$ \,is an identity for \,$\croix_r$\,.
\qed\\[1em]
For instance, we denote by \,$\entier\bBas{+} \,= \,\entier - \{0\}$ \,and we 
consider the graph\\[0.25em]
\hspace*{6em}$G \ = \ \{\ 0^n\ \fleche{0}\ 0^{n+1}\ |\ n \geq 0\ \} \,\cup
\,\{\ ui\ \fleche{0}\ u\ |\ u \in 0^*\entier\bBas{+}\!\!^* \,\wedge 
\,i \in \entier\bBas{+}\ \}$.\\[0.25em]
of vertex set \,$V_G \,= \,0^*\entier\bBas{+}\!\!^*$ \,and represented as 
follows:
\begin{center}
\includegraphics{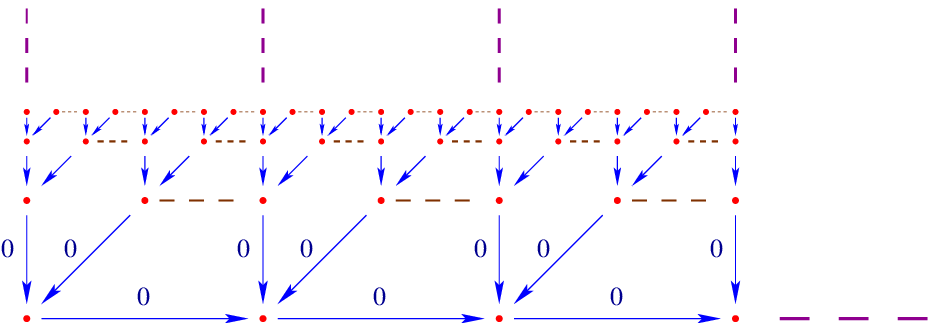}\\%magnitude 0.3
\end{center}
It is simple, deterministic, source-complete and loop-complete (and 
arc-symmetric). 
By Theorem~\ref{UnitCancel}, this graph is a generalized Cayley graph of a
left-cancellative magma having an identity. 
For instance, we complete \,$G$ \,into the graph\\[0.25em]
\hspace*{6em}$\InfSup{G} \ = \ G \,\cup 
\,\{\ 0^mu\ \fleche{0^nv}\ 0^{m+n}uv\ |\ m,n \geq 0 \,\wedge 
\,u,v \in \entier\bBas{+}\!\!^* \,\wedge \,0^nv \neq 0\ \}$\\[0.25em]
which remains simple, deterministic, source-complete, loop-complete, with the 
\,$1$-root \,$\varepsilon$.\\
By Proposition~\ref{EdgeProduct}, $G \,= \,{\cal C}(V_G,\{0\})$ \,for the 
left-cancellative magma \,$(V_G,\croix_{\varepsilon})$ \,of identity 
\,$\varepsilon$ \,with the edge-operation \,$\croix_{\varepsilon}$ \,of 
\,$\InfSup{G}$ \,defined for any \,$m,n \geq 0$, 
$u,v \in \entier\bBas{+}\!\!^*$ \,and \,$i \in \entier\bBas{+}$ \,by\\[0.25em]
\hspace*{6em}$0^mui \,\croix_{\varepsilon} \,0 \,= \,0^mu$ \ \ otherwise \ \ 
$0^mu \,\croix_{\varepsilon} \,0^nv \,= \,0^{m+n}uv$.\\[0.25em]
We will see that we can define \,$\InfSup{G}$ \,so that in addition, any 
vertex is an $1$-root \,{\it i.e.} \,$\InfSup{G}$ \,is complete 
(see Theorem~\ref{BoundedLeftQuasiCayley}).

\section{Generalized Cayley graphs of left-quasigroups}

We can now refine the previous characterization of generalized Cayley graphs 
from left-cancellative magmas to left-quasigroups 
(Theorem~\ref{LeftQuasiCayley}). 
These algebraic structures define the same family of finitely labeled 
generalized Cayley graphs (Theorem~\ref{BoundedLeftQuasiCayley}).\\[-0.5em]

A magma \,$(M,\cdot)$ \,is a {\it left-quasigroup} \,if for each 
\,$p,q \in \mathsf{M}$, there is a unique \,$r \in M$ \,such that 
\,$p{\cdot}r = q$ \,denoted by \,$r = p{\backslash}q$ \,the 
{\it left quotient} \,of \,$q$ \,by \,$p$.\\
This property ensures that each element of \,$M$ \,occurs exactly once 
in each row of the Cayley table for \,$\cdot$\,. For instance \,$\{a,b,c\}$ 
\,is a left-quasigroup for \,$\cdot$ \,defined by the following Cayley table:
\\[0.5em]
\hspace*{16em}\begin{tabular}{|c||c|c|c|}
\hline
$\cdot$ & $a$ & $b$ & $c$ \\
\hline\hline
$a$ & $a$ & $b$ & $c$ \\
\hline
$b$ & $b$ & $a$ & $c$ \\
\hline
$c$ & $c$ & $b$ & $a$ \\
\hline
\end{tabular}\\[0.5em]
Note that \,$\cdot$ \,is not associative since \,$c \cdot (b \cdot c) \,= \, a$
\,and \,$(c \cdot b) \cdot c \,= \, c$, and is not right-cancellative since 
\,$a{\cdot}b \,= \,c{\cdot}b$. 
The first figure in Section~\ref{CayMon} is a representation of \,${\cal C}(M)$ 
\,for the semigroup \,$M = \{a,b\}$ \,with \,$x{\cdot}y = y$ \,for any 
\,$x,y \in M$. 
This semigroup is also a left-quasigroup which is not right-cancellative.\\
Any left-quasigroup \,$M$ \,is left-cancellative. 
The converse is true for \,$M$ \,finite but is false in general: 
\,$(\entier,+)$ \,is cancellative but is not a left-quasigroup. Indeed,
\\[0.25em]
\hspace*{1em}$M$ \,is a left-quasigroup \ \
$\Longleftrightarrow$ \ \ $M$ \,is left-cancellative \,and 
\,$p{\cdot}M \,= \,M$ \ for any \,$p \in M$.\\[0.25em]
Let us refine Proposition~\ref{EdgeProduct} in the case where the graph is 
also complete.
\begin{proposition}\label{EdgeProductLeftQuasi}
Let \,$G$ \,be a simple, deterministic, complete and source-complete graph.\\
\hspace*{1em}For any vertex \,$r$, \,$(V_G,\croix_r)$ \,is a left-quasigroup.
\end{proposition}
\proof\mbox{}\\
Let \,$r$ \,be a vertex of \,$G$. 
As \,$G$ \,is complete, \,$r$ \,is an $1$-root.\\
By Proposition~\ref{EdgeProduct}, \,$(V_G,\croix_r)$ \,is a left-cancellative 
magma.\\
Let \,$s \in V_G$\,. 
It remains to check that \,$V_G \,\subseteq \,s \,\croix_r \,V_G$\,. 
Let \,$t \in V_G$\,.\\
As \,$G$ \,is complete, there exists \,$a \in A_G$ \,such that 
\,$s\ \fleche{a}_G\ t$.\\
As \,$G$ \,is source-complete, there exists \,$t' \in V_G$ \,such that 
\,$r\ \fleche{a}_G\ t'$.\\
By definition of \,$\croix_r$ \,we have \,$s\ \fleche{a}_G\ s\,\croix_r\,t'$. 
As \,$G$ \,is deterministic, we get \,$t \,= \,s\,\croix_r\,t'$.
\qed\\[1em]
Let us give a simple characterization of the generalized Cayley graphs for the 
left-quasigroups. For the previous left-quasigroup \,$M$, its graph 
\,${\cal C}(M)$ \,is the following:
\begin{center}
\includegraphics{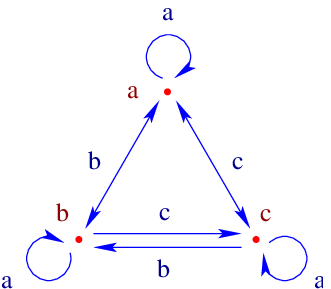}\\%magnitude 0.3
\end{center}
We begin by characterizing these graphs.
\begin{proposition}\label{CayleyLeftQuasi}
We have the following equivalences\,{\sf :}\\
{\bf a)} a graph is equal to \,${\cal C}\inter{M}$ \,for some left-quasigroup 
\,$M$ (resp. with a right identity),\\
{\bf b)} a graph is equal to \,${\cal C}\inter{M}$ \,for some left-quasigroup 
\,$M$ \,with a left identity (resp. identity),\\
{\bf c)} a graph is simple, deterministic, complete, source-complete (resp. 
and loop-complete).
\end{proposition}
\proof\mbox{}\\
{\bf b)} \,$\Longrightarrow$ \,{\bf a)}\,: \,immediate.\\[0.25em]
{\bf a)} \,$\Longrightarrow$ \,{\bf c)}\,: \,let \,$G \,= \,{\cal C}\inter{M}$ 
\,for some left-quasigroup \,$(M,\cdot)$ \,and injective mapping 
\,$\inter{\ }$.\\
By Facts~\ref{Magma} and \ref{CancelMagma}, \,$G$ \,is deterministic, 
source-complete and simple.\\
For any \,$p,q \in M$, we have 
\,$p\ \fleche{\interInd{p{\backslash}q}}_G\ q$ \,hence \,$G$ \,is complete.\\
If in addition \,$M$ \,has a right identity then, by Fact~\ref{LoopComp}, 
\,$G$ \,is loop-complete.\\[0.25em]
{\bf c)} \,$\Longrightarrow$ \,{\bf b)}\,: \,by 
Propositions \ref{EdgeProduct} and \ref{EdgeProductLeftQuasi}.
\qed\\[1em]
For instance the magma \,$M = \{0,1\}$ \,with \,$i \cdot j = 1-j$ \,for any 
\,$i,j \in \{0,1\}$, is a left-quasigroup without left and right identity 
element. 
By Propositions~\ref{EdgeProduct} and \ref{EdgeProductLeftQuasi}, the magma 
\,$N = \{0,1\}$ \,with the edge-operation \,$\croix_0$ \,of \,${\cal C}(M)$ 
\,defined by \,$i \,\croix_0 \,j = j$ \,for any \,$i,j \in \{0,1\}$ \,is a
left-quasigroup with \,$0$ \,and \,$1$ \,are left identities, and 
\,${\cal C}(M) \,= \,{\cal C}\inter{N}$ \,where \,$\inter{0} = 1$ \,and 
\,$\inter{1} = 0$.\\[0.25em]
We now extend Proposition~\ref{CayleyLeftQuasi} to the generalized Cayley 
graphs of left-quasigroups. 
To do this, we must recall and define basic graph notions.\\
Let \,$G$ \,be a graph. For any vertex \,$s$, its {\it out-degree} 
\,$\delta^+_G(s) \,= \,|G \,\cap \,\{s\}{\croix}A{\croix}V_G|$ \,is the number 
of edges of source \,$s$. The {\it out-degree} \,of \,$G$ \,is the cardinal 
\,$\Delta^+_G\ =\ {\rm sup}\{\ \delta^+_G(s)\ |\ s \in V_G\ \}$. 
We say that \,$G$ \,is of {\it bounded out-degree} \,when \,$\Delta^+_G$ \,is
finite.\\
We have \,$\Delta^+_G \,= \,|A_G|$ \,for \,$G$ \,deterministic and 
source-complete.
\begin{fact}\label{OutDegree}
For any vertex \,$s$ \,of a graph \,$G$, we have\\[0.25em]
\hspace*{2em}$\delta^+_G(s)\ \leq\ |A_G|$ \ for \,$G$ \,deterministic, and 
\,$|A_G|\ \leq\ \delta^+_G(s)$ \ for \,$G$ \,source-complete.
\end{fact}
In particular by Fact~\ref{Magma} and for any generalized Cayley graph \,$G$,\\
\hspace*{6em}$G$ \,is of bounded out-degree \ \ $\Longleftrightarrow$ \ \ 
$G$ \,is finitely labeled.\\
By removing the labeling of a graph \,$G$, we get the binary edge relation on 
\,$V_G$\,:\\[0.25em]
\hspace*{9em}$\fleche{}_G\ =\ \{\ (s,t)\ |\ \exists\ a \in A_G \ (s,a,t) \in G
\ \}$.\\[0.25em]
Let \,$R \,\subseteq \,V{\croix}V$ \,be a binary relation on a set \,$V$ 
\,{\it i.e.} \,is an unlabeled graph.\\
The {\it image} \,of \,$P \subseteq V$ \,by \,$R$ \,is the set 
\,$R(P) \,= \,\{\ t\ |\ \exists\ s \in P \ (s,t) \in R\ \}$.\\
So the out-degree of \,$s \in V$ \,is \,$\delta_R^+(s) \,= \,|R(s)|$ \,and
\,$\Delta^+_R\ =\ {\rm sup}\{\ \delta_R^+(s)\ |\ s \in V\ \}$ \,is the 
out-degree of \,$R$. 
For any graph \,$G$ \,and any vertex \,$s$, we have 
\,$\delta_{\rightarrow_G}^+(s) \leq \delta_G^+(s)$ \,hence 
\,$\Delta_{\rightarrow_G}^+ \leq \Delta_G^+$, and we have equalities for \,$G$ 
\,simple.\\
A relation \,$R$ \,is an {\it out-regular relation} \,if 
all the elements of \,$V$ \,have the same out-degree: 
\,$|R(s)| \,= \,|R(t)|$ \,for any \,$s,t \in V$.\\
Let us give simple conditions on \,$R$ \,so that its complement 
\,$V{\croix}V - R$ \,is out-regular.
\begin{lemma}\label{CoOutRegular}
Let \,$R \subseteq V{\croix}V$ \,and \,$S = V{\croix}V-R$ \,the complement of 
\,$R$ \,w.r.t. \,$V{\croix}V$.\\
\hspace*{6.5em}If \,$R$ \,is finite and out-regular \,then 
\,$S$ \,is out-regular.\\
\hspace*{6.5em}If \,$R$ \,is infinite and \,$\Delta^+_R < |V|$ \,then 
\,$S$ \,is out-regular.
\end{lemma}
\proof\mbox{}\\
{\bf i)} When \,$R$ \,is finite, we have \,$S(s) \ = \ |V| - |R(s)|$ \ for any 
\,$s \in V$.\\[0.25em]
In addition for \,$R$ \,out-regular and for any \,$s,t \in V$, 
\,$|R(s)| \,= \,|R(t)|$ \,hence \,$|S(s)| \,= \,|S(t)|$.\\[0.25em]
{\bf ii)} When \,$R$ \,is infinite with \,$\Delta^+_R < |V|$, we have 
\,$|S(s)| \,= \,|V|$ \ for any \,$s \in V$.\\
Hence \,$S$ \,is out-regular on \,$V$ \,with \,$\Delta^+_S \,= \,|V|$.
\qed\\[1em]
We say that a graph \,$G$ \,is {\it out-regular} \,if \,$\fleche{}_G$ \,is 
out-regular \,{\it i.e.} \,all the vertices have the same number of targets. 
For instance 
\,$G \,= \,\{s\ \fleche{a}\ t\,,\,t\ \fleche{a}\ s\,,\,t\ \fleche{b}\ s\}$ 
\,is out-regular since \,$\fleche{}_G(s) = \{t\}$ \,and 
\,$\fleche{}_G(t) = \{s\}$ \,while \,$\delta_G^+(s) = 1$ \,and 
\,$\delta_G^+(t) = 2$.\\
We also say that \,$G$ \,is {\it co-out-regular} \,if its unlabeled complement 
is out-regular \,{\it i.e.}\\
\hspace*{6em}$(V_G{\croix}V_G \,- \fleche{}_G) \ = \ \nofleche{}{}\,\!_G$ \ is 
an out-regular relation on \,$V_G$\,.\\
Under the assumption of the axiom of choice, we can characterize the 
generalized Cayley graphs of left-quasigroups.
\begin{theorem}\label{LeftQuasiCayley}
In ZFC set theory, the following graphs define the same family\,{\sf :}\\
{\bf a)} the generalized Cayley graphs of left-quasigroups (resp. with a right 
identity),\\
{\bf b)} the generalized Cayley graphs of left-quasigroups with a left 
identity (resp. an identity),\\
{\bf c)} the simple, deterministic, source-complete (resp. and loop-complete) 
co-out regular graphs.
\end{theorem}
\proof\mbox{}\\
{\bf b)} \,$\Longrightarrow$ \,{\bf a)}\,: \,immediate.\\[0.25em]
{\bf a)} \,$\Longrightarrow$ \,{\bf c)}\,: 
\,let \,$G \,= \,{\cal C}\inter{M,Q}$ \,for some left-quasigroup \,$M$ \,and 
\,$Q \subseteq M$.\\
By Proposition~\ref{CayleyLeftQuasi}, 
\,$G \,= \,{\cal C}\inter{M}^{|\,\interInd{Q}}$ \,remains simple, deterministic 
and source-complete.\\
For any \,$s \in M$, \,$\delta^+_{\rightarrow\!\!\!\!\!/\,_G}(s) \,= \,|M-Q|$ 
\,hence \,$G$ \,is co-out-regular.\\
If in addition \,$M$ \,has a right identity then, by Fact~\ref{LoopComp}, 
\,$G$ \,is loop-complete.\\[0.25em]
{\bf c)} \,$\Longrightarrow$ \,{\bf b)}\,: \,let \,$G$ \,be a graph which is 
simple, deterministic, co-out-regular, source-complete (resp. and 
loop-complete). 
The co-out-regularity of \,$G$ \,means that 
\,$|V_G - \fleche{}_G(s)| \,= \,|V_G - \fleche{}_G(t)|$ \,for any 
\,$s,t \in V_G$\,. 
Let \,$r$ \,be a vertex of \,$G$. Assume the axiom of choice. 
Let us apply the construction given in the proof of Theorem~\ref{LeftCancel}. 
As \,$G$ \,is co-out-regular, we can now take for each vertex \,$s$ \,a 
bijection \,$f_s$ \,from \,$V_G - \fleche{}_G(r)$ \,to 
\,$V_G - \fleche{}_G(s)$ \,and whose \,$f_r$ \,is the identity. 
As for the proof of Theorem~\ref{UnitCancel}, if \,$G$ \,is loop-complete and 
\,$r\ \nofleche{}{}\,\!_G\ r$, we add the condition that \,$f_s(r) = s$ \,for 
any \,$s \in V_G$\,.\\
Thus \,$\InfSup{G}$ \,remains simple, deterministic, source-complete (resp. 
loop-complete) and is in addition a complete graph. 
By Proposition~\ref{EdgeProductLeftQuasi}, $(V_G,\croix_r)$ \,is a 
left-quasigroup. By Proposition~\ref{EdgeProduct}, $r$ \,is a left-identity 
(resp. is an identity) and $G \,= \,{\cal C}\inter{V_G\,,\,\fleche{}_G(r)}$ 
\,with \,$\inter{s} = a$ \,for any \,$r\ \fleche{a}_G\ s$.
\qed\\[1em]
For instance, let \,$G \ = \ \{\ m\ \fleche{0}\ 0\ |\ m \geq 0\ \} \,\cup
\,\{\ m\ \fleche{1}\ m+1\ |\ m \geq 0\ \}$ \,be the graph that we had 
considered after Theorem~\ref{LeftCancel} and represented by
\begin{center}
\includegraphics{demiraz.eps}\\%magnitude 0.3
\end{center}
By adding edges, we transform \,$G$ \,into the following complete graph:
\\[0.25em]
\hspace*{1.5em}$\InfSup{G} \ = \ G 
\,\cup \,\{\ m\ \fleche{n}\ n-1\ |\ 2 \leq n \leq m+1\ \} \,\cup 
\,\{\ m\ \fleche{n}\ n\ |\ m \geq 0 \,\wedge \,n > m+1\ \}$\\[0.25em]
which remains simple, deterministic and source-complete.\\
By Propositions~\ref{EdgeProduct} and \ref{CayleyLeftQuasi}, 
$G \,= \,{\cal C}(\entier,\{0,1\})$ \,for the left-quasigroup 
\,$(\entier,\croix_0)$ \,of left-identity~\,$0$ \,with the edge-operation 
\,$\croix_0$ \,of \,$\InfSup{G}$ \,defined for any \,$m \geq 0$ \,by\\[0.25em]
\hspace*{3em}\begin{tabular}{rclcrcll}
$m \,\croix_0 \,0$ & $=$ & $0$ & ; & $m \,\croix_0 \,n$ & $=$ & $n-1$ & 
$\forall\ 2 \leq n \leq m+1$\\[0.25em]
$m \,\croix_0 \,1$ & $=$ & $m+1$ & ; & $m \,\croix_0 \,n$ & $=$ & $n$ & 
$\forall\ n > m+1$.
\end{tabular}\\[0.5em]
Another example is given by a graph \,$G$ \,of the following representation:
\begin{center}
\includegraphics{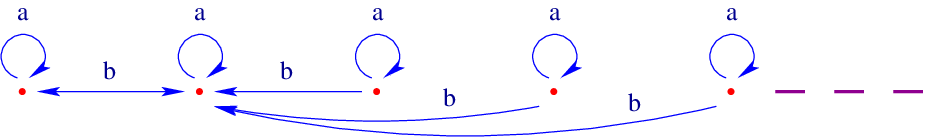}\\%magnitude 0.3
\end{center}
It is simple, deterministic, co-out-regular, source-complete and loop-complete.
\\
By Theorem~\ref{LeftQuasiCayley}, this graph is a generalized Cayley graph of 
a left-quasigroup with an identity. 
Indeed, by replacing \,$a$ \,by \,$0$ \,and \,$b$ \,by \,$1$, \,$G$ \,is 
isomorphic to the following graph:\\[0.25em]
\hspace*{2em}$H\ =\ \{\ m\ \fleche{0}\ m\ |\ m \geq 0\ \} \,\cup 
\,\{\ 1\ \fleche{1}\ 0\ \} \,\cup 
\,\{\ m\ \fleche{1}\ 1\ |\ m \geq 0 \,\wedge \,m \neq 1\ \}$.\\[0.25em]
We complete \,$H$ \,into the graph:\\[0.25em]
\hspace*{6em}\begin{tabular}{rcl}
$\InfSup{H}$ & $=$ & 
$\{\ m\ \fleche{0}\ m\ |\ m \geq 0\ \} \,\cup 
\,\{\ m\ \fleche{m}\ 0\ |\ m \geq 0\ \}$\\[0.35em]
& $\cup$ &$\{\ m\ \fleche{n}\ n\ |\ m \geq 0 \,\wedge \,n > 0 \,\wedge 
\,m \neq n\ \}$.
\end{tabular}\\[0.25em]
So \,$\InfSup{H}$ \,remains simple, deterministic, source-complete and 
loop-complete.\\
Furthermore $\InfSup{H}$ \,is complete \,{\it i.e.} \,any vertex is an 
\,$1$-root.\\
By Proposition~\ref{EdgeProduct}, the magma \,$(\entier,{\croix}_0)$ 
\,with the edge-operation \,$\croix_0$ \,of \,$\InfSup{H}$ \,{\it i.e.}\\
\hspace*{0.5em}$m \,\croix_0 \,0 \,= \,m$ \ ; \ $m \,\croix_0 \,m \,= \,0$ \ ; 
\ $m \,\croix_0 \,n \,= \,n$ \ for any \,$m,n \geq 0$ \,with \,$m \neq n$ 
\,and \,$n > 0$\\
is a left-quasigroup of identity element \,$0$.\\
Furthermore \,$G$ \,is isomorphic to \,${\cal C}\inter{\entier,\{0,1\}}$ 
\,with \,$\inter{0} = a$ \,and \,$\inter{1} = b$.\\[0.5em]
The co-out-regularity in Theorem~\ref{LeftQuasiCayley} can not be removed. 
For instance, consider the monoid \,$(\entier,+)$ \,which is is not a 
left-quasigroup. Its graph 
\,${\cal C}(\entier) \,= \,\{\ m\ \fleche{n}\ m+n\ |\ m,n \geq 0\ \}$ \,is 
simple, deterministic and source-complete. Furthermore we have 
\,$0\ \fleche{n}_{{\cal C}(\scriptsize\entier)}\ n$ \,for any \,$n \geq 0$ 
\,while there is no edge from \,$1$ \,to \,$0$. 
By Proposition~\ref{CayleyLeftQuasi}, this graph is not a generalized Cayley 
graph of a left-quasigroup.\\
By Lemma~\ref{CoOutRegular}, the co-out-regularity in 
Theorem~\ref{LeftQuasiCayley} can be removed for the graphs of bounded 
out-degree which coincides with the characterization of 
Theorem~\ref{LeftCancel}. In this case, we can also remove the assumption of 
the axiom of choice.
\begin{theorem}\label{BoundedLeftQuasiCayley}
For any finitely labeled graph \,$G$, the following properties are 
equivalent\,{\sf :}\\
{\bf a)} $G$ \,is a generalized Cayley graph of a left-cancellative magma 
(resp. with a right identity),\\
{\bf b)} $G$ \,is a gen. Cayley graph of a left-quasigroups with a left 
identity (resp. an identity),\\
{\bf c)} $G$ \,is simple, deterministic, source-complete (resp. and 
loop-complete).
\end{theorem}
\proof\mbox{}\\
{\bf b)} \,$\Longrightarrow$ \,{\bf a)}\,: \,immediate.\\[0.25em]
{\bf a)} \,$\Longrightarrow$ \,{\bf c)}\,: \,by 
Facts~\ref{Magma},\ref{CancelMagma},\ref{LoopComp}.\\[0.25em]
{\bf c)} \,$\Longrightarrow$ \,{\bf b)}\,: \,let \,$G$ \,be a simple, 
deterministic and source-complete graph of finite label set.\\
By Fact~\ref{OutDegree}, $G$ \,is of bounded out-degree. 
To each injective function \,$\ell : A_G\ \fleche{}\ A_G$\,, we associate a 
permutation \,$\overline{\ell}$ \,on \,$A_G$ \,extending \,$\ell$ \,{\it i.e.} 
\,$\overline{\ell}(a) = \ell(a)$ \,for every \,$a \in {\rm Dom}(\ell)$.\\
Let \,$r$ \,be a vertex of \,$G$. For each vertex \,$s$, we associate the 
injective function:\\[0.25em]
\hspace*{10em}$\ell_s \ = \ \{\ (a,b)\ |\ \exists\ t\ (r\ \fleche{a}_G\ t 
\,\wedge \,s\ \fleche{b}_G\ t)\ \}$.\\[0.25em]
We define the following graph:\\[0.25em]
\hspace*{3em}\begin{tabular}{rcl}
$\Infsup{G}$ & $=$ & $\{\ s\ \fleche{p}\ t\ |\ \exists\ a\ (r\ \fleche{a}_G\ p 
\,\wedge\ s\ \fleche{a}_G\ t)\ \}$\\[0.25em]
& $\cup$ & $\{\ s\ \fleche{p}\ t\ |\ \exists\ a \in A_G - {\rm Dom}(\ell_s) \ 
(r\ \fleche{a}_G\ t \,\wedge\ s\ \fleche{\overline{\ell_s}(a)}_G\ p)\ \}$
\\[0.25em]
 & $\cup$ & $\{\ s\ \fleche{t}\ t\ |\ t \in V_G - (\fleche{}_G(r) \,\cup 
\,\fleche{}_G(s))\ \}$.
\end{tabular}\\[0.25em]
For \,$G$ \,of vertex set \,$V_G \,= \,\{p,q,r,s,t,x\}$ \,with the following 
edges from \,$r$ \,and \,$s$\,:
\begin{center}
\includegraphics{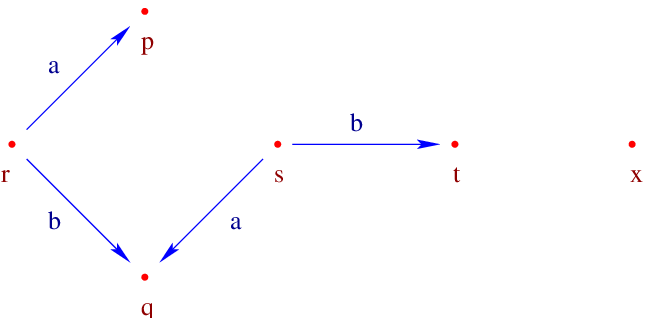}\\%magnitude 0.25
\end{center}
the graph \,$\Infsup{G}$ \,has the following edges from \,$s$\,:
\begin{center}
\includegraphics{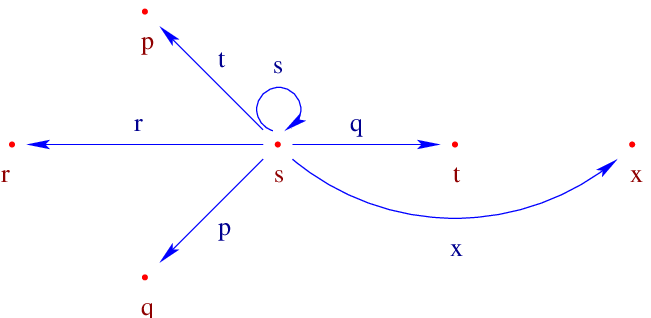}\\%magnitude 0.25
\end{center}
Thus \,$\Infsup{G}$ \,remains simple, deterministic, source complete with 
\,$V_{\InfsupInd{G}} \,= \,V_G \,= \,A_{\InfsupInd{G}}$\,.\\
Furthermore \,$\Infsup{G}$ \,is complete with 
\,$r\ \fleche{s}_{\InfsupInd{G}}\ s$ \,for any \,$s \in V_G$\,. 
By Proposition~\ref{EdgeProduct}, we have \,$\Infsup{G} = {\cal C}(V_G)$ for 
the left-cancellative magma $(V_G,\croix_r)$ of left identity $r$~with\\[0.25em]
\hspace*{10em}$s\ \fleche{t}_{\InfsupInd{G}}\ s \,\croix_r \,t$ \ for any 
\,$s,t \in V_G$\,.\\[0.25em]
Finally \,$G \,= \,{\cal C}\inter{V_G\,,\,\fleche{}_G(r)}$ \,with 
\,$\inter{s} = a$ \,for any \,$r\ \fleche{a}_G\ s$.\\[0.25em]
Now suppose that \,$G$ \,is loop-complete. We distinguish the two complementary 
cases below.\\
{\it Case 1}\,: \,all the vertices of \,$G$ \,have a loop of the same label.\\
\hspace*{2em}Then \,$\Infsup{G}$ \,remains loop-complete and by 
Proposition~\ref{EdgeProduct}, \,$r$ \,is an identity of \,$\croix_r$\,.\\
{\it Case 2}\,: \,$G$ \,has no loop.\\
\hspace*{2em}We take a new label \,$a \in A - A_G$ \,and we redefine 
\,$\Infsup{G}$ \,as being \,$\Infsup{G'}$ \,for\\
\hspace*{2em}$G' \ = \ G \,\cup \,\{\ s\ \fleche{a}\ s \mid s \in V_G\ \}$. 
We conclude by Case 1.
\qed\\[1em]
For the previous example, we have \ $\Infsup{H} \ = \ \InfSup{H}$.\\
For the penultimate example, we have\\[0.25em]
\hspace*{6em}\begin{tabular}{rcl}
$\Infsup{G}$ & $=$ & $\{\ m\ \fleche{1}\ m+1\ |\ m \geq 0\ \} \ \cup 
\ \{\ m\ \fleche{m+1}\ 1\ |\ m \geq 0\ \}$\\[0.25em]
 & $\cup$ & $\{\ m\ \fleche{n}\ n\ |\ m,n \geq 0 \,\wedge \,n \neq 1 \,\wedge 
\,n \neq m+1\ \}$.
\end{tabular}\\[0.25em]
For the last example of the previous section (after Theorem~\ref{UnitCancel}), 
we have\\[0.25em]
\hspace*{0.5em}\begin{tabular}{rcl}
$\Infsup{G}$ & $=$ & $\{\ 0^m\ \fleche{\varepsilon}\ 0^m\ |\ m \geq 0\ \} 
\,\cup \,\{\ 0^m\ \fleche{0}\ 0^{m+1}\ |\ m \geq 0\ \}$\\[0.25em]
 & $\cup$ & $\{\ 0^m\ \fleche{0^m}\ \varepsilon\ |\ m > 0\ \} \,\cup 
\,\{\ 0^m\ \fleche{0^{m+1}}\ 0\ |\ m > 1\ \} \,\cup 
\,\{\,0\ \fleche{00}\ \varepsilon\,\}$\\[0.25em]
 & $\cup$ & $\{\ 0^m\ \fleche{u}\ u\ |\
u \in 0^*\entier\bBas{+}\!\!^* - \{\varepsilon,0,0^m,0^{m+1}\}\ \}$\\[0.25em]
& $\cup$ & $\{\ ui\ \fleche{\varepsilon}\ ui\ |\ 
u \in 0^*\entier\bBas{+}\!\!^* \,\wedge \,i \in \entier\bBas{+}\ \} \,\cup 
\,\{\ ui\ \fleche{0}\ u\ |\ u \in 0^*\entier\bBas{+}\!\!^* \,\wedge 
\,i \in \entier\bBas{+}\ \}$\\[0.25em]
& $\cup$ & $\{\ ui\ \fleche{ui}\ \varepsilon\ |\ 
u \in 0^*\entier\bBas{+}\!\!^*-\{\varepsilon\} \,\wedge 
\,i \in \entier\bBas{+}\ \} \,\cup 
\,\{\ ui\ \fleche{u}\ 0\ |\ u \in 0^*\entier\bBas{+}\!\!^* \,\wedge 
\,i \in \entier\bBas{+}\ \}$\\[0.25em]
 & $\cup$ & $\{\ i\ \fleche{i}\ 0\ |\ i \in \entier\bBas{+}\ \} \,\cup 
\,\{\ ui\ \fleche{v}\ v\ |\ u,v \in 0^*\entier\bBas{+}\!\!^* \,\wedge 
\,i \in \entier\bBas{+} \,\wedge \,v \not\in \{\varepsilon,0,u,ui\}\ \}$.
\end{tabular}

\section{Generalized Cayley graphs of quasigroups}

We can now refine the previous characterization of generalized Cayley graphs 
from left-quasigroups (Theorem~\ref{LeftQuasiCayley}) to quasigroups 
(Theorem~\ref{QuasiCayley}).\\[-0.5em]

A magma \,$(M,\cdot)$ \,is a {\it quasigroup} \,if \,$\cdot$ \,obeys 
the {\it Latin square} property: for each \,$p,q \in \mathsf{M}$,\\
there is a unique \,$r \in \mathsf{M}$ \,such that \,$p{\cdot}r = q$ \,denoted 
by \,$r = p{\backslash}q$ \,the {\it left quotient} \,of \,$q$ \,by \,$p$,\\
there is a unique \,$s \in \mathsf{M}$ \,such that \,$s{\cdot}p = q$ \,denoted 
by \,$s = q/p$ \,the {\it right quotient} \,of \,$q$ \,by~$p$.\\[0.25em]
This property ensures that each element of \,$\mathsf{M}$ \,occurs exactly once 
in each row and exactly once in each column of the Cayley table for 
\,$\cdot$\,. The previous finite left-quasigroup is not a quasigroup. 
On the other hand, \,$(\{a,b,c\},\cdot)$ \,is a quasigroup with \,$\cdot$ 
\,defined by the Cayley table:\\[0.5em]
\hspace*{16em}\begin{tabular}{|c||c|c|c|}
\hline
$\cdot$ & $a$ & $b$ & $c$ \\
\hline\hline
$a$ & $a$ & $c$ & $b$ \\
\hline
$b$ & $c$ & $b$ & $a$ \\
\hline
$c$ & $b$ & $a$ & $c$ \\
\hline
\end{tabular}\\[0.5em]
Its Cayley graph \,${\cal C}(\{a,b,c\})$ \,is represented as follows:
\begin{center}
\includegraphics{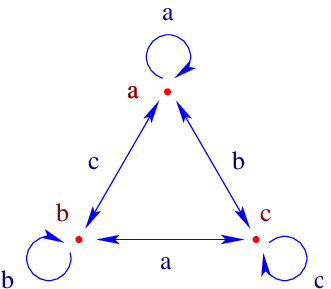}\\%magnitude 0.3
\end{center}
Note that \,$\cdot$ \,is not associative: \,$a \cdot (b \cdot c) \,= \, a$
\,and \,$(a \cdot b) \cdot c \,= \, c$. Furthermore,\\[0.25em]
\hspace*{1em}$M$ \,is a quasigroup \ \ 
$\Longleftrightarrow$ \ \ $M$ \,is cancellative \,and 
\,$p \cdot M \,= \,M \,= \,M \cdot p$ \ for any \,$p \in M$.\\[0.25em]
Let us refine Proposition~\ref{EdgeProductLeftQuasi} in the case where the 
graph is also co-deterministic and target-complete.
\begin{proposition}\label{EdgeProductQuasi}
Let \,$G$ \,be a graph which is simple, deterministic and co-deterministic,\\
\hspace*{0.5em}complete, source-complete and target-complete. 
For any vertex \,$r$, \,$(V_G,\croix_r)$ \,is a quasigroup.
\end{proposition}
\proof\mbox{}\\
Let \,$r$ \,be a vertex of \,$G$ \,and \,$\cdot$ \,be the edge-operation 
\,$\croix_r$\,.\\
By Proposition~\ref{EdgeProductLeftQuasi}, \,$(V_G,\cdot)$ \,is a 
left-quasigroup.\\
{\bf i)} Let us check that \,$(V_G,\cdot)$ \,is right-cancellative. 
Assume that \,$s \cdot t \,= \,s' \cdot t$.\\
As \,$G$ \,is complete, there exists \,$a$ \,such that $r\ \fleche{a}_G\ t$.\\
By definition of \,$\cdot$ \,we have \,$s\ \fleche{a}_G\ s \cdot t$ \,and
\,$s'\ \fleche{a}_G\ s' \cdot t \,= \,s \cdot t$.\\
As \,$G$ \,is co-deterministic, we get \,$s = s'$.\\[0.25em]
{\bf ii)} Let \,$t \in V_G$\,. 
Let us check that \,$V_G \,\subseteq \,V_G \cdot t$\,. Let \,$s \in V_G$\,.\\
As \,$G$ \,is complete, there exists \,$a \in A_G$ \,such that 
\,$r\ \fleche{a}_G\ t$.\\
As \,$G$ \,is target-complete, there exists \,$s' \in V_G$ \,such that 
\,$s'\ \fleche{a}_G\ s$.\\
By definition of \,$\cdot$ \,we have \,$s'\ \fleche{a}_G\ s' \cdot t$. 
As \,$G$ \,is deterministic, we get \,$s \,= \,s' \cdot t$.
\qed\\[1em]
Let us restrict Proposition~\ref{CayleyLeftQuasi} to the quasigroups.
\begin{proposition}\label{CayleyQuasi}
We have the following equivalences\,{\sf :}\\
{\bf a)} a graph is equal to \,${\cal C}\inter{M}$ \,for some quasigroup \,$M$ 
\,(resp. with a right identity),\\
{\bf b)} a graph is equal to \,${\cal C}\inter{M}$ \,for some quasigroup \,$M$ 
\,with a left identity (resp. an identity),\\
{\bf c)} a graph is simple, deterministic, co-deterministic, complete, 
source-complete,\\
\hspace*{1.2em}target-complete (resp. and loop-complete).
\end{proposition}
\proof\mbox{}\\
{\bf b)} \,$\Longrightarrow$ \,{\bf a)}\,: \,immediate.\\[0.25em]
{\bf a)} \,$\Longrightarrow$ \,{\bf c)}\,: \,let 
\,$G \,= \,{\cal C}\inter{M}$ \,for some quasigroup \,$(M,\cdot)$ \,and 
injective mapping \,$\inter{\ }$.\\
By Proposition~\ref{CayleyLeftQuasi}, \,$G$ \,is simple, deterministic, 
complete and source-complete.\\
By Fact~\ref{CancelMagma}, $G$ \,is co-deterministic.\\
For any \,$p,q \in M$, we have \,$p/q\ \fleche{\interInd{q}}_G\ p$ \,hence 
\,$G$ \,is target-complete.\\
If in addition \,$M$ \,has a right identity then, by Fact~\ref{LoopComp}, 
\,$G$ \,is loop-complete.\\[0.25em]
{\bf c)} \,$\Longrightarrow$ \,{\bf b)}\,: 
\,By Propositions \ref{EdgeProduct} and \ref{EdgeProductQuasi}.
\qed\\[1em]
For instance let us consider the division \,$\div$ \,on 
\,$\reel_+ \,= \,]0,+\infty[$. So \,$(\reel_+,\div)$ \,is a quasigroup of right 
identity \,$1$. Furthermore \,$\croix_1$ \,is the multiplication on 
\,$\reel_+$. Thus \,${\cal C}(\reel_+)$ \,for the quasigroup 
\,$(\reel_+,\div)$ \,is equal to \,${\cal C}\inter{\reel_+}$ \,for the group 
\,$(\reel_+,\croix_1)$ \,with \,$\inter{x} = \frac{1}{x}$ \,for any \,$x > 0$.
\\[0.25em]
We now adapt Theorem~\ref{LeftQuasiCayley} to the quasigroups. 
This will require a much more extensive development than what has been done 
with Theorem~\ref{LeftQuasiCayley}.\\
For any vertex \,$s$ \,of a graph \,$G$, its {\it in-degree} 
\,$\delta^-_G(s) \,= \,|G \,\cap \,V_G{\croix}A{\croix}\{s\}| \,= 
\,\delta^+_{G^{-1}}(s)$ \,is the number of edges of target \,$s$, and 
\,$\delta_G(s) = \delta^+_G(s) + \delta^-_G(s)$ \,is the {\it degree} \,of 
\,$s$.\\
The {\it in-out-degree} \,of \,$G$ \,is the cardinal\\[0.25em]
\hspace*{6em}$\Delta_G\ =\ {\rm sup}\bigl(\{\ \delta^+_G(s)\ |\ s \in V_G\ \} 
\,\cup \,\{\ \delta^-_G(s)\ |\ s \in V_G\ \}\bigr)$.\\[0.25em]
We say that a graph \,$G$ \,is of {\it bounded degree} \,when \,$\Delta_G$ 
\,is finite.\\[0.25em]
Let \,$R \,\subseteq \,V{\croix}V$ \,be a binary relation on a set \,$V$.\\
The in-degree of \,$s \in V$ \,is \,$\delta_R^-(s) \,= \,|R^{-1}(s)|$ \,for 
\,$R^{-1} \,= \{\ (t,s)\ |\ (s,t) \in R\ \}$ \,the {\it inverse} \,of \,$R$. 
The {\it in-out-degree} \,of \,$R$ \,is\\[0.25em]
\hspace*{9em}$\Delta_R\ =\ {\rm sup}\bigl(\{\ \delta^+_R(s)\ |\ s \in V\ \} 
\,\cup \,\{\ \delta^-_R(s)\ |\ s \in V\ \}\bigr)$.\\
A relation \,$R$ \,is a {\it regular relation} on \,$V$ \,if 
\,$|R(s)| \,= \,|R^{-1}(s)| \,= \,\Delta_R$ \,for any \,$s \in V$.\\
Let us apply Lemma~\ref{CoOutRegular} to \,$R$ \,and \,$R^{-1}$.
\begin{corollary}\label{CoRegular}
Let \,$R \subseteq V{\croix}V$ \,and \,$S = V{\croix}V-R$ \,the complement of 
\,$R$ \,w.r.t. \,$V{\croix}V$.\\
\hspace*{7.2em}If \,$R$ \,is finite and regular \,then \,$S$ \,is regular.\\
\hspace*{7.2em}If \,$R$ \,is infinite and \,$\Delta_R < |V|$ \,then \,$S$ \,is 
regular.
\end{corollary}
An {\it edge-labeling} \,of \,$R$ \,is a mapping \,$c : R\ \fleche{}\ A$ 
defining the respective graph and color set\\[0.25em]
\hspace*{3em}$R^c\ =\ \{\ (s,c(s,t),t)\ |\ (s,t) \in R\ \}$ \ \ and \ \ 
$c(R) \,= \,\{\ c(s,t) \mid (s,t) \in R\ \} \,= \,A_{R^c}$\,.\\[0.25em]
An {\it edge-coloring} \,of \,$R$ \,is an edge-labeling \,$c$ \,of \,$R$ 
\,such that \,$R^c$ \,is a deterministic and co-deterministic graph. 
In that case, we say that \,$R$ \,is \,$|c(R)|$-{\it edge-colorable} \,and we 
have \,$|c(R)| \geq \Delta_R$\,. 
We will give general conditions for a relation \,$R$ \,to be 
\,$\Delta_R$-edge-colorable.\\
An {\it undirected edge-coloring} \,of \,$R$ \,is an edge-labeling \,$c$ \,of 
\,$R$ \,such that two adjacent couples of \,$R$ \,have distinct colors: 
for any \,$(s,t)\,,\,(s',t') \in R$,\\[0.25em]
\hspace*{6em}if \ $(s,t) \neq (s',t')$ \,and \,$\{s,t\} \cap \{s',t'\} \neq 
\emptyset$ \ then \ $c(s,t) \neq c(s',t')$.\\[0.25em]
Any undirected edge-coloring is an edge-coloring.\\
Let \,$V' = \{\ s' \mid s \in V\ \}$ \,be a disjoint copy of \,$V$ \,{\it i.e.} 
\,$'$ \,is a bijection from \,$V$ \,to a disjoint set \,$V'$. 
We transform any relation \,$R \,\subseteq \,V{\croix}V$ \,into the relation
\\[0.25em]
\hspace*{12em}$R'\ =\ \{\ (s,t')\ |\ (s,t) \in R\ \}\ \subseteq\ V{\croix}V'$
\\[0.25em]
and any edge-labeling \,$c$ \,of \,$R$ \,into the edge-labeling \,$c'$ \,of 
\,$R'$ \,defined by\\[0.25em]
\hspace*{12em}$c'(s,t') \,= \,c(s,t)$ \ for any \,$(s,t) \in R$.\\[0.25em]
So \,$\Delta_R \,= \,\Delta_{R'}$ \,and for any edge-labeling \,$c$ \,of \,$R$,
\\[0.25em]
\hspace*{3em}\begin{tabular}{rcl}
$c$ \,is an edge-coloring of \,$R$ & $\Longleftrightarrow$ & $c'$ \,is an 
edge-coloring of \,$R'$\\
& $\Longleftrightarrow$ & $c'$ \,is an undirected edge-coloring of \,$R'$.
\end{tabular}\\[0.25em]
As \,$R' \,\subseteq \,V{\croix}V'$ \,with \,$V \cap V' = \emptyset$, \,$R'$ 
is a bipartite relation hence for \,$R'$ \,finite, and by K\"onig's theorem 
\cite{Ko}, $R'$ \,has an undirected \,edge-$\Delta_{R'}$-coloring . 
This implies that we have an edge-coloring of any finite relation \,$R$ \,using 
\,$\Delta_R$ \,colors.
\begin{lemma}\label{ColorRelation}
Any finite binary relation \,$R$ \,is \,$\Delta_R$-edge-colorable.
\end{lemma}
\proof\mbox{}\\
Instead of applying K\"onig's theorem to \,$R'$, we will adapt its standard 
proof directly to \,$R$.\\
Let \,$n \geq 0$, \,$R \,= \,\{(s_1,t_1),\ldots,(s_n,t_n)\}$ \,and 
\,$k = \Delta_R$\,.\\
By induction on \,$0 \leq i \leq n$, let us construct an edge-coloring \,$c_i$ 
\,of \,$R_i \,= \,\{(s_1,t_1),\ldots,(s_i,t_i)\}$ \,in 
\,$[k] \,= \,\{1,\ldots,k\}$.\\
For \,$i = 0$, the empty function \,$c_0$ \,is an edge-coloring of 
\,$R_0 = \emptyset$.\\
Let \,$0 \leq i < n$ \,and \,$c_i$ \,be an edge-coloring of \,$R_i$ \,in 
\,$[k]$. 
We denote by \,$s = s_{i+1}$\,, \,$t = t_{i+1}$\,,\\[0.25em]
\hspace*{3em}$O_s\ =\ \{\ c_i(s,q)\ |\ (s,q) \in {\rm Dom}(c_i)\ \}$ \ and \
$I_t\ =\ \{\ c_i(p,t)\ |\ (p,t) \in {\rm Dom}(c_i)\ \}$.\\
We distinguish the two complementary cases below.\\
{\it Case 1}\,: \,$O_s \cup I_t \,\subset \,\{1,\ldots,k\}$.
We extend \,$c_i$ \,to the edge-coloring \,$c_{i+1}$ \,of \,$R_{i+1}$ \,by 
defining\\[0.25em]
\hspace*{10em}$c_{i+1}(s,t)\ =\ {\rm min}\{\ j\ |\ j \not\in O_s \cup I_t\ \}$.
\\[0.25em]
{\it Case 2}\,: \,$O_s \cup I_t \,= \,\{1,\ldots,k\}$.\\
As \,$(s,t) \not\in {\rm Dom}(c_i)$, we have \,$|O_s| < k$ \,and \,$|I_t| < k$. 
So \,$\neg(O_s \subseteq I_t)$ \,and \,$\neg(I_t \subseteq O_s)$.\\
Thus there exists \,$a \in O_s - I_t$ \,and \,$b \in I_t - O_s$\,. 
In particular \,$a \neq b$.\\
As \,$R^{c_i}$ \,is deterministic and co-deterministic, there are unique \,$s'$ 
\,and \,$t'$ \,such that \,$c_i(s,t') = a$ \,and \,$c_i(s',t) = b$.
This is illustrated as follows:\\[0.25em]
\hspace*{10em}$t'\ \inverse{a}\ s\ \fleche{}\ t\ \inverse{b}\ s'$\\[0.25em]
where the labeled relation \,$\fleche{x}$ \,for any \,$x \in \{1,\ldots,k\}$ 
\,is defined by\\[0.25em]
\hspace*{12em}$\fleche{x}\ =\ \{\ (p,q) \in {\rm Dom}(c_i)\ |\ \,c_i(p,q) = x\ 
\}$.\\[0.25em]
Let us consider the chain in \,$R^{c_i}$ \,of maximal length and of the form
\\[0.25em]
\hspace*{12em}$s\ \fleche{a}\,\inverse{b}\,\fleche{a}\,\inverse{b}\,\ldots$
\\[0.25em]
As \,$b \not\in O_s$ \,and \,$R^{c_i}$ \,is deterministic and co-deterministic, 
this chain is finite.\\
As \,$a \not\in I_t$\,, \,$s'\ \fleche{b}\ t$ \,is not an edge of this chain.\\
We define another edge-labeling \,$c$ \,of \,$R_i$ \,by exchanging the labels
\,$a$ \,and \,$b$ \,for the edges of the chain: for any
\,$(p,q) \in {\rm Dom}(c_i)$,\\[0.25em]
\hspace*{6em}$c(p,q)\ =\ $
{$\left\{\ \begin{tabular}{ll}
$b$ & if \,$p\ \fleche{a}\ q$ \,is an edge of the chain\\
$a$ & if \,$p\ \fleche{b}\ q$ \,is an edge of the chain\\
$c_i(p,q)$ & otherwise.
\end{tabular}\right.$}\\[0.25em]
Thus \,$R_i^c$ \,remains deterministic and co-deterministic \,{\it i.e.} 
\,$c$ \,is an edge-coloring of \,$R_i$ \,with \,$c(s,t') = c(s',t) = b$. 
It remains to add \,$c(s,t) = a$ \,to get an edge-coloring of \,$R_{i+1}$ 
\,in~$[k]$.
\qed\\[1em]
Under the assumption of the axiom of choice (actually under the weaker 
assumption of the ultrafilter axiom), let us generalize 
Lemma~\ref{ColorRelation}. 
For this we use a coloring on the vertices instead on the edges. 
A {\it vertex-coloring} \,of \,$R \,\subseteq \,V{\croix}V$ \,is a mapping 
\,$c : V\ \fleche{}\ A$ \,such that \,$c(s) \neq c(t)$ \,for any 
\,$(s,t) \in R$, and in that case, we say that \,$R$ \,is 
\,$|c(V)|${\it -vertex-colorable}. 
Note that a relation with a reflexive pair has no vertex-coloring.\\
The {\it dual} \,of \,$R$ \,is the binary relation \,$D(R)$ \,on \,$R$ 
\,defined by\\[0.25em]
\hspace*{6em}\begin{tabular}{rcl}
$D(R)$ & $=$ & $\{\ ((r,s)\,,\,(r,t))\ |\ (r,s)\,,\,(r,t) \in R \ \wedge 
\ s \neq t\ \}$\\[0.25em]
 & $\cup$ & $\{\ ((s,r)\,,\,(t,r))\ |\ (s,r)\,,\,(t,r) \in R \ \wedge 
\ s \neq t\ \}$.
\end{tabular}\\[0.25em]
For any edge-labeling \,$c$ \,of \,$R$,\\[0.25em]
\hspace*{6em}$c$ \,is an edge-coloring of \,$R$ \ \ $\Longleftrightarrow$ \ \ 
$c$ \,is a vertex-coloring of \,$D(R)$.\\[0.25em]
Thus by Lemma~\ref{ColorRelation} and for any finite relation \,$R$, 
\,$D(R)$ \,has a \,$\Delta_R$-vertex-coloring.
We can apply the compactness theorem \cite{BE} to extend 
Lemma~\ref{ColorRelation} to any relation of bounded degree.
\begin{proposition}\label{ZFColorRelation}
In ZFC set theory, any bounded degree relation \,$R$ \,has a 
\,$\Delta_R$-edge-coloring.
\end{proposition}
\proof\mbox{}\\
Let \,$k$ \,be a positive integer and \,$R$ \,be a binary relation on a set 
\,$V$ \,with \,$\Delta_R = k$.\\
It is equivalent to show that \,$D(R)$ \,is \,$k$-vertex-colorable, or that 
\,$D(R)$ \,is vertex-colorable using at most \,$k$ \,colors.\\
By de Bruijn-Erd\"os theorem \cite{BE}, it is equivalent to check that any
finite subset of \,$D(R)$ \,is vertex-colorable with at most \,$k$ \,colors. 
Let \,$S \subseteq D(R)$ \,with \,$S$ \,finite. Let\\[0.25em]
\hspace*{9em}$P\ =\ \{\ s \in V\ |\ \exists\ t \ (s,t) \in V_S \,\wedge 
\,(t,s) \in V_S\ \}$\\[0.25em]
and \,$R_{|P} \,= \,R \,\cap P{\croix}P$ \,the induced relation of \,$R$ \,by 
\,$P$. 
So \,$S \,\subseteq \,D(R_{|P})$ \,which is finite.\\
By Lemma~\ref{ColorRelation}, \,$R_{|P}$ \,is edge-colorable using 
\,$\Delta_{R_{|P}} \,\leq \,\Delta_R \,= \,k$ \,colors.\\
Finally \,$D(R_{|P})$ \,hence \,$S$ \,are vertex-colorable using at most \,$k$ 
\,colors.
\qed\\[1em]
We now want to color a regular relation in a complete way. 
First we present a general way to extend an injection into a bijection 
avoiding given sets.
\begin{lemma}\label{Bijection}
Let \,$X,Y$ \,be equipotent well orderable infinite sets.\\
\hspace*{1em}Let an injection \,$p : P\ \fleche{}\ Y$ \,for some subset 
\,$P$ \,of \,$X$ \,with \,$|P| < |X|$.\\
\hspace*{1em}Let a sequence \,$(P_x)_{x  \in X-P}$ \,of subsets of \,$Y$ \,with 
\,$|P_x| < |Y|$ \,and such that\\
\hspace*{6em}$|\{\ x \in X-P\ |\ y \in P_x\ \}| \,< \,|X|$ \ for every 
\,$y \in Y-p(P)$.\\
\hspace*{1em}We can extend \,$p$ \,into a bijection \,$X\ \fleche{}\ Y$ \,such 
that \,$p(x) \not\in P_x$ \,for every \,$x \in X-P$.
\end{lemma}
\proof\mbox{}\\
Let \,$<_X$ \,be an initial well-ordering of \,$X$\,: \,$\forall\ x \in X,\ 
|\{\ x' \in X\ |\ x' < x\ \}| \,< \,|X|$.\\
Let \,$<_Y$ \,be an initial well-ordering of \,$Y$.\\
We define \,$p$ \,on \,$X-P$ \,by transfinite induction. Let \,$x \in X-P$.\\
Let us define \,$p(x)$ \,knowing \,$p(x')$ \,for any \,$x' <_X x$.\\
As \,$|P_x|\,,\,|P| \,< \,|X| \,= \,|Y|$ \,with \,$X$ \,infinite, the 
following subset of \,$Y$\\[0.25em]
\hspace*{9em}$Q_x \ = \ p(P) \,\cup \,P_x \,\cup \,\{\ p(x')\ |\ x' <_X x\ \}$
\\[0.25em]
is of cardinal \,$|Q_x| < |Y|$. So we can define\\[0.25em]
\hspace*{9em}$p(x) \,= \,{\rm min}_{<_Y}(Y-Q_x)$.\\[0.25em]
Thus \,$p$ \,is injective and \,$p(x) \not\in P_x$ \,for every \,$x \in X-P$.\\
Let us check that \,$p$ \,is surjective. 
Assume that \,${\rm Im}(p) \neq Y$. Let\\[0.25em]
\hspace*{9em}$\beta\ =\ {\rm min}_{<_Y}(Y-{\rm Im}(p))$.\\[0.25em]
As \,$|\{\ x \in X-P\ |\ \beta \in P_x\ \}| \,< \,|X|$, we can define\\[0.25em]
\hspace*{6em}$\alpha\ =\ {\rm min}\{\ x \in X-P\ |\ \,\beta \not\in P_x 
\,\wedge \,\beta <_Y p(x) \ \}$.\\[0.25em]
So \,$\alpha \in X-P$ \,and \,$\beta <_Y p(\alpha)$.\\
As \,$\beta \not\in {\rm Im}(p)$ \,and \,$\beta \not\in P_{\alpha}$\,, \,we have 
\,$\beta \not\in Q_{\alpha}$ \,hence \,$p(\alpha) \leq_Y \beta$ \ which is a 
contradiction.
\qed\\[1em]
A {\it complete} edge-coloring of a regular relation \,$R$ \,is an 
edge-coloring \,$c$ \,of \,$R$ \,such that \,$R^c$ \,is source-complete and 
target-complete. 
Under the assumption of the axiom of choice, we can color in a complete way any 
regular relation.
\begin{proposition}\label{CompleteColoring}
In ZFC set theory, any regular relation \,$R$ \,has a complete 
\,$\Delta_R$-edge-coloring.  
\end{proposition}
\proof\mbox{}\\
Let \,$R$ \,be a regular relation on a set \,$V$.\\
We distinguish the two complementary cases below.\\
{\it Case 1}\,: \,$\Delta_R \,< \,\aleph_0$.\\
\hspace*{1em}By Proposition~\ref{ZFColorRelation}, \,$R$ \,has a 
\,$\Delta_R$-edge-coloring \,$c$.\\
\hspace*{1em}So \,$R^c$ \,is a deterministic and co-deterministic graph with 
\,$|A_{R^c}| \,= \,\Delta_R \,= \,\Delta_{R^c}$\,.\\
\hspace*{1em}Thus \,$R^c$ \,is source-complete and target-complete \,{\it i.e.} 
\,$c$ \,is a complete edge-coloring.\\[0.25em]
{\it Case 2}\,: \,$\Delta_R \,\geq \,\aleph_0$.\\
\hspace*{1em}Under AC, it suffices to show the existence of a complete 
\,$\Delta_R$-edge-coloring for \,$R$ \,connected.\\
\hspace*{1em}Under AC and having \,$R$ \,connected, we have 
\,$|V_R| = |\Delta_R|$.\\
\hspace*{1em}Thanks to AC, let us consider an initial well-ordering \,$<$ \,of
\,$V_R$.\\
\hspace*{1em}By transfinite induction, let us define a complete 
\,$\Delta_R$-edge-coloring \,$c$ \,of \,$R$. Let \,$\lambda \in V_R$.\\
\hspace*{1em}Let us define \,$c$ \,on 
\,$\{\ (\lambda,\mu) \in R\ |\ \lambda \leq \mu\ \} \,\cup 
\,\{\ (\mu,\lambda) \in R\ |\ \lambda \leq \mu\ \}$\\
\hspace*{1em}knowing \,$c$ \,on 
\,$\{\ (\mu,\nu)\ |\ \mu < \lambda \,\vee \,\nu < \lambda\ \}$.\\
\hspace*{1em}First, we define \,$c(\lambda,\mu)$ \,for any 
\,$(\lambda,\mu) \in R$ \,with \,$\lambda \leq \mu$.\\
\hspace*{1em}This is illustrated below for \,$\rho < \lambda \leq \mu$.
\begin{center}
\includegraphics{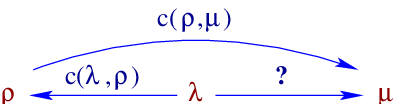}\\%magnitude 0.3
\end{center}
\hspace*{1em}As \,$R$ \,is regular, 
\,$R(\lambda)\ =\ \{\ \mu\ |\ (\lambda,\mu) \in R\ \}$ \,has cardinality 
\,$|R(\lambda)| \,= \,\Delta_R$\,. Let\\[0.25em]
\hspace*{1.5em}$P\ =\ \{\ \rho \in R(\lambda)\ |\ \rho < \lambda\ \}$ \ \ and \ 
\ $p\ : P\ \fleche{}\ \Delta_R$ \ with \ $p(\rho) = c(\lambda,\rho)$ \ for any 
\,$\rho \in P$.\\[0.25em]
\hspace*{1em}For any \,$\mu \in R(\lambda)-P$, we define\\[0.25em]
\hspace*{10em}$P_{\mu}\ =\ \{\ c(\rho,\mu)\ |\ (\rho,\mu) \in R \,\wedge 
\,\rho < \lambda\ \}$.\\[0.25em]
\hspace*{1em}By Lemma~\ref{Bijection}, we can extend \,$p$ \,into a bijection 
\,$R(\lambda)\ \fleche{}\ \Delta_R$ \,such that \,$p(\mu) \not\in P_{\mu}$ 
\,for\\
\hspace*{1em}any \,$\mu \in R(\lambda)-P$. It remains to define\\[0.25em]
\hspace*{10em}$c(\lambda,\mu) \,= \,p(\mu)$ \ for any \,$(\lambda,\mu) \in R$ 
\,and \,$\lambda \leq \mu$.\\[0.25em]
\hspace*{1em}Similarly, we define \,$c(\mu,\lambda)$ \,for any 
\,$(\mu,\lambda) \in R$ \,with \,$\lambda < \mu$.\\
\hspace*{1em}This is illustrated below for \,$\rho < \lambda < \mu$.
\begin{center}
\includegraphics{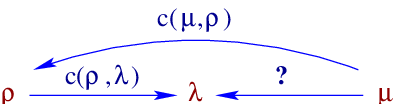}\\%magnitude 0.3
\end{center}
\qed\\[1em]
We say that a graph \,$G$ \,is {\it regular} \,if \,$\fleche{}_G$ \,is 
regular. For instance 
\,$\{s\ \fleche{a}\ t\,,\,t\ \fleche{a}\ s\,,\,t\ \fleche{b}\ s\}$ \,is a 
regular graph. 
We also say that \,$G$ \,is {\it co-regular} \,if its unlabeled complement 
is regular: \,$\nofleche{}{}\,\!_G$ \ is a regular relation on \,$V_G$\,.\\
Under the assumption of the axiom of choice, we can restrict 
Theorem~\ref{LeftQuasiCayley} to obtain a characterization of the generalized 
Cayley graphs of quasigroups.
\begin{theorem}\label{QuasiCayley}
In ZFC set theory, the following graphs define the same family\,{\sf :}\\
{\bf a)} the generalized Cayley graphs of quasigroups (resp. with a right 
identity),\\
{\bf b)} the generalized Cayley graphs of quasigroups with a left identity 
(resp. an identity),\\
{\bf c)} the simple, deterministic, co-deterministic, co-regular, 
source-complete, target-complete\\
\hspace*{1.2em}(resp. and loop-complete) graphs.
\end{theorem}
\proof\mbox{}\\
{\bf b)} \,$\Longrightarrow$ \,{\bf a)}\,: \,immediate.\\[0.25em]
{\bf a)} \,$\Longrightarrow$ \,{\bf c)}\,: \,let 
\,$G \,= \,{\cal C}\inter{M,Q}$ \,for some quasigroup \,$M$ \,and 
\,$Q \subseteq M$.\\
By Proposition~\ref{CayleyQuasi}, \,$G \,= \,{\cal C}\inter{M}^{|\,\interInd{Q}}$ 
\,remains simple, deterministic and co-deterministic, source and 
target-complete.\\
For any \,$s \in M$, \,$\delta^+_{\rightarrow\!\!\!\!\!/\,_G}(s) \,= 
\,\delta^-_{\rightarrow\!\!\!\!\!/\,_G}(s) \,= \,|M-Q|$ \,hence \,$G$ \,is 
co-regular.\\
If in addition \,$M$ \,has a right identity then, by Fact~\ref{LoopComp}, 
\,$G$ \,is loop-complete.\\[0.25em]
{\bf c)} \,$\Longrightarrow$ \,{\bf b)}\,: \,let \,$G$ \,be a graph which is 
simple, deterministic and co-deterministic, source and target-complete. 
So \,$G$ \,is regular with \,$\Delta_G \,= \,|A_G|$.\\
If \,$G$ \,is without loop (hence \,$G$ \,is loop-complete) then we take a new 
label \,$a \in A-A_G$ \,and we define 
\,$G' \,= \,G \,\cup \,\{\ s\ \fleche{a}\ s \mid s \in V_G\ \}$. 
If \,$G$ \,has at least one loop, we put \,$G' = G$.\\
Moreover, suppose also that \,$G$ \,is co-regular. By definition, the 
complement relation of \,$G'$\\[0.25em]
\hspace*{6em}$S \ = \ \{\ (s,t)\ |\ s,t \in V_G \,\wedge 
\,\{s\}{\croix}A_G{\croix}\{t\} \,\cap \,G' \,= \,\emptyset\ \}$\\[0.25em]
is a regular relation on \,$V_G$\,.\\
By Proposition~\ref{CompleteColoring}, \,$S$ \,has a complete 
\,$\Delta_S$-edge-coloring \,$c$ \,{\it i.e.} \,$S^c$ \,is a deterministic, 
co-deterministic, source and target-complete graph. 
By definition, \,$S^c$ \,is also simple.\\
Furthermore we can assume that \,$A_G \cap A_{S^c} \,= \,\emptyset$.\\
Let \,$H \,= \,G' \cup S^c$. Thus \,$H$ \,is source and target-complete. 
It is also complete, simple, deterministic and co-deterministic. 
Furthermore for \,$G$ \,loop-complete, $H$ \,is loop-complete.\\
Let \,$r$ \,be a vertex of \,$G$. 
By Proposition~\ref{EdgeProductQuasi}, $(V_G,\croix_r)$ \,is a quasigroup 
for the edge-operation \,$\croix_r$ \,of \,$H$. 
By Proposition~\ref{EdgeProduct}, $r$ \,is a left-identity 
(resp. is an identity) and 
\,$H \,= \,{\cal C}\inter{V_G\,,\,\fleche{}_H(r)}$ \,with 
\,$\inter{s} = a$ \,for any \,$r\ \fleche{a}_H\ s$. 
By label restriction to \,$A_G$\,, 
\,$G \,= \,{\cal C}\inter{V_G\,,\fleche{}_G(r)\,}$ \,is a 
generalized Cayley graph of \,$(V_G,\croix_r)$.
\qed\\[1em]
The co-regularity in Theorem~\ref{QuasiCayley} can not be removed. 
For instance, the following graph:\\[0.25em]
\hspace*{1em}\begin{tabular}{rcl}
PlusMinus & $=$ & $\{\ i + 2mj\ \fleche{j}\  i + (2m+1)j\ |\ m \geq 0 \,\wedge 
\,0 \leq i < j\ \} \,\cup \,\{\ i\ \fleche{0}\ i\ |\ i \geq 0\ \}$\\[0.25em]
 & $\cup$ & $\{\ i + (2m+1)j\ \fleche{j}\  i + 2mj\ |\ m \geq 0 \,\wedge 
\,0 \leq i < j\ \}$
\end{tabular}\\[0.25em]
is deterministic, co-deterministic, simple, source-complete and target-complete.
Furthermore it is not complete: there is no edge between $1$ and $2$ and more 
generally between \,$i + (2m+1)j$ \,and \,$i + (2m+2)j$ \,for any \,$m \geq 0$ 
\,and \,$0 \leq i < j$. Finally it is \,$0$-complete: \,$0\ \fleche{j}\ j$
\,for any \,$j \geq 0$.
By Proposition~\ref{CayleyQuasi}, \,PlusMinus \,is not a generalized Cayley 
graph of a quasigroup.\\
By Corollary~\ref{CoRegular}, the co-regularity in Theorem~\ref{QuasiCayley} 
can be removed for the graphs of bounded degree which corresponds by 
Fact~\ref{OutDegree} to the finitely labeled graphs \,$G$ \,when \,$G$ \,and 
\,$G^{-1}$ \,are deterministic and source-complete.
\begin{corollary}\label{BoundedQuasiCayley}
In ZFC set theory, a finitely labeled graph is a generalized Cayley graph of a 
quasigroup \,iff 
\,it is deterministic, co-deterministic, simple, source and target-complete.
\end{corollary}
For instance the following graph of all the cycles:\\[0.25em]
\hspace*{6em}{\it Cycles}
\,$= \,\{\ (m,n)\ \fleche{a}\ (m,n+1 \,({\rm mod} \,m))\ |\ m > n \geq 0\ \}$
\\[0.25em]
is by Corollary~\ref{BoundedQuasiCayley} a generalized Cayley graph of a 
quasigroup.\\
Let us summarize the characterizations obtained in ZFC theory for the 
\underline{finitely labeled} generalized Cayley graphs.\\[-0.5em]\mbox{}
\begin{center}
\includegraphics{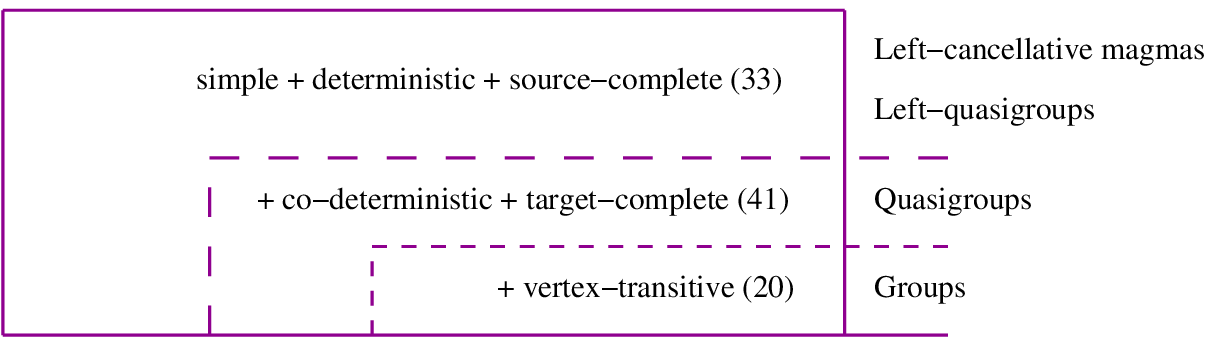}\\%magnitude 0.25
\end{center}%\mbox{}\\
For all the generalized Cayley graphs (not necessarily finitely labeled), we 
need the co-out-regularity for the left-quasigroups, and the co-regularity for 
the quasigroups.

\section{Decidability results}

We have given graph-theoretic characterizations of generalized Cayley graphs 
of various basic algebraic structures. 
These characterizations are adapted to decide whether a graph \,$G$ \,is a 
generalized Cayley graph, and if so, we got\\[0.25em]
\hspace*{8em}$G \,= \,{\cal C}\inter{V_G\,,\,\fleche{}_G(r)}$ \ with 
\ $\inter{s} = a$ \,for any \,$r\ \fleche{a}_G\ s$\\[0.25em]
for the operation on \,$V_G$ \,which is either the path-operation 
\,$\ast_r$ \,with \,$r$ \,a root, or the chain-operation 
\,$\overline{\ast}_r$\,, \,or the extended chain-operation 
\,$\overline{\ast}_P$ \,with \,$P$ \,a representative set of 
\,${\rm Comp}(G)$,  or the edge-operation \,$_{\InfSupInd{G}}\croix_r$ \,for 
some completion \,$\InfSup{G}$ \,of \,$G$ \,and for any vertex \,$r$.\\
We will show the effectiveness of these characterizations and their associated 
operations for a general family of infinite graphs. 
We restrict to the family of end-regular graphs of finite degree \cite{MS} 
which admits an external characterization by finite decomposition by 
distance which allows to decide the isomorphism problem, and an internal 
characterization as suffix graphs of word rewriting systems which have a 
decidable monadic second-order logic.

\subsection{End-regular graphs}

A {\it marked graph} \,is a couple \,$(G,P)$ \,of a graph \,$G$ \,with a 
vertex subset \,$P \subseteq V_G$\,. We extend the graph isomorphism to the 
marked graphs: \,$(G,P) \,\equiv \,(G',P')$ \,if 
\,$G \,\equiv_g \,G'$ \,for some isomorphism \,$g$ \,such that \,$g(P) = P'$, 
and we also write \,$(G,P) \,\equiv_g \,(G',P')$.\\
Let \,$G$ \,be a graph. 
The {\it frontier} \,${\rm Fr}_G(H)$ \,of \,$H \subseteq G$ \,is the set of 
vertices common to \,$H$ \,and \,$G-H$\,:\\[0.25em]
\hspace*{10em}${\rm Fr}_G(H) \,= \,V_H \,\cap \,V_{G-H}$\\[0.25em]
and we denote by \,${\rm End}_G(H)$ \,the set of {\it ends} \,obtained by 
removing \,$H$ \,in \,$G$\,:\\[0.25em]
\hspace*{6em}${\rm End}_G(H) \ = \ \{\ (C,{\rm Fr}_G(C))\ |\ 
C \in {\rm Comp}(G-H)\ \}$.\\[0.25em]
We say that \,$G$ \,is {\it end-regular} \,if there exists an increasing
sequence \,$H_0 \subseteq \ldots \subseteq H_n \subseteq \ldots$
\,of finite subgraphs \,$H_n$ \,of \,$G$ \,such that\\[0.25em]
\hspace*{6em}$G \,= \,\bigcup_{n \geq 0}H_n$ \ \ \ and \ \ \ 
$\bigcup_{n \geq 0}{\rm End}_G(H_n)$ \,is of finite index for \,$\equiv$\,.
\\[0.25em]
Note that any end-regular graph is finitely labeled and of finite or countable 
vertex set. 
Furthermore any end-regular graph of finite degree is of bounded degree. 
Moreover, every end-regular graph has only a finite number of non-isomorphic 
connected components.\\
We say that the previous sequence \,$(H_n)_{n \geq 0}$ \,is a {\it compatible 
decomposition} \,of \,$G$ \,if it also satisfies that two isomorphic ends have 
the same decomposition in the following sense: for any \,$m,n \geq 0$ \,and any 
\,$(C,P) \in {\rm End}_G(H_m)$ \,and \,$(D,Q) \in {\rm End}_G(H_n)$,\\
\hspace*{3em}if \ $(C,P) \,\equiv_g \,(D,Q)$ \ then 
\ $C \,\cap \,H_{m+p} \,\equiv_g \,D \,\cap \,H_{n+p}$ \ for every \,$p \geq 0$.
\\
Any end-regular graph has a compatible decomposition which can be finitely 
presented by a deterministic graph grammar \cite{Ca2}.\\
Any finite graph is end-regular.
Any regular tree (having a finite number of non isomorphic subtrees) is
end-regular. Except for the quater-grid and the graph {\it Cycles}, all other 
graphs in this article are end-regular. The following infinite graph:\\[0.25em]
\hspace*{6em}\begin{tabular}{rcl} 
$\Xi$ & $=$ & $\{\ ui\ \fleche{x}\ uxi\ |\ \,u \in \{a,b\}^* \,\wedge 
\,x \in \{a,b\} \,\wedge \,i \in \{0,1\}\ \}$\\[0.25em]
& $\cup$ & $\{\ ui\ \fleche{c}\ u(1-i)\ |\ u \in \{a,b\}^* \,\wedge 
\,i \in \{0,1\}\ \}$
\end{tabular}\\[0.25em]
of vertex set \,$\{a,b\}^*.\{0,1\}$ \,is represented below.
\begin{center}
\includegraphics{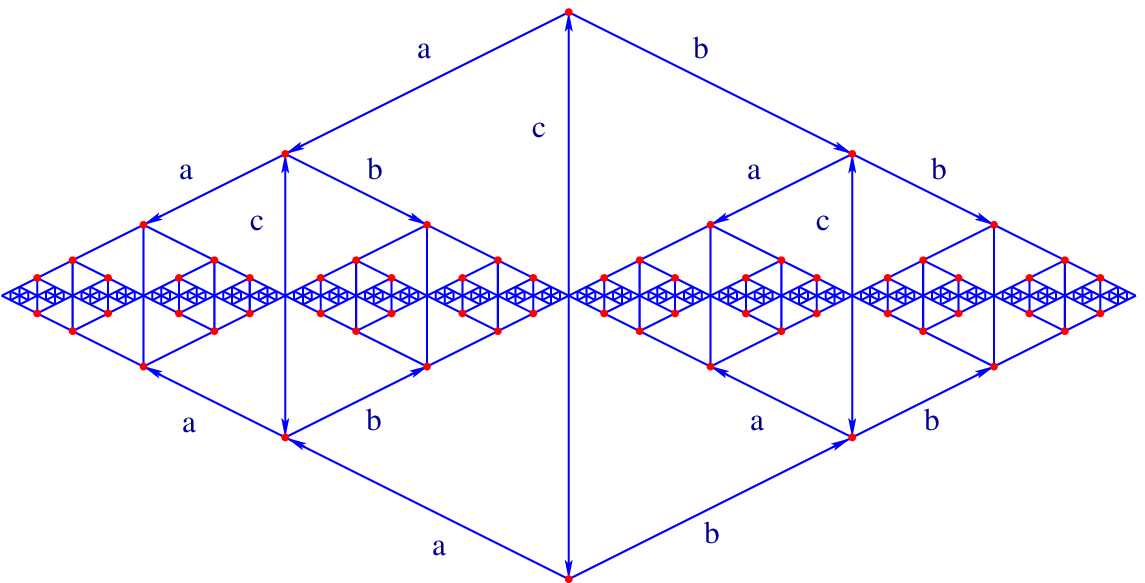}\\%magnitude 0.18
\end{center}
This graph is formed by two disjoint source-complete $\{a,b\}$-trees whose 
every node of a tree is connected by a $c$-edge to the corresponding node of 
the other tree. The graph \,$\Xi$ \,is end-regular since it is generated by the 
graph grammar \cite{Ca2} reduced to this unique rule:
\begin{center}
\includegraphics{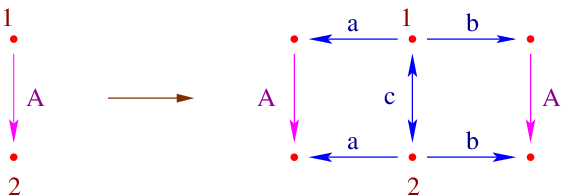}\\%magnitude 0.25
\end{center}
Note that \,$\Xi$ \,is strongly connected and of bounded degree, simple, 
deterministic and co-deterministic, source-complete but not target-complete, 
arc-symmetric but not symmetric. 
By Theorem~\ref{Monoid}, $\Xi$ \,is a Cayley graph of a cancellative monoid. 
Precisely \,$\Xi \,= \,{\cal C}\inter{V_{\Xi},\{a0,b0,1\}}$ \,with 
\,$\inter{a0} = a$, $\inter{b0} = b$, $\inter{1} = c$ \,for the cancellative 
monoid \,$(V_{\Xi},\ast_0)$ \,with the path-operation \,$\ast_0$ \,defined for 
any \,$u,v \in \{a,b\}^*$ \,and \,$i,j \in \{0,1\}$ \,by\\[0.25em]
\hspace*{9em}$ui \,\ast_0 \,vj \,= \,uvk$ \ with \ 
$k \,= \,i+j \,({\rm mod} \,2)$.\\[0.25em]
By Theorem~\ref{BoundedLeftQuasiCayley}, \,$\Xi$ \,is also a generalized Cayley 
graph of a left-quasigroup with an identity, namely 
\,$(\{a,b\}^*.\{0,1\},\cdot)$ \,for \,$\cdot$ \,the edge-operation \,$\croix_0$ 
\,of a completion \,$\Infsup{\Xi}$ \,that we can define for any 
$u \in \{a,b\}^*$, $x \in \{a,b\}$, $i \in \{0,1\}$ \,by\\[0.25em]
\hspace*{6em}$ui \,\cdot \,x0 \,= \,uxi \ \ ; \ \ ui \,\cdot \,uxi \,= \,x0$ \ 
for \,$ui \neq x0 \ \ ; \ \ x0 \,\cdot \,xx0 \,= \,0$\\[0.25em]
\hspace*{6em}$ui \,\cdot \,0 \,= \,ui \ \ ; \ \ 
ui \,\cdot \,1 \,= \,u(1-i) \ \ ; \ \ ui \,\cdot \,u(1-i) \,= \,1$\\[0.25em]
\hspace*{6em}$ui \,\cdot \,v \,= \,v$ \ otherwise.\\[0.25em]
The end-regularity of a graph can also be expressed on the vertices. 
A graph \,$G$ \,is {\it vertex-end-regular} \,if 
\,$V_G \,= \,\bigcup_{n \geq 0}V_n$ \,with\\[0.25em]
\hspace*{3em}$V_0 \subseteq \ldots \subseteq V_n \subseteq \ldots$ \,finite \ 
and \ $\bigcup_{n \geq 0}{\rm End}_G(G - G_{|V_n})$ \,is of finite index for 
\,$\equiv$\,.\\[0.25em]
This notion of vertex-end-regularity corresponds to the (edge-)end-regularity.
\begin{lemma}\label{RegularGraph}
A graph is end-regular if and only if it is vertex-end-regular.
\end{lemma}
\proof\mbox{}\\
$\Longleftarrow$\,: \,$G_n \,= \,G_{|V_n}$ \ suits.\\
$\Longrightarrow$\,: \,$V_n \,= \,V_{G_n}$ \ suits since
\ $G - G_{|V_n}\ =\
G - \bigl(G_n \,\cup \,\bigcup_{K \in {\rm Comp}(G - G_n)}G_{|{\rm Fr}_G(K)}\bigr)$.
\qed\\[1em]
We say that a graph \,$G$ \,is {\it end-regular by distance} \,from a vertex 
\,$r$ \,if it is vertex-end-regular for the sequence defined by 
\,$V_n \,= \,\{\ s\ |\ {\rm d}_G(r,s) \leq n\ \}$ \,for any \,$n \geq 0$. 
In that case, $G$ \,is connected and of finite degree. Furthermore the sequence 
\,$(H_n)_{n \geq 0}$ \,defined by\\[0.25em]
\hspace*{6em}$H_n \ = \ G_{|V_n} \ = 
\ \{\ (s,a,t) \in G\ |\ {\rm d}_G(r,s) \leq n \,\wedge \,{\rm d}_G(r,t) \leq n
\ \}$\\[0.25em]
is a compatible decomposition of \,$G$. 
The regularity by distance is a normal form for the connected end-regular 
graphs of finite degree \cite{Ca1,Ca2}.
\begin{proposition}\label{Distance}
For any connected graph \,$G$ \,of finite degree,\\
\hspace*{3em}\begin{tabular}{rcl}
$G$ \,is end-regular & $\Longleftrightarrow$ & 
$G$ \,is end-regular by distance from some vertex\\
 & $\Longleftrightarrow$ & 
$G$ \,is end-regular by distance from any vertex.
\end{tabular}
\end{proposition}
This normalization of the regularity by distance implies that for any 
end-regular graph \,$G$ \,of finite degree, the isomorphism problem 
is decidable: from any finite compatible decomposition of \,$G$, we can decide 
whether \,$s \,\simeq_G t$ \,by comparing by distance \,$G$ \,from \,$s$ 
\,with \,$G$ \,from \,$t$ \cite{Ca1}.
\begin{corollary}\label{IsoVertex}
For any end-regular graph \,$G$ \,of finite degree, \,$\simeq_G$ \,is decidable.
\end{corollary}
The representation of an end-regular graph \,$G$ \,by a graph grammar is an 
{\it external representation} of \,$G$, namely which is up to isomorphism: the 
vertices of \,$G$ \,are not taken into account. 
To obtain decidable logical properties on \,$G$, we need to give
an {\it internal representation} of \,$G$, namely by an isomorphic graph whose 
vertices are words.

\subsection{Suffix recognizable graphs}

Another way to describe the end-regular graphs of finite degree is through 
rewriting systems. A {\it labeled word rewriting system} \,$R$ \,over an 
alphabet \,$N$ \,is a finite \,$A$-graph of vertex set \,$V_R \subset N^*$ 
\,{\it i.e.} \,$R \subset N^*{\croix}A{\croix}N^*$ \,and \,$R$ \,is finite. 
Each edge \,$u\ \fleche{a}_R\ v$ \,is a {\it rule} \,labeled by \,$a$, of 
left hand side \,$u$ \,and right hand side \,$v$. 
The {\it suffix graph} \,of \,$R$ \,is the graph\\[0.25em]
\hspace*{6em}$N^*.R \ = \ \{\ wu\ \fleche{a}\ wv\ |\ (u,a,v) \in R \,\wedge 
\,w \in N^*\ \}$.\\[0.25em]
For instance let us consider the rewriting system \,$R$ \,over 
\,$N = \{a,b,0,1\}$ \,defined by\\[0.25em]
\hspace*{6em}$0\ \fleche{a}\ a0$ \hspace*{1em} $1\ \fleche{a}\ a1$ 
\hspace*{1em} $0\ \fleche{b}\ b0$ \hspace*{1em} $1\ \fleche{b}\ b1$ 
\hspace*{1em} $0\ \fleche{c}\ 1$ \hspace*{1em} $1\ \fleche{c}\ 0$\\[0.25em]
The connected component of the suffix graph \,$N^*.R$ \,of vertex \,$0$ \,is 
equal to the graph \,$\Xi$.\\
These suffix graphs give an internal representation of the end-regular graphs 
of finite degree~\cite{Ca1}.
\begin{proposition}\label{Suffix}
A connected graph of finite degree is end-regular if and only if it is 
isomorphic to a connected component of a suffix graph.
\end{proposition}
Any suffix graph \,$N^*.R$ \,can be obtained by a first order interpretation 
in the source-complete \,$N$-tree \,$T_N \,= \,\{\ u\ \fleche{a}\ ua\ |\ 
u \in N^* \,\wedge \,a \in N\ \}$ \,{\it i.e.}\\[0.25em]
\hspace*{10em}$N^*.R \ = \ \{\ u\ \fleche{a}\ v\ |\ T_N\ \models\ \phi_a(u,v)\ 
\}$\\[0.25em]
where for any \,$a \in A_R$\,, \,$\phi_a$ \,is the following first order 
formula\\[0.25em]
\hspace*{10em}$\phi_a(x,y) \,: \ \bigvee_{(u,a,v) \in R}\exists\ z\ 
(z\ \fleche{u}\ x \,\wedge \,z\ \fleche{v}\ y)$\\[0.25em]
and the path relation \,$x\ \fleche{a_1{\ldots}a_n}\ y$ \,for \,$n \geq 0$ 
\,and \,$a_1,\ldots,a_n \in A_R$ \,can be expressed by the first order 
formula\\[0.25em]
\hspace*{6em}$\exists\ z_0,\ldots,z_n\ (z_0\ \fleche{a_1}\ z_1 \,\wedge \ldots 
\wedge \,z_{n-1}\ \fleche{a_n}\ z_n \,\wedge \,z_0 = x \,\wedge \,z_n = y)$.
\\[0.25em]
As the existence of a chain between two vertices can be expressed by a monadic 
(second order) formula, any connected component of \,$N^*.R$ \,can be obtained
by a monadic interpretation in \,$T_N$\,. 
As \,$T_N$ \,has a decidable monadic theory \cite{Ra} and by 
Proposition~\ref{Suffix}, any connected end-regular graph has a decidable 
monadic theory. This is generalized \cite{AH,LS} to the 
{\it suffix recognizable graphs} \,over \,$N$ \,which are the graphs of the 
form\\[0.25em]
\hspace*{3em}$\bigcup_{i=1}^nW_i(U_i\ \fleche{a_i}\ V_i)$ \ where \,$n \geq 0$ 
\,and \,$U_1,V_1,W_1,\ldots,U_n,V_n,W_n \in {\rm Rec}(N^*)$\\[0.25em]
for \,${\rm Rec}(N^*)$ \,the family of recognizable (regular) languages over 
\,$N$. 
\begin{proposition}\label{MSO}
The suffix recognizable graphs \,over \,$N$ \,are the graphs obtained by 
monadic interpretations in \,$T_N$ \,hence have a decidable monadic second 
order theory.
\end{proposition}
In particular, we can decide whether a suffix recognizable graph is rooted or 
is connected. 
We can also decide first order properties like the simplicity which can be 
expressed by the following first order formula:\\[0.25em]
\hspace*{6em}$\forall\ x,y\ \bigwedge_a\,\bigl(x\ \fleche{a}\ y \ \ 
\Longrightarrow \ \ \neg\,\bigvee_{b \neq a}x\ \fleche{b}\ y\bigr)$\\[0.25em]
and this the same for the properties of being deterministic, co-deterministic, 
source-complete, target-complete, and loop-complete.\\
We still have to consider the decidability of the arc-symmetry and the
symmetry of end-regular graphs of finite degree. 
The suffix recognizable graphs form a strict extension of end-regular graphs 
that coincide for graphs of finite degree \cite{Ca2}.
\begin{proposition}\label{SuffixRec}
Any end-regular graph is isomorphic to a suffix recognizable graph.\\
Any suffix recognizable graph of finite degree is end-regular.
\end{proposition}
Note that a suffix recognizable graph over \,$N$ \,of finite degree is of the 
form\\[0.25em]
\hspace*{1em}$\bigcup_{i=1}^nW_i(u_i\ \fleche{a_i}\ v_i)$ \ where \,$n \geq 0$, 
\,$u_1,v_1,\ldots,u_n,v_n \in N^*$ \,and \,$W_1,\ldots,W_n \in {\rm Rec}(N^*)$.
\\[0.25em]
By Proposition~\ref{SuffixRec}, these graphs constitute an internal 
representation of end-regular graphs of finite degree. 
Note also that 
\,$\{\ A^m\ \fleche{a}\ A^{m+n}\ |\ m,n \geq 0\ \} \ = \ 
A^*(\varepsilon\ \fleche{a}\ A^*)$ \,is a suffix recognizable graph which is 
not an end-regular graph.\\
By monadic interpretation (or by a simple saturation method on graph grammars), 
the family of end-regular graphs is closed under accessibility.
\begin{corollary}\label{Access}
For any end-regular graph \,$G$ \,and any vertex \,$r$, \,$G_{{\downarrow}r}$ 
\,is end-regular.
\end{corollary}
By Corollaries~\ref{IsoVertex} and \ref{Access}, we can decide whether 
\,$s \downarrow_G t$ \,for \,$G$ \,end-regular of finite degree.
\begin{corollary}\label{IsoAccess}
For any end-regular graph \,$G$ \,of finite degree, \,$\downarrow_G$ is 
decidable.
\end{corollary}
The arc-symmetry of rooted graphs can be reduced to the 
accessible-isomorphism of a root with its successors.
\begin{lemma}\label{AccessTransitive}
A graph \,$G$ \,of root \,$r$ \,is arc-symmetric \,iff 
\ $r \downarrow_G s$ \,for any \,$r\ \fleche{}_G\ s$.
\end{lemma}
\proof\mbox{}\\
Let \,$G$ \,be a graph with a root \,$r$ \,such that
\,$r \downarrow_G s$ \,for any \,$r\ \fleche{}_G\ s$.\\
Let us check that \,$G$ \,is arc-symmetric \,{\it i.e.}
\,$r \downarrow_G s$ \,for any \,$r\ \fleche{}_G^*\ s$.\\
The proof is done by induction on \,$n \geq 0$ \,for \,$r\ \fleche{}_G^n\ s$.\\
For \,$n = 0$, we have \,$r = s$. For \,$n > 0$, let \,$t$ \,be a vertex such 
that \,$r\ \fleche{}_G^{n-1}\ t\ \fleche{}_G\ s$.\\
By induction hypothesis, we have \,$r \downarrow_G t$ \,{\it i.e.} 
\,$f(r) = t$ \,for some isomorphism \,$f$ \,from \,$G_{{\downarrow}r}$ \,to 
\,$G_{{\downarrow}t}$\,. 
As \,$t\ \fleche{}_G\ s$, there exists \,$r'$ \,such that 
\,$r\ \fleche{}_G\ r'$ \,and \,$f(r') = s$. So \,$r' \downarrow_G s$.\\
By hypothesis \,$r \downarrow_G r'$. 
By transitivity of \,$\downarrow_G$\,, we get \,$r \downarrow_G s$.
\qed\\[1em]
Let us transpose Lemma~\ref{AccessTransitive} to the symmetric graphs. 
The arc-symmetry of connected graphs can be reduced to the isomorphism of a 
vertex with its adjacent vertices.
\begin{lemma}\label{VertexTransitive}
A connected graph with a vertex \,$r$ \,is symmetric \,if and only if\\
\hspace*{12em}$r \,\simeq_G \,s$ \,for any 
\,$r\ \fleche{}_{G \,\cup \,G^{-1}}\ s$.
\end{lemma}
Let us apply Lemma~\ref{AccessTransitive} and \ref{VertexTransitive} with 
Corollaries~\ref{IsoVertex} and \ref{IsoAccess}.
\begin{corollary}\label{Dec2}
We can decide whether a rooted (resp. any) end-regular graph of finite degree 
is arc-symmetric (resp. symmetric).
\end{corollary}
In this corollary, we do not need the connected condition for the 
symmetry since any end-regular graph has only a finite number of 
non-isomorphic connected components. 
We can establish the effectiveness of previous Cayley graph characterizations 
for the regular graphs of finite degree.
This decidability result does not require the assumption of the axiom of choice.
\begin{theorem}\label{Reg1}
We can decide whether a suffix recognizable graph \,$G$ \,of finite degree is a 
Cayley graph of a left-cancellative monoid, of a cancellative monoid, of a 
group, and whether \,$G$ \,is a generalized Cayley graph of a left-quasigroup, 
of a quasigroup, of a group.\\
In the affirmative, \,$G \,= \,{\cal C}\inter{V_G\,,\,\fleche{}_G(r)}$ \,where 
\,$\inter{s} = a$ \,for any \,$r\ \fleche{a}_G\ s$ \,and with a computable 
suitable binary operation on \,$V_G$ \,and vertex \,$r$.
\end{theorem}
\proof\mbox{}\\
{\bf i)} By Proposition~\ref{MSO} and Corollary~\ref{Dec2}, we can decide 
whether \,$G$ \,is rooted, simple, arc-symmetric, deterministic 
(resp. and co-deterministic) \,{\it i.e.} \,by Theorem~\ref{LeftMonoid} 
(resp. Theorem~\ref{Monoid}) whether \,$G$ \,is a Cayley graph of a 
left-cancellative monoid (resp. cancellative monoid).\\
In the affirmative and by Proposition~\ref{PathProduct}, 
$(V_G,\ast_r)$ \,is a left-cancellative (resp. cancellative) monoid where \,$r$ 
\,is a root of \,$G$, and $G \,= \,{\cal C}\inter{V_G\,,\,\fleche{}_G(r)}$ 
\,where \,$\inter{s} = a$ \,for any \,$r\ \fleche{a}_G\ s$. 
It remains to check that the path-operation \,$\ast_r$ \,is computable.\\
We just need that \,$G$ \,is deterministic and arc-symmetric.\\
Let \,$s,t \in V_G$\,. By Proposition~\ref{PathProduct}, \,$s \ast_r t$ \,is a 
vertex that we can determine. The label set 
\,L$_G(r,t) \ = \ \{\ u \in A_G^*\ |\ r\ \fleche{u}_G\ t\ \}$ \,of the paths 
from \,$r$ \,to \,$t$ \,is an effective non empty context-free language 
\cite{Ca1}\,: we can construct a pushdown automaton recognizing 
\,${\rm L}_G(r,t)$ \,hence we can compute a word \,$u \in {\rm L}_G(r,t)$.\\
Thus \,$s \ast_r t$ \,is the target of the path from \,$s$ \,labeled by \,$u$ 
\,{\it i.e.} \,$s\ \fleche{u}_G\ s \ast_r t$.\\[0.25em]
{\bf ii)} By Proposition~\ref{MSO} and Corollary~\ref{Dec2}, we can decide 
whether \,$G$ \,is connected, symmetric, deterministic and co-deterministic 
\,{\it i.e.} \,by Theorem~\ref{CayleyGroup}, whether \,$G$ \,is a Cayley graph 
of a group. In the affirmative and by Proposition~\ref{ChainProduct}, it 
remains to check that the chain-operation \,$\overline{\ast}_r$ \,is computable 
where \,$r$ \,is any vertex of \,$G$. 
We have seen that \,$_G\overline{\ast}_r \,= \,_{\overline{G}}\ast_r$\,. 
As \,$r$ \,is a root of \,$\overline{G}$ \,which is deterministic and 
arc-symmetric and by (i), \,$\overline{\ast}_r$ \,is computable.
\\[0.25em]
{\bf iii)} As \,$G$ \,has a decidable first order theory, we can decide whether 
\,$G$ \,is simple, deterministic, source-complete \,{\it i.e.} by 
Theorem~\ref{BoundedLeftQuasiCayley}, whether \,$G$ \,is a generalized Cayley 
graph of a left-quasigroup. In the affirmative and by 
Propositions~\ref{EdgeProduct} and \ref{EdgeProductLeftQuasi}, it remains to 
check that the edge-operation \,$_{\InfsupInd{G}}\croix_r$ \,is computable where 
\,$r$ \,is any vertex and \,$\Infsup{G}$ \,is the completion of \,$G$ 
\,defined in the proof of Theorem~\ref{BoundedLeftQuasiCayley}. 
This edge-operation that we denote by \,$\cdot$ \,has been defined for any 
\,$s,t \in V_G$ \,by\\[0.25em]
\hspace*{10em}\begin{tabular}{ll}
$s\ \fleche{a}\ s \cdot t$ & for \ $r\ \fleche{a}_G\ t$\\[0.25em]
$r\ \fleche{a}\ s \cdot t$ & for \ $s\ \fleche{b}_G\ t$ \ and 
\ $(a,b) \in \overline{\ell_s}-\ell_s$\\[0.25em]
$s \cdot t \,= \,t$  & for 
\ $t \in V_G - (\fleche{}_G(r) \,\cup \,\fleche{}_G(s))$
\end{tabular}\\[0.25em]
where \,$\ell_s \ = \ \{\ (a,b)\ |\ \exists\ t\ (r\ \fleche{a}_G\ t 
\,\wedge \,s\ \fleche{b}_G\ t)\ \}$ \,and to each injective function 
\,$\ell : A_G\ \fleche{}\ A_G$\,, we have associated a permutation 
\,$\overline{\ell}$ \,on \,$A_G$ \,extending \,$\ell$. 
Thus \,$\cdot$ \,is computable.\\
Moreover we can check that \,$\cdot$ \,is an effective ternary 
suffix-recognizable relation.\\[0.25em]
{\bf iv)} As \,$G$ \,has a decidable first order theory, we can decide whether 
\,$G$ \,is simple, deterministic, co-deterministic, source and target-complete 
\,{\it i.e.} \,by Corollary~\ref{BoundedQuasiCayley} and under the assumption 
of the axiom of choice, whether \,$G$ \,is a generalized Cayley graph of a 
quasigroup.\\
Assume that \,$G$ \,is simple, deterministic, co-deterministic, source and 
target-complete.\\
In fact, we do not need the assumption of the axiom of choice: we will define 
a computable quasigroup operation on \,$V_G$\,.\\
Let \,$r$ \,be any vertex of \,$G$. 
By Proposition~\ref{EdgeProductQuasi} (and as for the proof of 
Theorem~\ref{QuasiCayley}), it is sufficient to define a complete graph 
\,$H \,\supseteq \,G$ \,of vertex set \,$V_G$ \,with the same properties 
as~\,$G$ \,such that \,$_H\croix_r$ \,is computable.\\
When \,$V_G$ \,is finite and by Lemma~\ref{ColorRelation}, such a completion 
\,$H$ \,is effective hence \,$_H\croix_r$ \,is computable. 
We have to deal with the case where \,$V_G$ \,is infinite.\\
The set \,$V_G$ \,is a regular language over some finite alphabet that we order 
totally.\\
For any integer \,$i \geq 0$, we can compute the \,$i$-th vertex \,$v_i$ \,by 
length-lexicographic order.\\
We can assume that the finite label set \,$A_G$ \,is disjoint of \,$\entier$. 
Let us define a mapping \,$T$ \,from \,$\entier\croix\entier$ \,into 
\,$A_G \cup \entier$ \,such that the completion of \,$G$ \,is 
\,$H\ =\ \{\ v_i\ \fleche{T(i,j)}\ v_j\ |\ i,j \geq 0\ \}$.\\
By Lemma~\ref{Bijection} and Proposition~\ref{CompleteColoring}, $T$ \,is 
defined for any \,$i,j \geq 0$ \,by\\[0.25em]
\hspace*{1.5em}{$T(i,j) \ = \ \left\{\begin{tabular}{ll}
$a$ & if \ $(v_i,a,v_j) \in G$\\[0.25em]
min$\bigl(\entier - \{T(i,0),\ldots,T(i,j-1),T(0,j),\ldots,T(i-1,j)\}\bigr)$ & 
otherwise.
\end{tabular}\right.$}\\[0.25em]
Thus \,$T$ \,is computable hence also \,$_H\croix_r$\,.\\[0.25em]
{\bf v)} As \,$G$ \,has a decidable first order theory and by 
Corollary~\ref{Dec2}, we can decide whether \,$G$ \,is simple, symmetric, 
deterministic and co-deterministic \,{\it i.e.} \,by Theorem~\ref{MainFour} 
and under the assumption of the axiom of choice, whether \,$G$ \,is a 
generalized Cayley graph of a group.\\
Assume that \,$G$ \,is simple, symmetric, deterministic and co-deterministic.\\
In fact, we do not need the hypothesis of the axiom of choice: we will define 
a computable extended chain-operation.\\
As \,$G$ \,is a suffix-recognizable graph of finite degree over some alphabet 
\,$N$, it is of the form
\\[0.25em]
\hspace*{0em}$G\ =\ 
\bigcup_{i=1}^nW_i(u_i\ \fleche{a_i}\ v_i)$ \ where \,$n \geq 0$, 
\,$u_1,v_1,\ldots,u_n,v_n \in N^*$ \,and \,$W_1,\ldots,W_n \in {\rm Rec}(N^*)$.
\\[0.25em]
Thus \,$V_G \,= \,\bigcup_{i=1}^nW_i.\{u_i,v_i\}$ \,is a regular language and 
we denote by\\[0.25em]
\hspace*{10em}${\rm In}_G \,= \,\{u_1,\ldots,u_n\}$ \ \ and \ \ 
${\rm Out}_G \,= \,\{v_1,\ldots,v_n\}$.\\[0.25em]
Let \,$<_{_{ll}}$ \,be the length-lexicographic order extending a linear order on 
\,$N$.\\[0.25em]
Given languages \,$L,M \subseteq N^*$, the {\it left residual} and the 
{\it right residual} \,of \,$L$ \,by \,$M$ \,are the respective languages:
\\[0.25em]
\hspace*{3em}$M^{-1}L \,= \,\{\ v\ |\ \exists\ u \in M\ uv \in L\ \}$ \ \ and 
\ \ $L\,M^{-1} \,= \,\{\ u\ |\ \exists\ v \in M\ uv \in L\ \}$.\\[0.25em]
Let \,$C$ \,be a connected component of \,$G$. 
Thus \,$V_C$ \,is a regular language over \,$N$ \cite{Bu}.\\
The left residual of \,$C$ \,by a word \,$w \in N^*$ \,is the graph 
\,$w^{-1}C \ = \ \{\ u\ \fleche{a}\ v\ |\ wu\ \fleche{a}_C\ wv\ \}$.\\
Let \,$w \in V_C\,({\rm In}_G \cup {\rm Out}_G)^{-1}$ \,of minimal length. 
Thus \,$w$ \,is prefix of any vertex of \,$C$ \,and is denoted by $_{_C}w$. 
Let \,$z_{_C} \in {\rm In}_G \cup {\rm Out}_G$ \,be the minimal word by 
length lexicographic order such that \,$_{_C}w.z_{_C} \in V_C$\,. 
Note that such a representative \,$_{_C}w.z_{_C}$ \,of \,$V_C$ \,can be of 
non-minimal length and the words \,$_{_C}w\,,\,z_{_C}$ \,are computable from 
any vertex of \,$C$. These words form the following representative set of 
\,${\rm Comp}(G)$\,:\\[0.25em]
\hspace*{10em}$P \ = \ \{\ _{_D}w.z_{_D}\ |\ D \in {\rm Comp}(G)\ \}$.\\[0.25em]
This set is an effective regular language. 
Indeed, let \,$D$ \,be a connected component of \,$G$ \,with 
\,$z_{_C} \,= \,z_{_D}$\,. Thus \,$(_{_C}w)^{-1}C \,= \,(_{_D}w)^{-1}D$ \,and we 
denote by \,$W_{z_{_C}} \,= \,(_{_C}w)^{-1}V_C$\,.\\
The set \,$Z\ =\ \{\ z_{_D}\ |\ D \in {\rm Comp}(G)\ \}\ \subseteq\ 
{\rm In}_G \cup {\rm Out}_G$ \,is finite hence\\[0.25em]
\hspace*{3em}$P \ = \ \bigcup_{z \in Z}\,(V_G \cap N^*z) - N^*(W_z - \{z\})$ \
is an effective regular language.\\[0.25em]
Let \,Rk$(u) \,= \,|\{\ v \in P \mid v <_{_{ll}} u\ \}|$ \,be the rank of 
\,$u \in P$ \,according to \, $<_{_{ll}}$ \ {\it i.e.} \,$u$ \,is the 
\,Rk$(u)$-word in \,$P$ \,by \,$<_{_{ll}}$\,. 
We have a group \,$(P,+)$ \,for \,$u+v$ \,defined for any \,$u,v \in P$ \,by
\\[0.25em]
\hspace*{6em}${\rm Rk}(u+v) \ = \ {\rm Rk}(u) + {\rm Rk}(v) \ 
\,({\rm mod}\ |P|)$ \ \ for \,$P$ \,finite,\\[0.25em]
and for \,$P$ \,countable, we consider the bijection 
\,$\norme{\ } : P\ \fleche{}\ \relatif$ \,defined for any \,$u \in P$ \,by
\\[0.25em]
\hspace*{6em}{$\norme{u} \ = \ \left\{\begin{tabular}{ll}
$\frac{{\rm Rk}(u)}{2}$ & if \ ${\rm Rk}(u)$ \,is even,\\[0.25em]
$-\frac{{\rm Rk}(u)+1}{2}$ & if \ ${\rm Rk}(u)$ \,is odd
\end{tabular}\right.$}\\[0.25em]
and we define \,$u+v \in P$ \,by \,$\norme{u+v} \,= \,\norme{u} + \norme{v}$.\\
Let \,$x \in V_G$\,. It is connected to the unique 
\,$v_x \,= \,_{_D}w.z_{_D} \in P$ \,for \,$D \in {\rm Comp}(G)$ \,and
\,$x \in V_D$\,. The label set \,${\rm L}_x\ =\ 
\{\ u \in (A_G \,\cup \,\overline{A_G} )^*\ |\ v_x\ \fleche{u}_G\ x\ \}$ \,of 
the chains between \,$v_x$ \,and \,$x$ \,is an effective context-free 
language.\\
As \,$G$ \,is symmetric, the extended chain-operation \,$x \cdot y$ \,has been 
defined by \,$v_{x + y}\ \fleche{\ell_x\ell_y}_G\ x \cdot y$ \,with 
\,$\ell_x \in {\rm L}_x$ \,and \,$\ell_y \in {\rm L}_y$\,. 
Thus \,$\cdot$ \,is an effective group operation.
\qed\\[1em]
We can consider the generalization of Theorem~\ref{Reg1} to all the 
suffix-recognizable graphs (allowing vertices of infinite degree) which form
the first level of a stack hierarchy for which any graph has a decidable
monadic theory \cite{CW}. 
To extend Theorem~\ref{Reg1} to any graph of this hierarchy, we have to decide
on the arc-symmetry (resp. symmetry) when these graphs are deterministic 
(resp. and co-deterministic).\\
The decidability result given by Theorem~\ref{Reg1} is a first application of
the Cayley graph characterizations presented in this paper. 
Another application is to describe differently a generalized Cayley graph 
by defining another operation on its vertex set. 
A trivial example is given by the quasigroup \,$(\relatif,-)$ \,of right
identity \,$0$. 
Its Cayley graph \,${\cal C}(\relatif)$ \,is strongly connected, symmetric, 
deterministic and co-deterministic. Its path-operation from \,$0$ \,is 
\,$\ast_0 \,= \,+$ \,hence by Theorem~\ref{RestrictedGroup}, it is equal to 
\,${\cal C}\inter{\relatif}$ \,for the group \,$(\relatif,+)$ \,with 
\,$\inter{n} = -n$ \,for any \,$n \in \relatif$. 
In particular for any \,$P \subseteq \relatif$, the generalized Cayley graph 
\,${\cal C}\inter{\relatif,P}$ \,of the quasigroup \,$(\relatif,-)$ \,is equal 
to the generalized Cayley graph \,${\cal C}\inter{\relatif,-P}'$ \,of the group 
\,$(\relatif,+)$ \,with \,$\inter{-n}' = \inter{n}$ \,for any \,$n \in P$. 
Similarly by Theorem~\ref{BoundedLeftQuasiCayley}, any finitely labeled 
generalized Cayley graph \,$G$ \,of a left-cancellative magma is a 
generalized Cayley graph of a left-quasigroup and its operation is computable 
for \,$G$ \,end-regular.

\section{Conclusion}

We obtained simple graph-theoretic characterizations for Cayley graphs of 
elementary algebraic structures. We have shown the effectiveness of these
characterizations for infinite graphs having a structural regularity. 
This is only a first approach in the structural description and its 
effectiveness of Cayley graphs of algebraic structures.
% Quoi de plus naturel de caractériser des graphes par de simples propriétés
% structurelles. 
% Pourquoi a-t-il fallu attendre plus d'un siècle pour donner une description
% structurelle élémentaire des graphes de Cayley des groupes ? de monoïdes ? ...
% Il y a tant de travail intéressant à faire sur le sujet.
% Ce n'est vraiment qu'un début.
\newpage\noindent
\bibliographystyle{alpha}

\end{document}